\documentclass[review, a4paper,12pt]{elsarticle}
\usepackage[left=0.5in, right=0.5in, top=1in, bottom=1in]{geometry}

\usepackage{lineno,hyperref}

\usepackage{amsmath}
\usepackage{amssymb}
\usepackage{bbm}
\usepackage{graphicx}
\usepackage{accents}
\usepackage{amsthm}
\usepackage{color}
\usepackage{graphics}
\graphicspath{ {./figures/} }
\usepackage{caption}
\usepackage{subcaption}
\usepackage{epsfig}
\usepackage{cancel}
\usepackage{epstopdf}
\usepackage{hyperref}
\usepackage{xcolor}
\hypersetup{
colorlinks,
linkcolor={blue!50!black},
citecolor={blue!50!black},
urlcolor={blue!80!black}
}
\usepackage{natbib}
\usepackage{stmaryrd} 

\newcommand{\mbf}[1]{\mathbf{#1}}
\newcommand{\bbm}[1]{\mathbbm{#1}}
\newcommand{\mcal}[1]{\mathcal{#1}}
\newcommand{\bsym}[1]{\boldsymbol{#1}}
\newcommand{\bbbm}[1]{\bsym{\bbm{#1}}}

\newcommand{\wt}[1]{\widetilde{#1}}
\newcommand{\wh}[1]{\widehat{#1}}
\newcommand{\pd}[2]{\frac{\partial #1}{\partial #2}}

\newcommand{\fd}[2]{\frac{\mathrm{d} #1}{\mathrm{d} #2}}

\newtheorem{remark}{Remark}

\DeclareMathOperator{\CGleft}{\bsym{b}}
\DeclareMathOperator{\CGright}{\bsym{C}}
\DeclareMathOperator{\Dgrad}{\bsym{F}}

\DeclareMathOperator{\bod}{\mcal{B}}
 
\DeclareMathOperator{\strPK}{\mbf{P}} 
\DeclareMathOperator{\strC}{\bsym{\sigma}} 

\DeclareMathOperator{\elecR}{\bbbm{E}} 
\DeclareMathOperator{\elecC}{\bbbm{e}} 
\DeclareMathOperator{\disR}{\bbbm{D}} 
\DeclareMathOperator{\disC}{\bbbm{d}} 
\DeclareMathOperator{\potR}{\bbbm{A}} 

\DeclareMathOperator{\Curl}{\text{Curl}}

\DeclareMathOperator{\vr}{\varrho}
\DeclareMathOperator{\wtr}{\wt{\varrho}}
\DeclareMathOperator{\wte}{\wt{\eta}}

\DeclareRobustCommand{\rchi}{{\boldsymbol{\mathpalette\irchi\relax}}}
\newcommand{\irchi}[2]{\raisebox{\depth}{$#1\chi$}} 

\usepackage{mdframed} 
\newmdtheoremenv[%
linecolor=gray,leftmargin=0,%
rightmargin=0,
backgroundcolor=gray!10,%
innertopmargin=0pt,%
ntheorem]{sidenote}{Note}[section]

\modulolinenumbers[1]

\journal{Journal}

\usepackage{numcompress}

\begin{document}

\begin{frontmatter}

\title{Coupled electro-elastic deformation and instabilities of a toroidal membrane}


\author[GLA]{Zhaowei Liu}

\author[GLA]{Andrew McBride}

\author[UTTAR]{Basant Lal Sharma}

\author[GLA,FAU]{Paul Steinmann}

\author[GLA]{Prashant Saxena\corref{corr}}
\cortext[corr]{Corresponding author}
\ead{prashant.saxena@glasgow.ac.uk}

\address[GLA]{Glasgow Computational Engineering Centre, James Watt School of Engineering, University of Glasgow, Glasgow, G12 8LT, United Kingdom}

\address[UTTAR]{Department of Mechanical Engineering, Indian Institute of Technology Kanpur, Kanpur, Uttar Pradesh-208016, India}

\address[FAU]{Chair of Applied Mechanics, University of Erlangen--Nuremberg, Paul-Gordan-Str. 3, D-91052, Erlangen, Germany}

\begin{abstract}
%

We analyse here the problem of large deformation of dielectric elastomeric membranes under coupled electromechanical loading. 
Extremely large deformations (enclosed volume changes of $100$ times and greater) of a toroidal membrane are studied by the use of a variational formulation that accounts for the total energy due to mechanical and electrical fields.
A modified shooting method is adopted to solve the resulting system of coupled and highly nonlinear ordinary differential equations.
We demonstrate the occurrence of limit point, wrinkling, and symmetry-breaking buckling instabilities in the solution of this problem.
Onset of each of these ``reversible'' instabilities depends significantly on the ratio of the mechanical load to the electric load, thereby providing a control mechanism for state switching. 
\end{abstract}


\end{frontmatter} 

\linenumbers

\section{Introduction}
Thin electroelastic structures made from electroactive polymers \cite{pelrine2000high} find wide use in engineering applications including artificial muscles~\cite{bar2001electroactive}, soft grippers~\cite{araromi2014rollable, kofod2007energy, lau2017dielectric}, and energy generators~\cite{Moretti2019}.
 In this work we use the theory of nonlinear electroelasticity \cite{dorfmann2005nonlinear} to analyse the large deformation of a toroidal electroelastic membrane inflated by a mechanical pressure and actuated by an electric potential difference applied across its thickness.
Extreme deformations induce limit point, wrinkling, and symmetry-breaking buckling instabilities in the membrane. {These are systematically studied in this contribution}.

\subsection{Nonlinear electroelasticity}
Developments in the theory of electroelasticity date back to the classic work of \citet{toupin1956elastic}. 
By combining the theory of continuum mechanics and electrostatics, a framework was established for analysing the nonlinear response of isotropic dielectric materials. 
 \citet{toupin1963dynamical} extended his seminal work by deriving the governing equations for the dynamics of elastic dielectrics. 
 \citet{eringen1963foundations} followed by formulating the governing equations in an alternative way and applied his method to the problem of an incompressible thick-walled cylindrical tube subjected to a radial electric field. 
 \citet{tiersten1978perturbation} simplified the formulations proposed by~\citet{toupin1956elastic} and~\citet{eringen1963foundations}, thereby making a fundamental contribution to nonlinear electroelastic material modelling. 
 He also made significant contributions to the theory of piezoelectricity~\cite{tiersten1978perturbation, tiersten1981electroelastic}. 
 
 Interest in the development of nonlinear theories of electroelasticity was renewed owing to the development and industry adoption of dielectric elastomers that can undergo large deformations and nonlinear electroelastic coupling.
 \citet{mcmeeking2005electrostatic} used the principle of virtual work to derive simpler governing equations for quasi-electrostatics in Eulerian form. 
Concurrently,  \citet{dorfmann2005nonlinear, dorfmann2006nonlinear} developed a general Lagrangian formulation and constitutive relations within the framework of continuum mechanics to simulate the finite deformation of electroelastic materials coupled to electric fields. 
 Rate-dependent theories to account for dissipation owing to viscoelasticity in dielectric elastomers were developed by \citet{Ask2012} and \citet{Saxena2014a}.
 Variational formulations of electroelasticity to enable the development of computational methods were presented by \citet{vu2007numerical, Liu2013a},
  and have been applied to analyse stability by computation of higher variations by \citet{bustamante2009nonlinear}, and \citet{saxena2019equilibrium}.
%
 A {comprehensive} review of the theory of nonlinear electroelasticity and its applications is presented in  \citep{dorfmann2017nonlinear}.

\subsection{Dielectric elastomers}

Electroactive polymers are materials that can undergo deformation due to an applied electric field.
Dielectric elastomers are one of the most commonly used electroactive polymers.
They are composed of a soft elastomer sandwiched between two compliant electrodes.
Application of a potential difference between the two electrodes results in a large deformation in the elastomer due to the electrostatic forces generated by the opposite electric charges \cite{pelrine2000high}. 

This principle has been widely used in the design of sensing and actuating systems. For example, \citet{bar2001electroactive} explored their use as artificial muscles.
\citet{Moretti2019} developed wave energy generators based on the inflation of dielectric elastomers. 
\citet{kofod2006self} presented the principle of self-organized dielectric elastomer minimum energy structures (DEMESs) and developed a gripper~\cite{kofod2007energy}.
\citet{araromi2014rollable} applied DEMESs as a gripper to capture debris in space. 
\citet{lau2017dielectric} developed a dielectric elastomer finger for grasping and pinching highly deformable objects.
For a detailed review on grippers made of dielectric elastomers, see~\citet{shintake2018soft}.
\citet{Ozsecen2010} developed haptic interfaces, \citet{Michel2008} performed a feasibility study for a bionic propulsion system, and \citet{O'Halloran2008} explored sensing systems based on dielectric elastomers. {We refer to the review papers by \citet{suo2010theory} and \citet{lu2020mechanics} for further detailed discussion on dielectric elastomers.}

{
\subsection{Biological and engineering applications with toroidal membranes}
\label{sec: torus applications}
Biological membranes with a toroidal shape also naturally exist in human body. One of the most important cells in the human body, Erythrocyte (red blood cell), is a small dielectrophoretic (DEP) electromagnetic field (EMF) driven cell~\cite{purnell2018bio}. It has a unique toroidal shape in order to increase the surface area-to-volume ratio which increases the diffusion of oxygen and carbon dioxide through their cell membrane. Toroidal structures are common in DNA-binding enzymes~\cite{hingorani1998toroidal} and bacteriophages~\cite{leo2011toroidal}. Further discussion on natural biological toroidal structures can be found in~\cite{di2014dielectric}. Topologically, a torus can be considered as the simplest example of a genus 1 orientable surface which appeals to mathematicians and engineers. Toroidal membranes and shells are widely applied in engineering such as tyres, air springs~\cite{de2018simulation}, soft grippers and inflatable actuators. \citet{zang2020bionic} designed a bionic toroidal soft gripper in order to catch objects with arbitrary shapes and sizes. \citet{adams2018water} designed an electro-pneumatic device using a string of inflatable toroidal membranes for water pipe inspection.
Thus, analysing the mechanics of large deformation in toroidal membranes can not only aid the engineering design of such devices but also develop a better understanding of certain biological processes at the cellular level.
}
 
\subsection{Instabilities in nonlinear membranes}
\label{sec: membrane instability}
Nonlinear membranes are widely applied in engineering structures and naturally appear in the form of biological tissues. 
Air bags, diaphragm valves, balloons, skin tissue, and cell walls are examples of nonlinear membranes. 

Inflation can cause large deformation in the membranes resulting in instabilities.
A well-known instability phenomenon of inflating membranes is the limit point.
This is a critical point after which the membrane appears to lose stiffness to inflation and undergoes very large inflation with a small increase in pressure.
This phenomenon is also called snap-through bifurcation and has been well studied~\citep[see e.g.][]{benedict1979determination, carroll1987pressure, khayat1992inflation, muller2002inflating,tamadapu2013geometrical}.  
Computation of accurate pressure-volume characteristics in this case require a path-following scheme due to the non-uniqueness of solution \cite{Reddy2017}.

In-plane deformation of membrane can also result in wrinkling which is a form of localised buckling.
An ideal membrane is a structure with negligible bending stiffness and can only sustain tensile loading.
If any part of the membrane structure experiences compression, it  undergoes local out-of-plane deformation to avoid the in-plane compressive stresses.
Tension field theory developed by \citet{Pipkin1986} and \citet{Steigmann1990} is a widely used tool to model wrinkles in nonlinear elastic membranes.
It assumes zero bending stiffness and an infinitely continuous distribution of wrinkles orientated in the direction of the positive principal stress.
To avoid a contribution to the energy by compressive stresses, a relaxed energy function is used that constrains the stress tensor to be positive semi-definite.
As a result the amplitude and wavelength of wrinkles cannot be computed by using this theory.
This theory has been applied to study wrinkles in axisymmetric hyperelastic membranes \cite{Li1995a, Li1995b} and to model wrinkles in skin during wound closure \cite{Swain2015}, to name a few applications.
A generalisation of the tension field theory has been attempted (although without a rigorous mathematical proof) for the case of electroelasticity by \citet{de2010pull, DeTommasi2011, Greaney2019}, and for the case of magnetoelasticity by \citet{Reddy2017, reddy2018instabilities, saxena2019magnetoelastic}. 
\citet{Wong2006} developed an analytical method to quantify the location, amplitude, and wavelengths of linear elastic membranes.
%
%
\citet{nayyar2011stretch} and \citet{10.1115/1.4031243} developed a nonlinear finite element method for simulating stretch-induced wrinkling of hyperelastic thin sheets. 




While wrinkling is a localised buckling, the membrane structure can also experience a global buckling on account of large deformations.
In structures with a geometrical symmetry, this instability manifests as a bifurcation from the symmetric principal solution and leads to a loss of symmetry.
The theory of elastic buckling, developed by \citet{koiter1970stability} and~\citet{budiansky1974theory}, provides methods to evaluate the critical point of such instability.
This typically requires  checking the sign of the second variation of the total potential energy to determine the stability state.
\citet{Chaudhuri2014} studied the perturbed deformations of inflated hyperelastic circular membranes, \citet{Venkata2020} analysed buckling of hyperelastic toroidal membranes, \citet{Xie2016} analysed the shape bifurcations of a dielectric elastomeric sphere through a direct perturbation approach, and \citet{Reddy2017, reddy2018instabilities, saxena2019magnetoelastic} analysed shape bifurcations of magnetoelastic membranes.

\subsection{Electroelastic instability}

Experimental investigation of dielectric elastomeric membranes have revealed all the three instabilities discussed in Section~\ref{sec: membrane instability}.
%
%
An experimental investigation of electroelastic membranes by \citet{kollosche2012complex} demonstrates an interplay between the limit point and wrinkling instabilities due to coupling effects.
\citet{li2013giant} studied large voltage-induced deformation of dielectric elastomers. 
In addition to wrinkling and limit point, they also demonstrate symmetry-breaking and bulge formation in the inflation of a circular membrane.
Careful experimental investigations on the rate-dependent behaviour of dielectric elastomers have shown a relation between the viscoelastic reponse and the breakdown limit \cite{Ahmad2020, Hossain2014, MEHNERT2019103797}.
\citet{Zhang2016} and \citet{Mao2018} presented a controlled experimental procedure to produce wrinkles in dielectric elastomer membranes.

Theoretical and computational procedures to model these electroelastic instabilities  have been developed largely for bulk media with some recent works towards the analysis of membranes.
\citet{zhao2007method} analysed the instability of dieletric elastomers  to guide the design of actuator configurations and materials.  \citet{rudykh2012snap} presented a method to use snap-through instability of thick-wall electroactive balloons to design actuators.
\citet{miehe2015computational} developed an algorithm for finite element computations of both structural and material stability analysis in electroelasticity.
\citet{dorfmann2014instabilities} studied the critical stretch corresponding to loss of stability of a thick electroelastic plate by perturbation of the equilibrium equations. 
\citet{dorfmann2014nonlinear} also investigated radial deformations of a thick-walled spherical shell using the nonlinear electroelastic theory.
\citet{Melnikov2018} presented a mathematical approach to study bifurcation of a finitely deformed thick-walled cylindrical tube.
A more complete set of references can be found in the comprehensive review on the instability of soft dielectrics by \citet{Dorfmann2017c}.

\citet{Xie2016} undertook a bifurcation analysis of a spherical dielectric membrane under inflation and also derived post-buckling solutions.
\citet{Greaney2019} used a modified tension field theory to analyse wrinkling and pull-in instabilities in dielectric membranes.
For the related problem of magnetoelastic membranes, \citet{Reddy2017} recently demonstrated new instability phenomena of an additional limit point and reversal of wrinkling location in an inflating toroidal magnetoelastic membrane.
In addition to the engineering and biological applications discussed in Section \ref{sec: torus applications}, these results provided a partial motivation to study instabilities in a toroidal electroelastic membranes.
In this contribution we, for the first time, present a novel analysis of the interaction between limit point, wrinkling, and buckling instabilities in a toroidal membrane.
For the wrinkling analysis we apply the tension field theory of \citet{Steigmann1990} generalised to electroelasticity by \citet{de2010pull} and \citet{Greaney2019}. We extend the aforementioned theoretical works by providing the numerical details to apply the tension field theory and thereby deliver physically meaningful predictions of wrinkled solutions post-instability. These predictions are compared to primary solutions that ignore the inability of a membrane to sustain compressive states.
%

\subsection{Organisation of the manuscript}
This contribution is organised as follows:
Section~\ref{sec: kinematics} introduces the kinematics of the deformation. 
Section~\ref{sec:energy_consideration} formulates the equilibrium equations using the first variation of the total potential energy functional that is composed of mechanical and electrical contributions. The Mooney-Rivlin model~\cite{mooney1940theory, rivlin1948large} is adopted for the hyperelastic energy density and  the electrical contribution is accounted for via the energy density function by coupling the electric displacement to the deformation tensor. 
Three instabilities (snap-through, wrinkling and loss of symmetry) are analysed and discussed in detail.
Computations of wrinkling using the energy relaxation method are presented in Section~\ref{sec:wrinkling_instability_analysis}. Section~\ref{sec:loss_symmetry} describes the second variation analysis of the energy function to consider the loss of symmetry in the circumferential direction of the torus. 
Section~\ref{sec:numerical_results} presents several numerical examples to elucidate the theory. Conclusions are presented in Section~\ref{sec: conclusions}.
{
\subsection{Notation}
\paragraph{Brackets}
Three types of brackets are used. Square brackets $[\quad]$ are used to clarify the order of operations in an algebraic expression. Curly brackets $\{\quad\}$ define a set and circular brackets $(\quad)$ are used to define the parameters of a function. 
If brackets are used to denote an interval then $(\quad)$ stands for an open interval and $[\quad]$ a closed interval.
\paragraph{Symbols}
A variable typeset in a normal weight font represents a scalar. A bold weight font denotes a vector or a second-order tensor. An upper-case bold letter denotes a vector or tensor in the reference configuration and a lowercase bold letter denotes a vector or tensor in the current (deformed) configuration. A tensor directly enclosed by square brackets, for example $[\mbf{A}]$, denotes the matrix representation of the tensor in a selected coordinate system.  A subscript denotes the partial derivative with respect to the field. For example, consider a function $A\big(a,b(a),c(a)\big)$. $A_b$ denotes the partial derivative of $A$ with respect to $b$, which is equivalent to $\pd{A}{b}$. $\fd{A}{a}$ is the full derivative with respect to $a$, given by
\begin{equation*}
\fd{A}{a} = \pd{A}{a} + \pd{A}{b}\pd{b}{a} + \pd{A}{c}\pd{c}{a}.
\end{equation*}
\paragraph{Functions}
 $\text{det}(\mbf{A})$ denotes the determinant of the second-order tensor $\mbf{A}$. $\text{diag}(a,b,c)$ denotes a second-order tensor with only diagonal entries $a,b$ and $c$. }

\section{Kinematics}
\label{sec: kinematics}
\begin{figure}
\centering
\begin{subfigure}[b]{0.55\linewidth}
\centering
	\includegraphics[width=\linewidth]{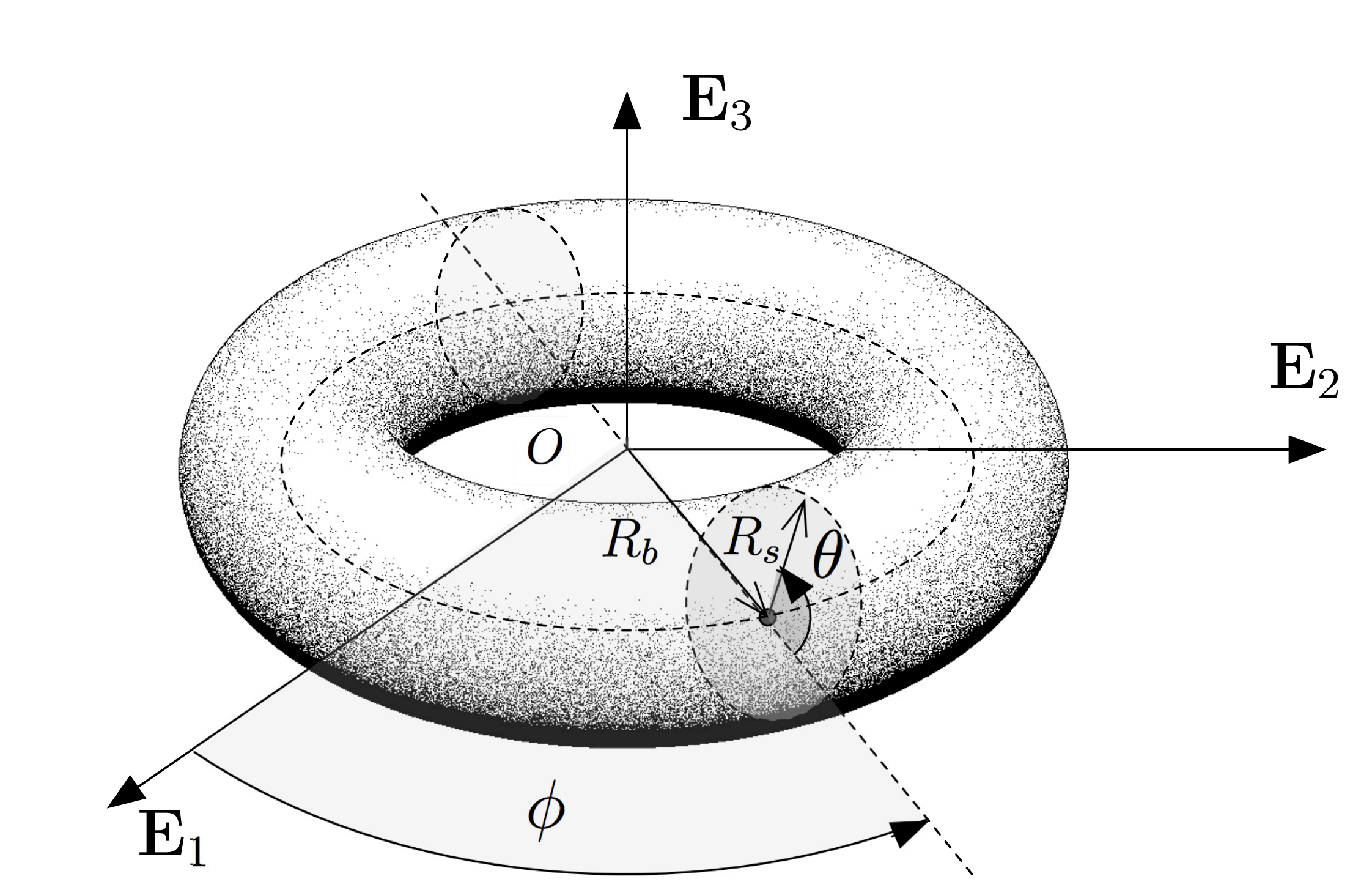}
	\caption{}
	\label{fig:toroidal_membrane_reference}
\end{subfigure}
\begin{subfigure}[b]{0.4\linewidth}
\centering
	\includegraphics[width=\linewidth]{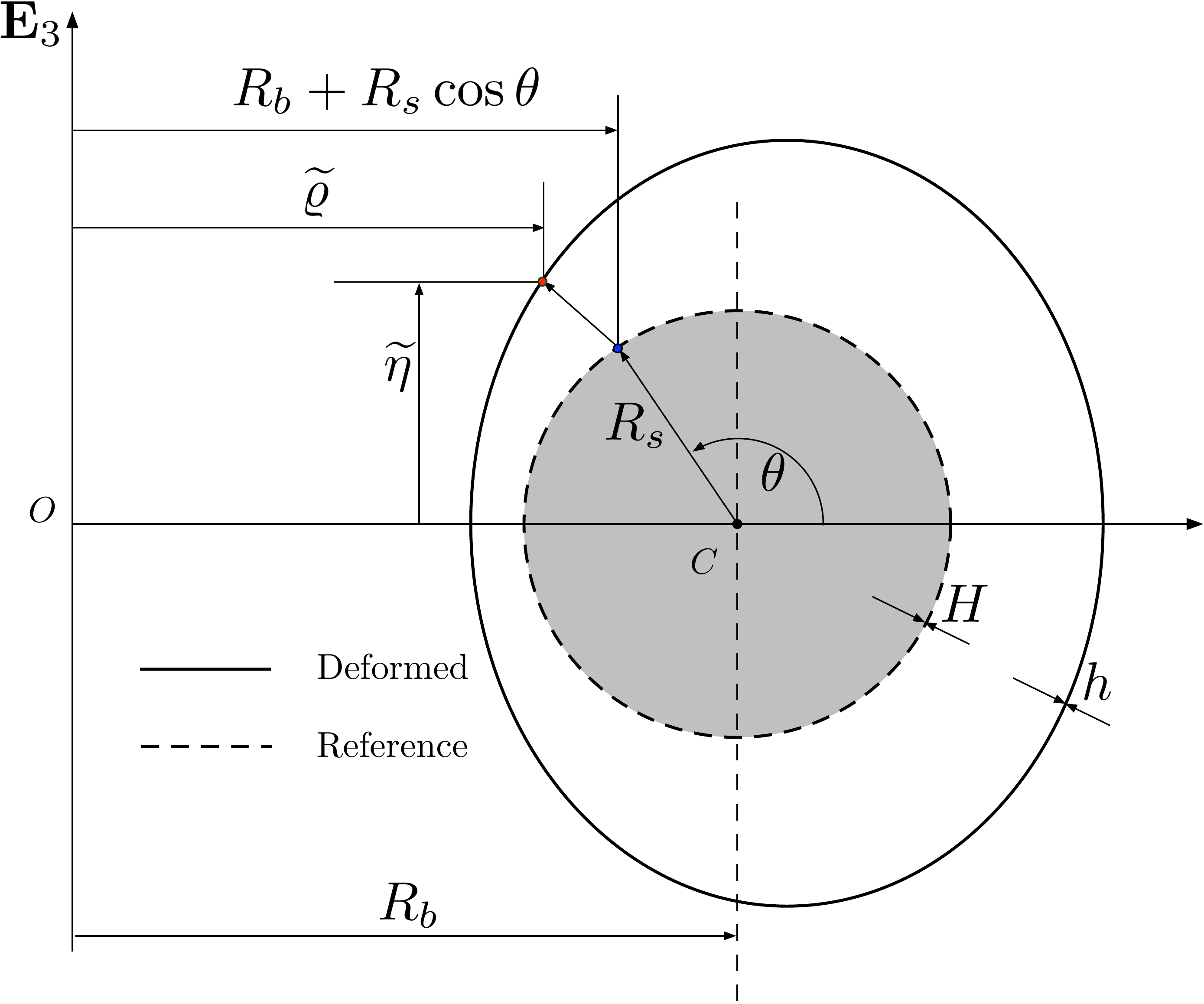}
	\caption{}
	\label{fig:}
\end{subfigure}
\begin{subfigure}[b]{0.35\linewidth}
\centering
	\includegraphics[width=\linewidth]{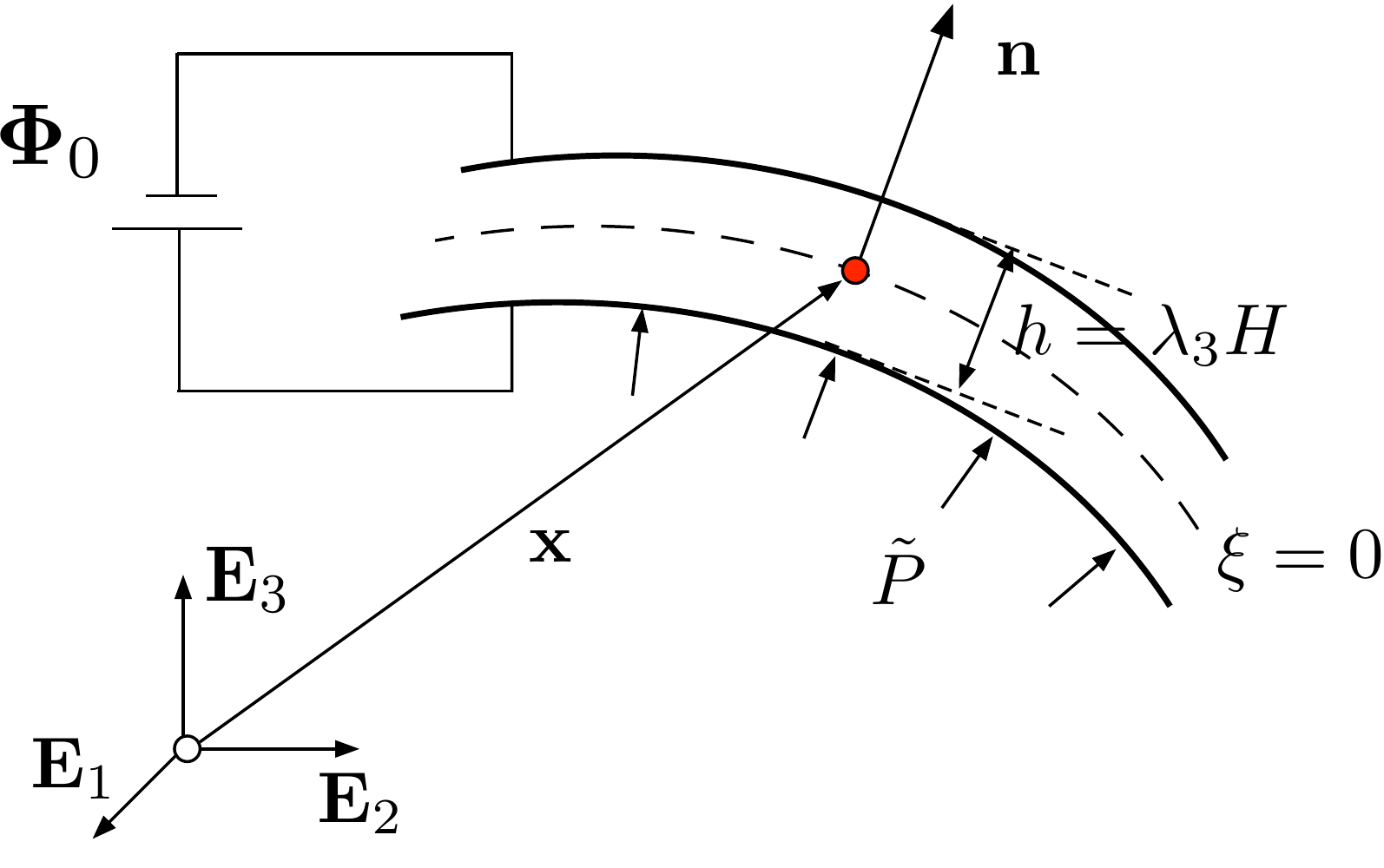}
	\caption{}
	\label{fig:flesh_thedeformed_meembrane}
\end{subfigure}
	\caption{(a) The reference configuration of a toroidal membrane with circular cross-section highlighted. (b) The reference and deformed configurations of a cross section of the toroidal membrane. (c) Close-up of part of the deformed membrane.}
	\label{fig:toroidal_geometry}
\end{figure}

Consider the toroidal membrane in Figure \ref{fig:toroidal_geometry} with  major and minor radii $R_b$ and $R_s (<R_b)$, respectively. 
The initial thickness of the membrane $H$ is assumed constant, where $H/R_s \ll 1$.
The geometry and kinematics of this system are similar to previously studied problems \cite{Reddy2017, Venkata2020}.
The torus is inflated by an internal pressure $\wt{P}$ and an electric potential difference $\Phi_0$ is applied across its thickness. The membrane is assumed to be incompressible. 
\subsection{Reference configuration}
The position vector $\mathbf{X}$ of a point in the undeformed toroidal membrane  is given by
\begin{align}\label{eqn:x_position}
\mathbf{X}(\theta, \phi, \xi) = &\left[R_b + [R_s + \xi]\cos\theta\right]\cos\phi\, \mathbf{E}_1 + \left[R_b + [R_s + \xi]\cos\theta\right]\sin\phi\, \mathbf{E}_2 \nonumber \\
 &+ \left[R_s + \xi\right]\sin\theta\, \mathbf{E}_3 \, ,
\end{align}
where $\xi$ is the distance of the point from the mid-surface (defined by $\xi = 0$) of the membrane along the radius. 
The set $\{\mathbf{E}_i\}$ of orthonormal vectors  is the basis corresponding to the $(Y^1, Y^2, Y^3)$ coordinate system with origin $O$. 
The components of the covariant metric tensor with respect to the local (curvilinear) system $(\theta, \phi, \xi)$ (see Figure \ref{fig:toroidal_geometry})  for the membrane in its reference (perfect torus) configuration,  are given by
\begin{equation}\label{eqn:metric_undef}
[\mbf G] =
\begin{bmatrix}
R_s ^2 & 0 & 0\\
0 & [R_b + R_s \cos \theta]^2 & 0\\
0 & 0 & 1\\
\end{bmatrix} ,
\end{equation}
with determinant $G = \text{det}\mbf G = R_s^2 [R_b + R_s \cos \theta]^2$. 
Here, we have used the small thickness assumption ($H \ll R_s < R_b$).

\subsection{Deformed configuration}
Let the position vector of the deformed mid-surface be given by $\mbf{x}$ (corresponding to $\mathbf{X}$ in the reference configuration) with the unit outward normal vector $\mbf{n}$.
It can be shown that 
\begin{equation}
 \mbf{x} = \wtr \cos \phi \mbf{E}_1 + \wtr \sin \phi \mbf{E}_2 + \wte \mbf{E}_3 \, , \label{eqn: deformed position vector}
\end{equation}
where $\wtr: [0,2\pi) \times [0,2\pi) \to \bbm{R}$ and $\wte: [0,2\pi) \times [0,2\pi) \to \bbm{R}$ are functions depending on $\theta$ and $\phi$ that describe the position of points on the mid-surface as shown in Figure \ref{fig:flesh_thedeformed_meembrane}. 
Components of the three-dimensional covariant metric tensor for the deformed configuration are given by
\begin{equation}
 [\mbf{\wt{g}}] = \begin{bmatrix}
           \wtr_\theta ^2 + \wte_\theta ^2 & \wtr_\theta\wtr_\phi +\wte_\theta\wte_\phi  & 0 \\
           \wtr_\theta\wtr_\phi +\wte_\theta\wte_\phi & \wtr_\phi^2 + \wtr^2 + \wte_\phi^2 & 0 \\
           0 & 0 & \lambda_3^2
          \end{bmatrix} ,
\end{equation}
where $\lambda_3 = h/H$ with $h$ being the thickness of the torus in the deformed configuration. 
We further introduce the dimensionless quantities
\begin{equation}
 \gamma = \frac{R_s}{R_b}, \quad \vr = \frac{\wtr}{R_b}, \quad \eta = \frac{\wte}{R_b}, \label{eqn: non-dimensionalization}
\end{equation}
where the parameter $\gamma$ describes the aspect ratio of the undeformed torus.
The two principle stretches $\lambda_1$ and $\lambda_2$ can be expressed using the reference and deformed covariant metric tensors as
\begin{align}
\lambda_1^2 &=  \mathcal{P} + \mathcal{Q}, \nonumber \\
\lambda_2^2 &=  \mathcal{P} - \mathcal{Q} ,
\label{eqn:lambda1_2}
\end{align}
where
\begin{align*}
\mathcal{P} &= \frac{1}{2} \left[ \frac{\vr^2_\theta + \eta^2_\theta}{\gamma^2} + \frac{\vr^2_\phi + \eta^2_\phi + \vr^2}{[1+\gamma \cos \theta]^2} \right] , \\
\mathcal{Q} &= \frac{1}{2} \sqrt{\left[ \frac{\vr^2_\theta + \eta^2_\theta}{\gamma^2} - \frac{\vr^2_\phi + \eta^2_\phi + \vr^2}{[1+\gamma \cos \theta]^2} \right]^2 + 4\left[ \frac{\vr_\theta \vr_\phi + \eta_\theta \eta_\phi}{\gamma [ 1 + \gamma \cos \theta]} \right]^2}.
\end{align*}

The deformation map $\rchi$, from the undeformed configuration to the deformed configuration, is defined via 
$$
\mbf{x} = \rchi(\mbf{X}) ,
$$
and the corresponding deformation gradient is defined by
$$
\Dgrad(\mbf{X}) := \text{Grad} \,\rchi(\mbf{X}).
$$
{ Henceforth, for notational convenience, it is assumed that $\mbf{X}$ and $\mbf{x}$ are related as above to map the evaluation at $\mbf{x}$ of quantities defined on the deformed configuration to their counterparts at $\mbf{X}$ on the undeformed configuration.}
%
The incompressibility constraint is given by 
$$
J = \mathrm{det}(\Dgrad) = 1,
$$
as a consequence of which, the third principle stretch follows as
\begin{equation}
\lambda_3^2 =\frac{1}{\lambda_1^2\lambda_2^2} = \frac{\gamma^2[1+\gamma \cos \theta]^2}{[\vr_\theta \eta_\phi - \vr_\phi \eta_\theta]^2 + \vr^2[\vr^2_\theta + \eta^2_\theta]}.
\label{eqn:lambda3}
\end{equation}
The right Cauchy--Green deformation tensor $\CGright = \Dgrad^\top \Dgrad$ is given in the local coordinate system of the torus by
\begin{equation}
 [\CGright] = \mathrm{diag} \left( \lambda_1^2 , \lambda_2^2, \lambda_3^2 \right). 
 \label{eq: CG right expr}
\end{equation}

\begin{remark}
If the solution is  symmetric with respect to $\mbf{E}_3$ along the $\phi$ (azimuthal) direction (which happens to be the case for the principal solution as shown in Section~\ref{sec:equilibrium}.),
the first two principle stretches can be simplified 
as
\begin{equation}
\lambda_1 = \frac{1}{\gamma} \left[ \vr_\theta ^2 + \eta_\theta ^2 \right]^{1/2}, \quad \lambda_2 = \frac{\vr}{1 + \gamma \cos \theta} .
 \label{eqn: lambda1 lambda2 definitions}
\end{equation}
\end{remark}

\section{Electroelastic energy based variational formulation and equations of equilibrium}
\label{sec:energy_consideration}
This section formulates the  equations of electroelastic equilibrium using the first variation of the total potential energy functional. Section~\ref{sec:Electrostatics} briefly presents the  equations for electrostatics. Thereafter the total potential energy of the system under an applied pressure and an electric field is given. Section~\ref{sec:equilibrium} considers the first variation of the total potential energy and the resulting three governing equations. Section~\ref{sec:energy density} decomposes the energy density function into an elastic energy density function and an electric contribution. A Mooney--Rivilin constitutive model is employed for the elastic energy density. This yields the governing equations presented in Section~\ref{sec:governing_equations}. A numerical method is proposed in Section~\ref{sec:numerical_method} to solve the resulting system of nonlinear ordinary differential equations (ODEs).

\subsection{Electrostatics}
\label{sec:Electrostatics}
Maxwell's equations for electrostatics are given by
\begin{align}
\text{Curl} \elecR = \mbf{0}, && \text{and} && \text{Div} \disR = 0, \label{eqn: maxwell}
\end{align}
where $\elecR$ is the electric field in the reference configuration and $\disR$ is the electric displacement in the reference configuration assuming the free charge density in the volume is zero. Equation~\eqref{eqn: maxwell}$_2$ motivates the introduction of an electric vector potential $\potR$ defined as
\begin{equation}
\disR = \text{Curl} \potR. \label{eqn: A definition}
\end{equation}
%
The referential vectors $\elecR$ and $\disR$ can be expressed as the pull-backs of the electric field and displacement in the current (deformed) configuration, $\elecC$ and $\disC$,  as~\citep{Dorfmann2014b}
\begin{align}
 \elecR  = \Dgrad^{\top} \elecC && \text{and} &&  \disR = J \Dgrad^{-1} \disC.
\end{align}
Within the electroelastic solid, the constitutive relation
\begin{equation}
 \elecR = \Omega_{\disR}
 \label{eqn: constitutive}
\end{equation}
relates $\elecR$ and $\disR$ via the total energy density $\Omega$ of the material. 
{In free space, $\elecC$ and $\disC$ are related through the electric permittivity of vacuum $\varepsilon_0$ as
\begin{equation}
 \elecC = \varepsilon_0^{-1} \disC .
\end{equation}
}
\subsection{Potential energy functional}
\label{sec:potental_energy}
The toroidal membrane occupies the region $\bod_0$ in the reference configuration and its total internal energy density per unit volume $\Omega$ is parameterised by the deformation gradient $\Dgrad$ and the referential electric displacement $\disR$.
Under an applied pressure $\wt{P}$, the total potential energy of the system  can be written as
\begin{equation}
 \mathfrak{E} (\rchi, \potR) = \int\limits_{\bod_0} \Omega (\Dgrad, \disR) \mathrm{d}v_0  - \int\limits_{V_0}^{V_0 + \Delta V} \wt{P} \mathrm{d}V.
 \label{pot_func_A}
\end{equation}
We note here that the first integral is over the region occupied by the solid membrane while the second integral of pressure work is over the volume of fluid enclosed by the inflated membrane (that is the region lying in the interior of the torus). 
Hence {the differentials are distinguished as $\mathrm{d}v_0$ and $\mathrm{d}V$}.
Since the electric potential is specified on the inner and outer surfaces of the torus, 
we assume that the electric field does not leak out and therefore 
there is no contribution to the energy in the region outside the membrane given by $\bbm{R}^3 \setminus\bod_0$.

As the membrane is considered to be thin $(H/R_s \ll 1)$ with negligible bending stiffness, the deformation field $\rchi$ is determined by the functions $\wtr$ and $\wte$ which describe the geometry of the mid-surface of the membrane (see Figure \ref{fig:toroidal_geometry}b).
The potential energy functional $\mathfrak{E}$ \eqref{pot_func_A} can therefore be reparametrised as
\begin{equation}
 E(\wtr, \wte, \potR) = H \int\limits_{0}^{2 \pi} \int\limits_{0}^{2 \pi} \Omega (\Dgrad, \disR) \sqrt{G} \, \mathrm{d}\theta \, \mathrm{d}\phi  - \frac{1}{2} \int\limits_{0}^{2 \pi}  \int\limits_{0}^{2 \pi} \wt{P} \wtr^2 \wte_\theta  \mathrm{d}\theta \, \mathrm{d}\phi . \label{eqn: final E functional to be minimised}
\end{equation}
The modification of the second term for the pressure work is based on the calculations provided in Appendix A (supporting file). 
\subsection{First variation and equations of equilibrium}
\label{sec:equilibrium}
At equilibrium, the total potential energy of the inflated toroidal membrane will be stationary, that is
{
\begin{align}
 \delta E&\equiv \delta E((\wtr, \wte, \potR); (\delta\wtr, \delta\wte, \delta\potR)) = 0, \label{eqn:first_variation_full}
\\
\text{with }
 \delta E &= H \int\limits_{0}^{2 \pi} \int\limits_{0}^{2 \pi} \bigg[ \big[ \Omega_{\wtr} \delta \wtr + \Omega_{\wtr_\theta } \delta \wtr_\theta  + \Omega_{\wte_\theta } \delta \wte_\theta + \Omega_{\wtr_\phi } \delta \wtr_\phi  + \Omega_{\wte_\phi } \delta \wte_\phi + \Omega_{\disR} \cdot \delta \disR \big]  \sqrt{G} \nonumber \\
 &- \frac{1}{2H} \wt{P} \big[ 2 \wtr \wte_\theta  \delta \wtr + \wtr^2 \delta \wte_\theta  \big]  \bigg] \, \mathrm{d}\theta \, \mathrm{d}\phi = 0.
 \label{eqn:first_variation_full}
\end{align}
}
As shown in Figure~\ref{fig:toroidal_membrane_reference}, the geometry of the toroidal membrane is {axisymmetric with respect to $\mbf{E}_3$.} 
Thus, the principle solutions $\wtr$ and $\wte$ should  be constant along the $\phi$ direction, which implies $\wtr_\phi = \wte_\phi = 0$. Thus Equation~\eqref{eqn:first_variation_full} can be simplified as
{
\begin{align}
 \delta E = {2 \pi} H  \int\limits_{0}^{2 \pi} &\bigg[ \big[ \Omega_{\wtr} \delta \wtr + \Omega_{\wtr_\theta } \delta \wtr_\theta  + \Omega_{\wte_\theta } \delta \wte_\theta  + \Omega_{\disR} \cdot \delta \disR \big]  \sqrt{G} \nonumber \\
 &- \frac{1}{2H} \wt{P} \big[ 2 \wtr \wte_\theta  \delta \wtr + \wtr^2 \delta \wte_\theta  \big]  \bigg] \, \mathrm{d}\theta  = 0. \label{eqn: eq18}
\end{align}
}

The rotational symmetry of the torus leads to the periodic boundary conditions
\begin{equation}
 \delta \wtr|_{\theta = 0} = \delta \wtr|_{\theta = 2 \pi} , \quad \delta \wte|_{\theta = 0} = \delta \wte|_{\theta = 2 \pi} , \label{eqn: rot symm bc}
\end{equation}
while a perturbation of equation \eqref{eqn: A definition} gives
%
$\delta \disR =  \Curl{\delta \potR}$.
Thus we can express the variations of all the quantities in equation \eqref{eqn: eq18} in terms of the variations $\delta \wtr, \delta \wte$ and $\delta \potR$, as desired. 
Using integration by parts on the terms containing $\delta \wtr_\theta$ and $\delta \wte_\theta$,  the vector identity
\begin{equation}
  [\nabla \times \mbf{u}] \cdot \mbf{v} = \nabla \cdot [\mbf{u} \times \mbf{v}] + [\nabla \times \mbf{v}] \cdot \mbf{u}, \quad \quad \mbf{u}, \mbf{v} \in \bbm{R}^3   \label{eqn: curl identity}
\end{equation}
for the terms containing $\Curl{\delta \potR}$, and the arbitrariness of the variations results in the following Euler-Lagrange  equations for this system
\begin{subequations} \label{eqn: gov rho eta tilde}
\begin{align}
 \sqrt{G}\Omega_{\wtr} - \fd{}{ \theta} \left( \sqrt{G} \Omega_{\wtr_\theta } \right) - \frac{\wt{P} \wtr \wte_\theta }{H} = 0, \label{eqn:gov_1} \\
 \fd{}{ \theta} \left( \sqrt{G} \Omega_{\wte_\theta } - \frac{\wt{P} \wtr^2}{2H}  \right) = 0 , \label{eqn:gov_2} \\
 \Curl \elecR  = \mbf{0}, \label{eqn:gov_3}
\end{align}
\end{subequations}
for $\theta \in (0, 2 \pi]$ along with the boundary conditions \eqref{eqn: rot symm bc} and a specified potential difference $\Phi_0$ across the torus thickness.

For the thin-walled membrane considered,
the electric field can be reasonably approximated as \citep{Xie2016}
\begin{equation}
 \elecC = \frac{\Phi_0}{h} \mbf{n}. \label{eqn: elec from phi}
\end{equation}
Thus, from~\eqref{eqn: constitutive} the electric field in reference configuration is given by
\begin{equation}
\elecR = \frac{\Phi_0}{\lambda_3 H} \Dgrad^\top \mbf{n},\label{eqn: elec from phi conf}
\end{equation}
which also satisfies the governing equation \eqref{eqn:gov_3}.

\subsection{Energy density function}
\label{sec:energy density}

The total energy density $\Omega$ can be further decomposed into an elastic energy density $\wh{W}$ and an electric contribution as follows, see~\cite{dorfmann2006nonlinear}
\begin{equation}
 \Omega (\Dgrad, \disR) = \wh{W} (\Dgrad) + \beta [\CGright \disR] \cdot \disR  \, , \label{eqn: energy decomposition}
\end{equation}
where $\beta$ is a positive material constant representing electroelastic coupling and the term $[\CGright \disR] \cdot \disR$ is a scalar invariant. 
This allows the electric field in the reference configuration to be expressed as
\begin{equation}
 \elecR = \Omega_{\disR} = 2 \beta \CGright \disR. \label{eqn: elec from energy}
\end{equation}
On comparing equations \eqref{eqn: elec from phi} and \eqref{eqn: elec from energy}, the electric displacement in the reference configuration can be expressed as
\begin{equation}
 \disR = \frac{\Phi_0}{2 \beta \lambda_3 H} \Dgrad^{-1} \mbf{n} = \frac{\Phi_0}{2 \beta  H} \CGright^{-1} \mbf{N} , \label{eqn: D expression for problem}
\end{equation}
where $\mbf{n}$ and $\mbf{N}$ are unit vectors along the outward normal in current (deformed) and reference configurations, respectively.
Here Nanson's relation is employed to relate the normal vectors as  $\mbf{n} da = J \Dgrad^{-\top} \mbf{N} dA$ with $da = \lambda_1 \lambda_2 dA$. 

We consider an incompressible Mooney--Rivlin constitutive model for the elastic energy density $\wh{W}$ given in terms of two material constants $C_1$ and $C_2$ where
\begin{subequations}\label{eqn:What}
\begin{gather}
\wh{W}(I_1, I_2) = C_1[I_1 - 3] + C_2[I_2 - 3] , \label{eqn:What part a}\\
\text{with} \qquad I_1 = \lambda_1^2 + \lambda_2^2 + \lambda_3^2 \qquad \text{and} \qquad I_2 = \lambda_1^{-2} + \lambda_2^{-2} + \lambda_3^{-2} . \label{eqn:What part b}
\end{gather}
\end{subequations}
The energy density $\Omega$ in Equation~\eqref{eqn: energy decomposition} can therefore be expressed as
\begin{equation}
\Omega =  C_1[I_1 - 3] + C_2[I_2 - 3]  + \beta [\CGright \disR] \cdot \disR .
\label{eqn:energy_density_function}
\end{equation}
The derivatives of $\Omega$ required for the subsequent derivations and computations are tabulated in Appendix B (supporting file).

\subsection{Governing equations}
\label{sec:governing_equations}
We introduce the following dimensionless versions of pressure $P$ and electric loading $\mcal{E}$ to simplify our calculations
\begin{align}
 P = \frac{\wt{P} R_b}{C_1 H} , \quad \quad \quad  \mcal{E} = \frac{\Phi_0^2}{C_1 \beta H^2} .
\end{align}
 Further introducing a dimensionless parameter $\alpha = C_2/C_1$ , the ordinary differential equations~\eqref{eqn:gov_1} and~\eqref{eqn:gov_2} can be written in dimensionless form for $\theta \in [0, \pi]$ as
\begin{subequations}
\begin{align}
 &\frac{\mathrm{d}}{\mathrm{d} \theta} \left( [1 + \gamma \cos \theta] \bigg[  \frac{2 \varrho_\theta }{\gamma^2}  \left[ 1 + \alpha \lambda_2^2 \right] \left[ 1 - \frac{1}{\lambda_1^4 \lambda_2^2} \right] -  \frac{\rho_\theta  \mcal{E}\lambda_2^2}{2 \gamma^2} \bigg] \right) \nonumber \\
 &-   2\lambda_2  \left[ 1+\alpha \lambda_1^2 \right] \left[ 1 - \frac{1}{\lambda_1^2 \lambda_2^4} \right]  + \frac{\mcal{E}\lambda_2 \lambda_1^2}{2 }   + \frac{P \vr \eta_\theta }{\gamma} &= 0, \label{eqn: final principal coupled ODE 1} \\
 &\frac{\mathrm{d}}{\mathrm{d} \theta} \left( [1 + \gamma \cos \theta] \Bigg[ \frac{2\eta_\theta }{\gamma^2} \left[ 1 + \alpha \lambda_2^2 \right] \left[ 1 - \frac{1}{\lambda_1^4 \lambda_2^2} \right] - \frac{ \eta_\theta  \mcal{E} \lambda_2^2}{ 2 \gamma^2} \Bigg] - \frac{P \vr^2}{2 \gamma}  \right) &= 0, \label{eqn: final principal coupled ODE 2}
 \end{align}
 \label{eqn: final principal coupled ODEs}
\end{subequations}
along with the boundary conditions, accounting for the additional assumption of a reflection symmetry of the system with respect to the $Y^1- Y^2$ plane, given by
\begin{subequations}
\begin{align}
\vr_\theta &= 0,\quad \eta = 0 \quad \text{at} \quad \theta = 0 \label{eqn:boundary_condition1}
\\
 \text{ and } \vr_\theta &= 0,\quad \eta = 0 \quad \text{at}\quad\theta = \pi.
 \label{eqn:boundary_condition2}
\end{align}
 \label{eqn:boundary_condition}
\end{subequations}

\subsection{Numerical solution procedure}
\label{sec:numerical_method}
The governing equations~\eqref{eqn: final principal coupled ODE 1} and~\eqref{eqn: final principal coupled ODE 2} are coupled nonlinear second-order ordinary differential equations. 
These can be converted to a system of first-order ODEs by defining
\begin{equation}
 y_1 = \vr, \quad y_2 = \vr_\theta , \quad y_3 = \eta, \quad y_4 = \eta_\theta,
\end{equation}
and rewriting the system as
\begin{equation}
 \underbrace{\begin{bmatrix}
  1 & 0 & 0 & 0 \\
  0 & A_1 & 0 & A_2 \\
  0 & 0 & 1 & 0 \\
  0 & B_1 & 0 & B_2
 \end{bmatrix}}_{\mbf{A}}
 \underbrace{ \begin{bmatrix}
  y_1' \\
  y_2' \\
  y_3' \\
  y_4'
 \end{bmatrix}}_{\mbf{y}}
 =
\underbrace{ \begin{bmatrix}
  y_2 \\
  -A_3 \\
  y_4 \\
  -B_3
 \end{bmatrix}}_{\mbf{b}},
 \label{eqn:matrix_ode_system}
\end{equation}
where $(\bullet)'$ denotes the derivative with respect to $\theta$, together with boundary conditions
\begin{equation}
 y_2 = 0,  \quad y_3= 0\quad \text{at} \quad \theta = 0\quad \text{and} \quad \theta = \pi.
 \label{eqn:boundary_conds}
\end{equation}
The components of matrices $\mbf{A}$ and $\mbf{b}$ are derived in Appendix D (supporting file).

\begin{figure}[]
\centering
	\includegraphics[width=0.6\linewidth]{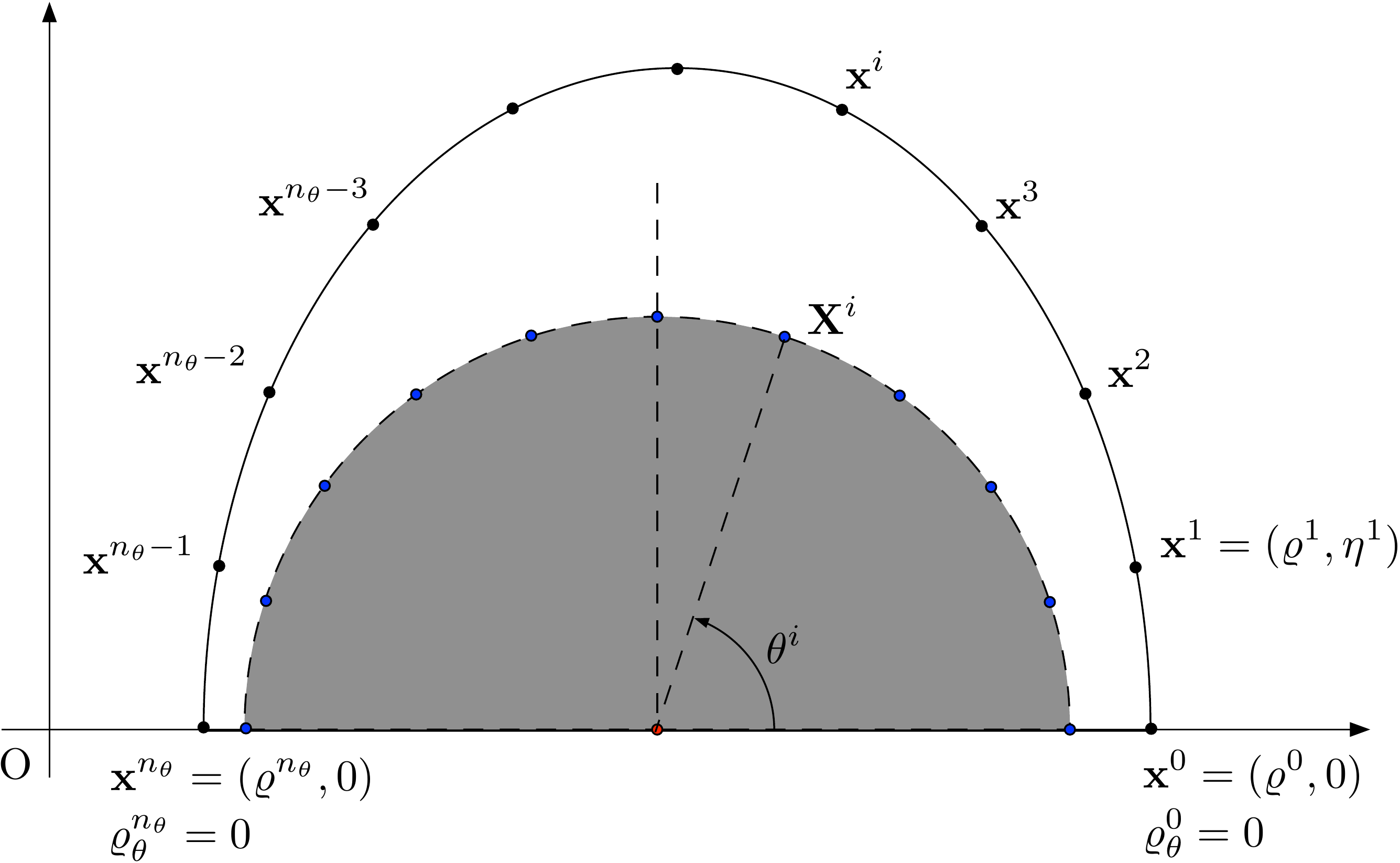}
	\caption{ Half of the cross-section is discretised into $n_\theta$ segments with respect to $\theta$ and the boundary conditions~\eqref{eqn:boundary_condition} are applied at $\theta = 0$ and $\pi$.}
	\label{fig:discretisation_cross_section}
\end{figure}

We discretise the reference configuration of the membrane into $n_\theta$ segments as shown in Figure~\ref{fig:discretisation_cross_section}. The $n_\theta+1$ discretised points in the deformed configuration are denoted by $\mbf{x}^i(\theta)$, where $i = 0,1,2,\cdots,n_\theta$.
%
%
For each $\theta^i$, the matrix $\mbf{A}$ and the right hand vector $\mbf{b}$ can be computed given the mechanical and electrical loads $P$ and $\mcal{E}$ that are independent of $\theta$. 
$P$ is only present in the vector $\mbf{b}$ as the mechanical load does not affect the material ``stiffness''. 
However, $\mcal{E}$ appears in both the matrix $\mbf{A}$ and the vector $\mbf{b}$ since the electrical load affects not only  the boundary conditions but also changes the material behaviour. 

\begin{figure}
\centering
	\includegraphics[width=0.7\linewidth]{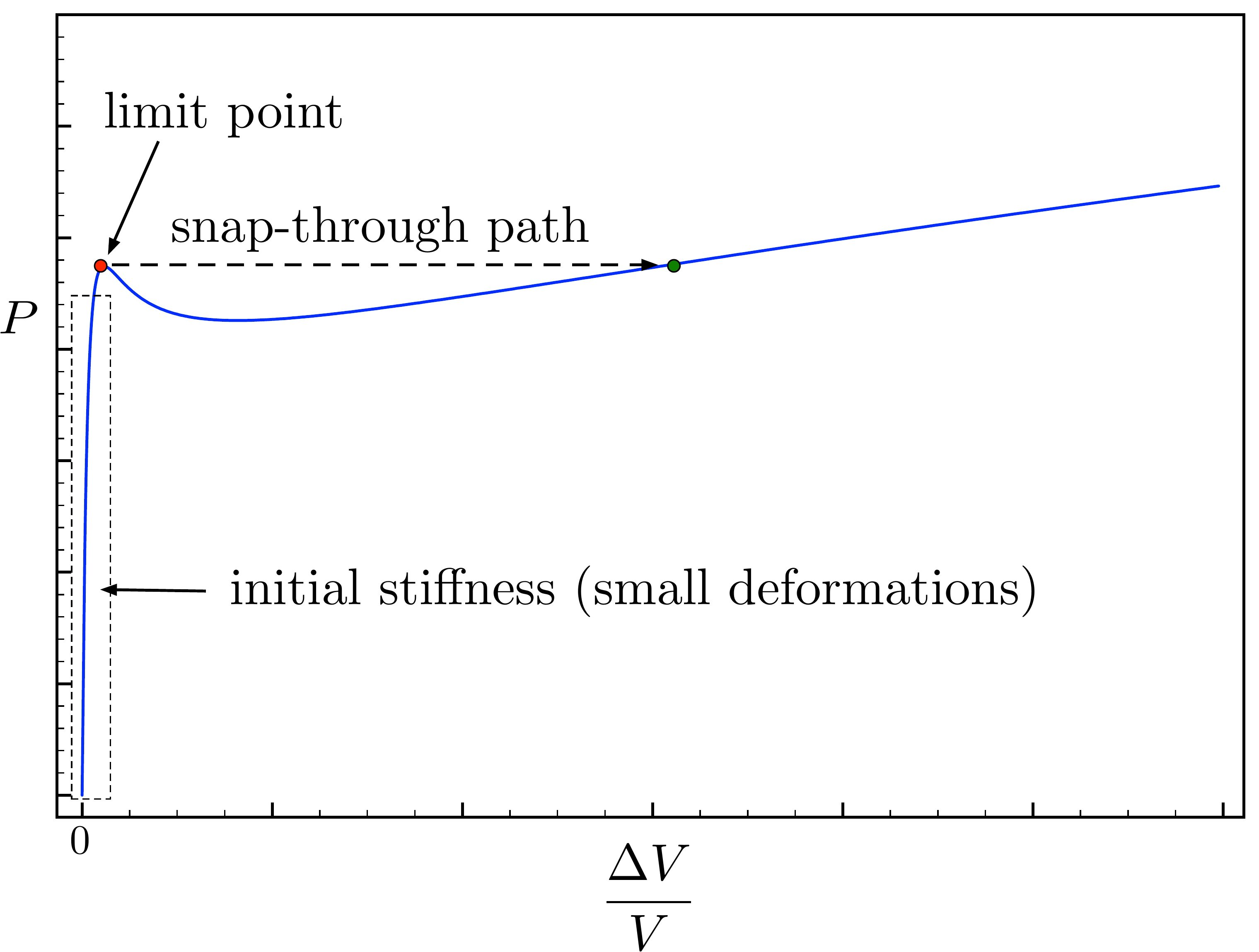}
		\caption{{A sketch of the initial stiffness and snap-through path for an ideal electroelastic toroidal membrane under large
inflation. For a membrane with imperfections, snap-through may occur prior to reaching the limit point.} }
	\label{fig:snap-through}
\end{figure}

The system \eqref{eqn:matrix_ode_system} and \eqref{eqn:boundary_conds} is solved using a variation of the shooting method~\cite{morrison1962multiple} employing arc-control~\cite{Reddy2017}.
We use the $\texttt{ode45}$ solver in \texttt{Matlab}~\cite{MATLAB:2018b} for the numerical approximation of the ODEs.
$y_2^0$ and $y_3^0$ are zero as per the boundary conditions~\eqref{eqn:boundary_conds} and a value of $y_1^0$ is chosen. To solve the system of equation, a reasonable initial guess for the variables $y_4^0$ and $P$ is required. 
Then the function $\texttt{ode45}$ can be used to solve for the unknowns $y_1^i,y_2^i,y_3^i, y_4^i$ for any $i = 0, 1, 2, 3, \cdots, n_\theta$. 
 A cost function $f_c = \sqrt{[{y_2^{n_\theta}}]^2+ [{y_3^{n_\theta}}]^2 }$ is introduced to ensure satisfaction of the boundary conditions; this is required to be minimised as part of the shooting method. Then the function $\texttt{fminsearchbnd}$~\cite{fminsearch} is used to search for a pair of $y_4^0$ and $P$ such that the cost function $f_c$ is less than a tolerance $\epsilon = 10^{-6}$. Once $f_c < \epsilon$, the values of $y_4^0$ and $P$ are assumed sufficiently converged and the values of $y_1^i,y_2^i,y_3^i, y_4^i$ for all $i =0, 1, 2, 3, \cdots, n_\theta$ are computed. 
 The arc-length control (by specifying $y_1^0$ and computing $P$) helps in the evaluation of solutions between the snap-through path, see Figure~\ref{fig:snap-through}.

Since the solutions are computed using a minimisation procedure, convergence is dependent upon the initial guess of $P$ and $y_4^0$. In general, the values of $y_4^0$ and $P$ from the previous load step can be used as initial guesses for the updated deformed profile, thereby allowing the inflation of the toroidal membrane to be simulated. However, in the initial stages of the inflation of the membrane, a relatively large pressure increase results in a very small deformation. This is due to the high initial stiffness as shown in Figure~\ref{fig:snap-through}. In this initial region, the gradient of $P$ with respect to volume change is very high and convergence of the method using $\texttt{fminsearchbnd}$ is problematic. To simulate the inflation in this region, one should seek solutions close to the limit point and then solve for a less deformed membrane profile by decreasing $\mathrm{d}y_1^0$ with a small decrement.  Convergence is also improved by introducing a scaling coefficient $\kappa =| \mathrm{d} P/\mathrm{d}y_1^0|$, where the change of pressure $ \mathrm{d} P$ is estimated based on the previous solutions. Then, instead of searching for $y_4^0$ and $P$ to achieve $f_c = 0$, one searches for $P$ and $\kappa y_4^0$. Following this approach, the convergence of the  $\texttt{fminsearchbnd}$ is greatly improved and the method is robust.
A computer code employing the above-described scheme is available at~\cite{Liu2020github}.
\section{Wrinkling instability analysis}
\label{sec:wrinkling_instability_analysis}
A membrane structure can only sustain tensile loading and has no resistance to compressive stress. 
When an in-plane compressive stress is about to occur in a membrane structure, the membrane tends to develop localised out-of-plane deformation to lower the energy.
This phenomenon is known as wrinkling. 
For the problem considered here, when the toroidal membrane inflates and gradually undergoes increasing deformation,
the tensile stresses in inner regions ($\theta \approx \pi$) are gradually reduced.
Eventually, this will lead to wrinkling in the inner regions of the toroidal membrane. 
Post wrinkle formation, the principle governing equations~\eqref{eqn: final principal coupled ODE 1} and~\eqref{eqn: final principal coupled ODE 2} are no longer descriptors of the state of the system. 
An energy relaxation method is then adopted to modify the governing equations and compute new solutions valid for the post-wrinkling regime.

\subsection{In-plane stress components}
The total Piola stress tensor $\strPK$ for the incompressible case is given by \cite{dorfmann2006nonlinear}
\begin{equation}
 \strPK = \Omega_{\Dgrad} - p \Dgrad^{- \top} ,
\end{equation}
where $p$ is a Lagrange multiplier associated with the constraint of incompressibility.
The total Cauchy stress is (as $J=1$)
\begin{equation}
 \strC = \strPK \Dgrad^{\top} = \Omega_{\Dgrad} \Dgrad^\top - p \bsym{i},
\end{equation}
where $\bsym{i}$ is the second-order unit tensor in the current configuration. Let the local orthonormal coordinate system in the membrane mid-surface be given by $\{ \mbf{n}, \mbf{t}_1, \mbf{t}_2 \}$, where the unit normal $\mbf{n}$ is defined in Appendix A (supporting file) and $\mbf{t}_1$ and $\mbf{t}_2$ are unit vectors in the tangent plane of the membrane.
Following the discussion in Section \ref{sec: kinematics}, the tensor $\Dgrad$ and subsequently the Cauchy stress $\strC$ are represented by diagonal matrices in this local basis.
We can compute the in-plane stress components as
\begin{equation}
 s_{11} = \big[ \strC \mbf{t}_1 \big] \cdot \mbf{t}_1, \quad s_{22} = \big[ \strC \mbf{t}_2 \big] \cdot \mbf{t}_2, \quad s_{12} = s_{21} = \big[ \strC \mbf{t}_2 \big] \cdot \mbf{t}_1 = 0,
\end{equation}
Using the balance of traction at the inner boundary,
\begin{equation}
 \strC \mbf{n} + \wt{P} \mbf{n} = \mbf{0},
\end{equation}

One determines the Lagrange multiplier $p$ as
\begin{equation}
 p = \wt{P} + \big[  \Omega_{\Dgrad} \Dgrad^\top \mbf{n} \big] \cdot \mbf{n} ,
\end{equation}
using which one can write the two in-plane stress components as
\begin{align}
 s_{11} & =  \big[  \Omega_{\Dgrad} \Dgrad^\top \mbf{t}_1 \big] \cdot \mbf{t}_1 - \wt{P} - \big[  \Omega_{\Dgrad} \Dgrad^\top \mbf{n} \big] \cdot \mbf{n} , \nonumber \\
 s_{22} & = \big[  \Omega_{\Dgrad} \Dgrad^\top \mbf{t}_2 \big] \cdot \mbf{t}_2 - \wt{P} - \big[  \Omega_{\Dgrad} \Dgrad^\top \mbf{n} \big] \cdot \mbf{n} .
 \label{eqn:stress_components}
\end{align}
Substituting in the specific form of $\Omega$ from equations \eqref{eqn: energy decomposition} and \eqref{eqn:What part a}, and making use of \eqref{eqn: D expression for problem}, yields
\begin{equation}
 \Omega_{\Dgrad} \Dgrad^\top = 2 C_1 \CGleft + 2 C_2 \big[ I_1 \CGleft - \CGleft^2 \big] + \frac{\Phi_0^2}{2 \beta \lambda_3^2 H^2} \mbf{n} \otimes \mbf{n},
\end{equation}
allowing one to rewrite the in-plane principal stress components as
\begin{subequations}
\begin{align}
 s_{11} 
 &= C_1 \Bigg[ - \frac{P H}{R_b} + 2 \lambda_1^2 + 2 \alpha \lambda_1^2  \bigg[ \lambda_2^2 + \frac{1}{\lambda_1^2 \lambda_2^2} \bigg] - \frac{2}{\lambda_1^2 \lambda_2^2} - 2 \alpha \bigg[ \frac{1}{\lambda_1^2} + \frac{1}{\lambda_2^2} \bigg] - \frac{\mcal{E} \lambda_1^2 \lambda_2^2}{2 } \Bigg]  ,\\
 s_{22} 
 &= C_1 \Bigg[ - \frac{P H}{R_b} + 2 \lambda_2^2 + 2 \alpha \lambda_2^2 \bigg[ \lambda_1^2 + \frac{1}{\lambda_1^2 \lambda_2^2} \bigg]  - \frac{2}{\lambda_1^2 \lambda_2^2} - 2 \alpha \bigg[ \frac{1}{\lambda_1^2} + \frac{1}{\lambda_2^2} \bigg] - \frac{\mcal{E} \lambda_1^2 \lambda_2^2}{2 } \Bigg] . \label{eqn: s11 s22 expressions}
\end{align}
\end{subequations}
Note that the thinness assumption of the membrane, $ H/R_s \ll 1$, nullifies the influence of the pressure term on the in-plane membrane stresses \citep{Crandall1972}. 
A constant value $H/R_s = \gamma^{-1} 10^{-4}$ is thus adopted in all numerical examples.

\subsection{Energy Relaxation}
\label{sec:energy_relaxation}
Adopting the tension field theory developed by \cite{Pipkin1986, Steigmann1990} for hyperelastic membranes and later extended to electroelastic membranes by \cite{DeTommasi2011, Greaney2019}, we introduce the concept of a {\it generalised natural state}.
In the current problem, the key kinematic variables in the total energy density function $\Omega(\Dgrad, \disR)$ are $\{\lambda_1, \lambda_2, \mcal{E}\}$ and hence $\Omega = \wh{\Omega}(\lambda_1, \lambda_2, \mcal{E})$.

If the membrane is mechanically stretched along one direction with stretch $\lambda_1>1$ (resp.\ $\lambda_2>1$) and an electric load $\mcal{E}$ is applied, then the value of $\lambda_2$ (resp.\ $\lambda_1$) that sets the principal stress component $s_{22}$ (resp.\ $s_{11}$) to zero is denoted as the {\it natural width} $w_1$.
This natural {width} is given by
\begin{align}
 w_1(\lambda_1, \mcal{E}) := \lambda_2^* && \text{such that} && s_{22} (\lambda_1, \lambda_2^*, \mcal{E}) = 0,
 \label{eqn:s22=0}
\end{align}
where ${\lambda_2^*}^2$ can be expressed as a function of $\lambda_1$ and $\mcal{E}$ as
\begin{align}
 {\lambda^\ast_2}^2(\lambda_1, \mcal{E}) = \frac{\frac{PH}{R_b}\lambda_1 + \sqrt{\frac{P^2 H^2}{R_b^2} {\lambda_1^2}- 4\alpha\mcal{E}\lambda_1^4 + 16\alpha^2\lambda_1^4 + 32\alpha {\lambda_1^2}- 4\mcal{E}{\lambda_1^2}+{16}}}{4\lambda_1 + 4\alpha\lambda_1^3 - \mcal{E} \lambda_1^3}.
 \label{eqn:relaxed_stress}
\end{align}
The relaxed energy is then given by
\begin{align}
\Omega^\ast  (\lambda_1,\lambda^\ast_2) = C_1[I^\ast_1 -3] + C_2[I^\ast_2-3] + \beta [ \CGright \disR]\cdot\disR,
\label{eqn:RE_energy_density}
\end{align}
where
\begin{align}
I^\ast_1 (\lambda_1,\lambda^\ast_2)= \lambda_1^2 +  {\lambda^\ast_2}^2 + \frac{1}{\lambda_1^2  {\lambda^\ast_2}^2}, \quad
I^\ast_2 (\lambda_1,\lambda^\ast_2)= \frac{1}{\lambda_1^2} + \frac{1}{ {\lambda^\ast_2}^2} + {\lambda_1^2  {\lambda^\ast_2}^2},
\end{align}
and
\begin{align}
\beta [ \CGright \disR] \cdot\disR = \frac{C_1\mcal{E}}{4} {\lambda_1^2  {\lambda^\ast_2}^2}.
\end{align}

Since ${\lambda^\ast_2}$ is a function of $\lambda_1$ and $\mcal{E}$ as shown in~\eqref{eqn:relaxed_stress}, $I^\ast_1$ and $I^\ast_2$ are only functions of $\lambda_1$ and $\mcal{E}$. Similarly, $\beta [ \CGright \disR] \cdot\disR$ is a function of $\lambda_1$ and $\mcal{E}$.
Hence
\begin{align}
\Omega^\ast(\lambda_1,\mcal{E}) = C_1[I^\ast_1 -3]+C_2[I^\ast_2-3] + \beta [ \CGright \disR] \cdot\disR.
\end{align}
The governing equations~\eqref{eqn: final principal coupled ODEs} thus become
\begin{align}
[1 + \gamma \cos \theta]\fd{}{\theta} \left( \pd{\Omega^\ast}{\vr_\theta} \right) -\gamma\sin \theta \pd{\Omega^\ast}{\vr_\theta}- \pd{\Omega^\ast}{\vr} [1 + \gamma \cos \theta] + \frac{\wt{P} R_b \vr \eta_\theta}{\gamma H} &= 0 ,\nonumber\\
[1 + \gamma \cos \theta]\fd{}{\theta} \left( \pd{\Omega^\ast}{\eta_\theta} \right) -\gamma\sin \theta \pd{\Omega^\ast}{\eta_\theta} -\fd{}{ \theta} \left(\frac{\wt{P} R_b \vr^2}{2 \gamma H} \right) &= 0.
\label{eqn:RE_ODEs}
\end{align}
These can be rearranged into a matrix form as in~\eqref{eqn:matrix_ode_system}. The reformulation is given in Appendix E (supporting file). The same shooting method described in Section~\ref{sec:numerical_method} is used to solve these modified equations.

\section{Loss of symmetry in the $\phi$ direction}
\label{sec:loss_symmetry}
This section considers the loss of symmetry of the principal solution obtained in Section~\ref{sec:energy_consideration} due to a perturbation along the outer equator which is an instability of the toroidal membrane. 
If the second variation of the potential energy functional vanishes, the solution bifurcates to a lower energy branch that is no longer symmetric.
\subsection{Second variation of the potential energy functional}
\label{sec:second_variation}
 For the analysis of the critical point $(\rchi, \potR)$ of instability,  one seeks $\Delta \rchi := (\Delta \wtr, \Delta \wte)$ and $\Delta \potR$ such that the following bilinear functional vanishes
\begin{equation}
 \delta^2 E[\rchi, \potR; (\delta \rchi, \delta \potR), (\Delta \rchi, \Delta \potR)]
 = 0.  \label{eqn: second variation condition}
\end{equation}

The principal solution has no dependence on the $\phi$ coordinate, but one might be interested in perturbations along the $\phi$ direction.
Hence, we derive a more general Taylor expansion of the functional $E$ in \eqref{eqn: final E functional to be minimised} considering $\wtr$ and $\wte$ to have dependence on both $\theta$ and $\phi$, i.e. $\wtr(\theta, \phi)$ and $\wte(\theta, \phi)$. This yields
\begin{align}
 &E [ \wtr + \delta \wtr, \wte + \delta \wte, \potR + \delta \potR] = H \int\limits_{0}^{2 \pi} \int\limits_{0}^{2 \pi} \bigg[ \Omega + \Omega_{\wtr} \delta \wtr + \Omega_{\wtr_\theta} \delta \wtr_\theta + \Omega_{\wtr_\phi} \delta \wtr_\phi + \Omega_{\wte_\theta} \delta \wte_\theta  \nonumber \\
 & +  \Omega_{\wte_\phi} \delta \wte_\phi + \Omega_{\disR} \cdot \delta \disR + \frac{1}{2} \bigg[ \Omega_{\wtr \wtr} \delta \wtr \, \delta \wtr + \Omega_{\wtr_\theta \wtr_\theta} \delta \wtr_\theta \, \delta \wtr_\theta + \Omega_{\wtr_\phi \wtr_\phi} \delta \wtr_\phi \, \delta \wtr_\phi  \nonumber \\
 &+ \Omega_{\wte_\theta \wte_\theta} \delta \wte_\theta \, \delta \wte_\theta + \Omega_{\wte_\phi \wte_\phi} \delta \wte_\phi \, \delta \wte_\phi + \delta \disR \cdot \left[ \Omega_{\disR \disR}  \delta \disR  \right] + 2 \Omega_{\wtr \wtr_\theta} \delta \wtr \, \delta \wtr_\theta + 2 \Omega_{\wtr \wtr_\phi} \delta \wtr \, \delta \wtr_\phi \nonumber \\
 & + 2 \Omega_{\wtr \wte_\theta} \delta \wtr \, \delta \wte_\theta + 2 \Omega_{\wtr \wte_\phi} \delta \wtr \, \delta \wte_\phi + 2  \delta \disR \cdot \Omega_{\wtr \disR} \delta \wtr +  2 \Omega_{\wtr_\theta \wtr_\phi} \delta \wtr_\theta \, \delta \wtr_\phi +  2 \Omega_{\wtr_\theta \wte_\theta} \delta \wtr_\theta \, \delta \wte_\theta \nonumber \\
 &  + 2 \Omega_{\wtr_\theta \wte_\phi} \delta \wtr_\theta \, \delta \wte_\phi + 2  \delta \disR \cdot \Omega_{\wtr_\theta \disR} \delta \wtr_\theta +   2 \Omega_{\wtr_\phi \wte_\theta} \delta \wtr_\phi \, \delta \wte_\theta + 2 \Omega_{\wtr_\phi \wte_\phi} \delta \wtr_\phi \, \delta \wte_\phi \nonumber \\
 & + 2  \delta \disR \cdot \Omega_{\wtr_\phi \disR} \delta \wtr_\phi + 2 \Omega_{\wte_\theta \wte_\phi} \delta \wte_\theta \delta \wte_\phi + 2  \delta \disR \cdot \Omega_{\wte_\theta \disR} \delta \wte_\theta + 2  \delta \disR \cdot \Omega_{\wte_\phi \disR} \delta \wte_\phi
   \bigg]  \Bigg] \sqrt{G} \, d\theta \, d\phi \nonumber \\
 & - \frac{1}{2} \int\limits_{0}^{2 \pi}  \int\limits_{0}^{2 \pi} \wt{P} \left[[\wtr+\delta\wtr]^2 [\wte_\theta + \delta\wte_\theta]\right] d\theta \, d\phi. \label{eqn: general taylor expansion of E}
\end{align}
Application of integration by parts  and making use of the vector identity \eqref{eqn: curl identity}, the second variation is expressed as
\begin{align}
& \delta^2 E[\rchi, \potR; (\delta \rchi, \delta \potR), (\Delta \rchi, \Delta \potR)] = - H \int\limits_{0}^{2 \pi} \int\limits_{0}^{2 \pi} \Bigg[ \nonumber \\
& \bigg[ \sqrt{G}\left[ - \Omega_{\wtr \wtr} \Delta \wtr - \Omega_{\wtr \wtr_\theta} \Delta \wtr_\theta -  \Omega_{\wtr \wtr_\phi} \Delta \wtr_\phi - \Omega_{\wtr \wte_\theta} \Delta \wte_\theta - \Omega_{\wtr \wte_\phi} \Delta \wte_\phi - \Omega_{\wtr \disR} \cdot \Delta \disR \right] \nonumber \\ 
& +  \fd{}{\theta}\Big( \left[ \Omega_{\wtr_\theta \wtr_\theta} \Delta \wtr_\theta +  \Omega_{\wtr \wtr_\theta} \Delta \wtr + \Omega_{\wtr_\theta \wtr_\phi} \Delta \wtr_\phi + \Omega_{\wtr_\theta \wte_\theta} \Delta \wte_\theta + \Omega_{\wtr_\theta \wte_\phi} \Delta \wte_\phi + \Omega_{\wtr_\theta \disR} \cdot \Delta \disR \right] \sqrt{G} \Big) \nonumber \\
& + \fd{}{\phi}\Big(\left[ \Omega_{\wtr_\phi \wtr_\phi} \Delta \wtr_\phi +  \Omega_{\wtr \wtr_\phi} \Delta \wtr + \Omega_{\wtr_\theta \wtr_\phi} \Delta \wtr_\theta + \Omega_{\wtr_\phi \wte_\theta} \Delta \wte_\theta + \Omega_{\wtr_\phi \wte_\phi} \Delta \wte_\phi + \Omega_{\wtr_\phi \disR} \cdot \Delta \disR \right] \sqrt{G} \Big) \bigg] \delta \wtr \nonumber \\
& + \bigg[ \fd{}{\theta}\Big( \left[  \Omega_{\wte_\theta \wte_\theta} \Delta \wte_\theta + \Omega_{\wtr \wte_\theta} \Delta \wtr + \Omega_{\wtr_\theta \wte_\theta} \Delta \wtr_\theta +  \Omega_{\wtr_\phi \wte_\theta} \Delta \wtr_\phi + \Omega_{\wte_\theta \wte_\phi} \Delta \wte_\phi + \Omega_{\wte_\theta \disR } \cdot \Delta \disR \right] \sqrt{G} \Big) \nonumber \\
& + \fd{}{\phi}\Big( \left[ \Omega_{\wte_\phi \wte_\phi} \Delta \wte_\phi + \Omega_{\wtr \wte_\phi} \Delta \wtr + \Omega_{\wtr_\theta \wte_\phi} \Delta \wtr_\theta +  \Omega_{\wtr_\phi \wte_\phi} \Delta \wtr_\phi + \Omega_{\wte_\theta \wte_\phi} \Delta \wte_\theta + \Omega_{\wte_\phi \disR } \cdot \Delta \disR \right] \sqrt{G} \Big) \bigg] \delta \wte \nonumber \\
& - \Curl \bigg[ \Omega_{\disR \disR} \Delta \disR + \Omega_{\wtr \disR} \Delta \wtr + \Omega_{\wtr_\theta \disR} \Delta \wtr_\theta + \Omega_{\wtr_\phi \disR} \Delta \wtr_\phi + \Omega_{\wte_\theta \disR} \Delta \wte_\theta + \Omega_{\wte_\phi \disR} \Delta \wte_\phi \bigg] \cdot \delta \potR \sqrt{G} \Bigg] d \theta \, d \phi \nonumber \\
& - \int\limits_{0}^{2 \pi}  \int\limits_{0}^{2 \pi} \wt{P} \bigg[ \left[  \wtr\Delta\wte_\theta + \wte_\theta \Delta \wtr \right]\delta\wtr - \left[ \wtr_\theta \Delta\wtr + \wtr\Delta\wtr_\theta\right]\delta\wte  \bigg]  d\theta \, d\phi.
\end{align}
 The Euler-Lagrange equations corresponding to~\eqref{eqn: second variation condition} are derived in dimensionless form as

\begin{align}
 &\left[- \Omega_{\vr \vr} \Delta \vr - \Omega_{\vr \vr_\theta} \Delta \vr_\theta -  \Omega_{\vr \vr_\phi} \Delta \vr_\phi - \Omega_{\vr \eta_\theta} \Delta \eta_\theta - \Omega_{\vr \eta_\phi} \Delta \eta_\phi - \Omega_{\vr \disR} \cdot \Delta \disR \right] [1+\gamma\cos\theta] \nonumber \\ 
& + \fd{}{\theta} \Big( \left[\Omega_{\vr_\theta \vr_\theta} \Delta \vr_\theta +  \Omega_{\vr \vr_\theta} \Delta \vr + \Omega_{\vr_\theta \vr_\phi} \Delta \vr_\phi + \Omega_{\vr_\theta \eta_\theta} \Delta \eta_\theta + \Omega_{\vr_\theta \eta_\phi} \Delta \eta_\phi + \Omega_{\vr_\theta \disR} \cdot \Delta \disR\right]  [1+\gamma\cos\theta]  \Big) \nonumber \\
& + \fd{}{\phi} \Big(\left[ \Omega_{\vr_\phi \vr_\phi} \Delta \vr_\phi +  \Omega_{\vr \vr_\phi} \Delta \vr + \Omega_{\vr_\theta \vr_\phi} \Delta \vr_\theta + \Omega_{\vr_\phi \eta_\theta} \Delta \eta_\theta + \Omega_{\vr_\phi \eta_\phi} \Delta \eta_\phi + \Omega_{\vr_\phi \disR} \cdot \Delta \disR \right] [1+\gamma\cos\theta] \Big) \nonumber \\
& + \frac{ C_1 P}{\gamma} \Big[  \vr\Delta\eta_\theta + \eta_\theta \Delta \vr \Big] = 0 , 
 \label{eqn: bifur gov 1}
\end{align}

\begin{align}
 & \fd{}{\theta} \Big(\left[ \Omega_{\eta_\theta \eta_\theta} \Delta \eta_\theta + \Omega_{\vr \eta_\theta} \Delta \vr + \Omega_{\vr_\theta \eta_\theta} \Delta \vr_\theta +  \Omega_{\vr_\phi \eta_\theta} \Delta \vr_\phi + \Omega_{\eta_\theta \eta_\phi} \Delta \eta_\phi + \Omega_{\eta_\theta \disR } \cdot \Delta \disR\right] [1+\gamma\cos\theta] \Big) \nonumber \\
& + \fd{}{\phi}  \Big(\left[ \Omega_{\eta_\phi \eta_\phi} \Delta \eta_\phi + \Omega_{\vr \eta_\phi} \Delta \vr + \Omega_{\vr_\theta \eta_\phi} \Delta \vr_\theta +  \Omega_{\vr_\phi \eta_\phi} \Delta \vr_\phi + \Omega_{\eta_\theta \eta_\phi} \Delta \eta_\theta + \Omega_{\eta_\phi \disR } \cdot \Delta \disR \right][1+\gamma\cos\theta] \Big) \nonumber \\
 &- \frac{C_1 P}{\gamma}  \Big[ \vr \Delta \vr_\theta + \vr_\theta \Delta \vr \Big]  = 0, 
 \label{eqn: bifur gov 2}
\end{align}

\begin{align}
 \Curl \bigg[\Omega_{\disR \disR} \Delta \disR + \Omega_{\vr \disR} \Delta \vr + \Omega_{\vr_\theta \disR} \Delta \vr_\theta + \Omega_{\vr_\phi \disR} \Delta \vr_\phi + \Omega_{\eta_\theta \disR} \Delta \eta_\theta + \Omega_{\eta_\phi \disR} \Delta \eta_\phi \bigg] = \mbf{0} , \label{eqn: bifur gov 3}
\end{align}
with periodic boundary conditions along $\theta$ and $\phi$ given by
\begin{align}
 \{ \Delta \vr, \Delta \eta, \Delta \disR, \Delta \vr_\theta, \Delta \vr_\phi,  \Delta \eta_\theta, \Delta \eta_\phi\} \bigr|_{\phi = 0} = \{ \Delta \vr, \Delta \eta, \Delta \disR, \Delta \vr_\theta, \Delta \vr_\phi,  \Delta \eta_\theta, \Delta \eta_\phi\} \bigr|_{\phi = 2 \pi}\, , \\
  \{ \Delta \vr, \Delta \eta, \Delta \disR, \Delta \vr_\theta, \Delta \vr_\phi,  \Delta \eta_\theta, \Delta \eta_\phi\} \bigr|_{\theta = 0} = \{ \Delta \vr, \Delta \eta, \Delta \disR, \Delta \vr_\theta, \Delta \vr_\phi,  \Delta \eta_\theta, \Delta \eta_\phi\} \bigr|_{\theta = 2 \pi}\, .
\end{align}
Since $\vr_\phi$ and $\eta_\phi$  vanish at the bifurcation point, $\Omega_{\vr \vr_\phi}$, $\Omega_{\vr \eta_\phi}$, $\Omega_{\vr_\theta \vr_\phi}$, $\Omega_{\vr_\theta \eta_\phi}$, $\Omega_{\vr_\phi \disR}$ and $\Omega_{\eta_\phi \disR}$ are also zero as shown in Appendix C (supporting file). Rearranging equations~\eqref{eqn: bifur gov 1} and~\eqref{eqn: bifur gov 2} gives
\begin{align}
&\left[[1+ \gamma\cos\theta] \left[-\Omega_{\vr \vr} + \fd{\Omega_{\vr \vr_\theta}}{\theta} \right]  - \gamma\sin\theta \Omega_{\vr \vr_\theta} + \frac{ C_1 P}{\gamma} \eta_\theta \right] \Delta \vr \nonumber \\
&+ \left[[1+\gamma\cos\theta]\left[- \Omega_{\vr \vr_\theta} + \fd{\Omega_{\vr_\theta \vr_\theta}}{\theta}\right] -\gamma\sin\theta \Omega_{\vr_\theta \vr_\theta} \right] \Delta \vr_\theta \nonumber \\
&+ \left[[1+\gamma\cos\theta]\Omega_{\vr_\theta \vr_\theta} \right]\Delta \vr_{\theta \theta} +\left[[1+\gamma\cos\theta]\left[ \fd{\Omega_{\vr_\phi \vr_\phi}}{\phi} \right] \right] \Delta \vr_{\phi} \nonumber \\ 
&+ \left[[1+\gamma\cos\theta]\Omega_{\vr_\phi \vr_\phi} \right] \Delta \vr_{\phi \phi}  + \left[[1+ \gamma\cos\theta] \left[-\Omega_{\vr \eta_\theta} + \fd{\Omega_{\vr_\theta \eta_\theta}}{\theta} \right]  - \gamma\sin\theta \Omega_{\vr_\theta \eta_\theta}  + \frac{C_1 P}{\gamma} \vr \right] \Delta \eta_\theta \nonumber \\
&+ \left[ [1+\gamma\cos\theta] \Omega_{\vr_\theta \eta_\theta} \right]\Delta \eta_{\theta\theta} +\left[ [1+\gamma\cos\theta] \left[ \fd{\Omega_{\vr_{\phi} \eta_{\phi}}}{\phi} \right]\right]\Delta\eta_{\phi} + \left[ [1+\gamma\cos\theta] \Omega_{\vr_\phi \eta_\phi}\right]\Delta\eta_{\phi \phi} \nonumber \\
&+ \left[ [1+\gamma\cos\theta]\left[ - \Omega_{\vr \disR} +\fd{\Omega_{\vr_\theta \disR}}{\theta} \right] -\gamma\sin\theta\Omega_{\vr_\theta \disR} \right]\cdot \Delta \disR + \left[[1+\gamma\cos\theta]\Omega_{\vr_\theta \disR}\right]\cdot\fd{\Delta \disR}{\theta} = 0
\label{eqn:gov_eq_1_explicitly}
\end{align}

\begin{align}
&\left[ [1+\gamma\cos\theta] \fd{\Omega_{\vr\eta_\theta}}{\theta} -\gamma\sin \theta \Omega_{\vr \eta_\theta} -\frac{C_1 P}{\gamma}\vr_\theta \right]\Delta \vr  \nonumber \\
&+ \left[[1+\gamma\cos\theta]\left[\Omega_{\vr \eta_\theta} + \fd{\Omega_{\vr_\theta \eta_\theta}}{\theta}\right] - \gamma\sin\theta \Omega_{\vr_\theta \eta_\theta} -\frac{C_1 P}{\gamma}\vr \right] {\Delta \vr}_{\theta} \nonumber \\
&+ \left[[1+\gamma\cos\theta]\Omega_{\vr_\theta \eta_\theta} \right]\Delta \vr_{\theta \theta} + \left[[1+\gamma\cos\theta]\fd{\Omega_{\vr_\phi \eta_\phi}}{\phi}  \right] \Delta \vr_{\phi} \nonumber \\
&+  \left[[1+\gamma\cos\theta]\Omega_{\vr_\phi \eta_\phi} \right] \Delta \vr_{\phi \phi} + \left[ [1+\gamma\cos\theta] \fd{\Omega_{\eta_\theta \eta_\theta}}{\theta}  - \gamma\sin\theta\Omega_{\eta_\theta \eta_\theta} \right]\Delta \eta_\theta  \nonumber \\
&+ \left[ [1+\gamma\cos\theta] \Omega_{\eta_\theta \eta_\theta} \right] \Delta \eta_{\theta\theta} + \left[ [1+\gamma\cos\theta]\fd{\Omega_{\eta_{\phi} \eta_{\phi}}}{\phi} \right]\Delta\eta_{\phi}  \nonumber \\
&+ \left[ [1+\gamma\cos\theta] \Omega_{\eta_\phi \eta_\phi}\right]\Delta\eta_{\phi \phi} + \left[ [1+\gamma\cos\theta]\fd{\Omega_{\eta_\theta \disR}}{\theta} -\gamma\sin\theta \Omega_{\eta_\theta \disR} \right]\cdot \Delta \disR \nonumber \\
&+ \left[[1+\gamma\cos\theta] \Omega_{\eta_\theta \disR}\right]\cdot\fd{\Delta \disR}{\theta} = 0
\label{eqn:gov_eq_2_explicitly}
\end{align}
The third governing equation~\eqref{eqn: bifur gov 3} physically means that the curl of the perturbed electrical field is zero, i.e.\ $\text{Curl}(\Delta \elecR) = \mathbf{0}$. As the electrical field does not change in the normal directions, i.e.\ $\fd{}{\mathbf{N}}(\Delta \elecR) = \mathbf{0}$, \eqref{eqn: bifur gov 3} reduces to
\begin{equation}
[\Omega_{\disR \disR} \Delta \disR + \Omega_{\vr \disR} \Delta \vr + \Omega_{\vr_\theta \disR} \Delta \vr_\theta + \Omega_{\vr_\phi \disR} \Delta \vr_\phi + \Omega_{\eta_\theta \disR} \Delta \eta_\theta + \Omega_{\eta_\phi \disR} \Delta \eta_\phi]\cdot \mbf{N}  = {0}.
\label{eqn:new gov 3}
\end{equation}
\subsection{Bifurcation of solution along the $\phi$ coordinate}
To study {prismatic-type} bifurcation of the solution along the $\phi$ coordinate, we consider the following perturbations of the fields superimposed upon the principal solution
\begin{align}
 \Delta \vr = \mathcal{F} \, \text{exp} ({im \phi}), \quad 
 \Delta \eta = \mathcal{G} \, \text{exp}({im \phi}), \quad
 \Delta \disR = \frac{\Phi_0}{\beta H} \mathcal{H} \, \text{exp} ({im \phi})\mbf{N} , \quad m \in \bbm{Z}^+. \label{eqn: ansatz perturb}
\end{align}
Here we have assumed $\mathcal{F}, \mathcal{G}$ and $\mathcal{H}$ are constants and, invoking the thin-membrane assumption, we restrict the solution $\Delta \disR$ to along the thickness direction $\mbf{N}$. 
Upon substitution of \eqref{eqn: ansatz perturb}, the equations~\eqref{eqn:gov_eq_1_explicitly}, \eqref{eqn:gov_eq_2_explicitly} and~\eqref{eqn:new gov 3} can be expressed in terms of three unknown variables $\mathcal{F}$, $\mathcal{G}$ and $\mathcal{H}$ as
\begin{align}
& \underbrace{\Bigg[ [1+\gamma\cos\theta] \left[-\Omega_{\vr \vr} + \fd{\Omega_{\vr \vr_\theta }}{\theta} \right]  - \gamma \sin\theta \Omega_{\vr \vr_\theta}+ \frac{C_1 P}{\gamma} \eta_\theta - m^2 [1+\gamma\cos\theta]\Omega_{\vr_\phi \vr_\phi} \Bigg]}_{M_{11}} \mathcal{F} \\
& \underbrace{- m^2 \left[ [1+\gamma\cos\theta]\Omega_{\vr_\phi \eta_\phi} \right]}_{M_{12}}\mathcal{G} + \underbrace{\frac{\Phi_0}{\beta H}  \Bigg[  \bigg[ [1+\gamma\cos\theta] \big[ - \Omega_{\vr \disR}+ \fd{\Omega_{\vr_\theta \disR}}{\theta} \big] - \gamma\sin\theta \Omega_{\vr_\theta \disR}  \bigg] \cdot \mbf{N}  \Bigg]}_{M_{13}} \mathcal{H} = 0 , \label{eqn:bifur_gov_1}\\
 &\underbrace{\bigg[[1+\gamma\cos\theta] \fd {\Omega_{\vr \eta_\theta }}{\theta} -\gamma\sin\theta \Omega_{\vr \eta_\theta }  - \frac{ C_1 P}{\gamma}  \vr_\theta - m^2 [1+\gamma\cos\theta]\Omega_{\vr_\phi \eta_\phi}  \bigg]}_{M_{21}} \mathcal{F} \nonumber \\
&\underbrace{-m^2\left[[1+\gamma\cos\theta]\Omega_{\eta_\phi \eta_\phi}\right]}_{M_{22}} \mathcal{G} + \underbrace{\frac{\Phi_0}{\beta H}\Bigg[ \bigg[ [1+\gamma\cos\theta] \fd{ \Omega_{\eta_\theta \disR}}{ \theta} - \gamma\sin\theta \Omega_{\eta_\theta \disR} \bigg] \cdot \mbf{N} \Bigg] }_{M_{23}} \mathcal{H} = 0 , \label{eqn:bifur_gov_2}
\end{align}
\begin{equation}
\underbrace{\Omega_{\vr \disR} \cdot\mbf{N}}_{M_{31}}\mathcal{F} +\underbrace{ [\frac{\Phi_0}{\beta H} \Omega_{\disR \disR}]}_{M_{33}} \mathcal{H} = 0. \label{eqn:bifur_gov_3}
\end{equation}
Finally the governing system of equations~\eqref{eqn:bifur_gov_1},~\eqref{eqn:bifur_gov_2} and~\eqref{eqn:bifur_gov_3} can be expressed in a matrix form as
\begin{equation}
\underbrace{\begin{bmatrix}
   M_{11} & M_{12}  & M_{13}  \\
   M_{21} & M_{22}  & M_{23}  \\
   M_{31} & 0 & M_{33} \\
 \end{bmatrix}}_{\mbf{M}}
 \begin{bmatrix}
	\mathcal{F} \\
   \mathcal{G} \\
   \mathcal{H} \\
 \end{bmatrix}
 =
 \begin{bmatrix}
  0\\
  0 \\
  0 \\
 \end{bmatrix}.
 \label{eqn:matrix_system}
\end{equation}
If $\text{det}(\mbf{M}) = 0$, the system of equations has a solution, which means the toroidal membrane could lose stability due to a bifurcation in the $\phi$ direction.

\section{Numerical examples}
\label{sec:numerical_results}
Section~\ref{sec:principle_solution} presents numerical results obtained by solving the governing equations~\eqref{eqn: final principal coupled ODE 1} and~\eqref{eqn: final principal coupled ODE 2} with associated boundary conditions~\eqref{eqn:boundary_condition}. These results are only valid before the torus loses stability. 
Recall that as the membrane inflates, there are three kinds of possible instabilities considered here. They are snap-through, wrinkling, and bifurcation in the $\phi$ direction. 
If wrinkling happens first, the energy relaxation method is adopted and the governing equation are modified as described in Section~\ref{sec:energy_relaxation}.
The new solutions obtained with the energy relaxation method  are computed in Section~\ref{sec:wrinkling_solution} and compared with the principle solutions .
If the bifurcation happens first, the toroidal membrane loses its symmetry along the $\phi$ direction at the critical point as demonstrated in Section~\ref{sec: loss of symmetry}. 
The principle solution becomes unstable after the bifurcation point.

\subsection{Principal solution and limit point instability}
\begin{figure}
\centering
	\includegraphics[width=0.6\linewidth]{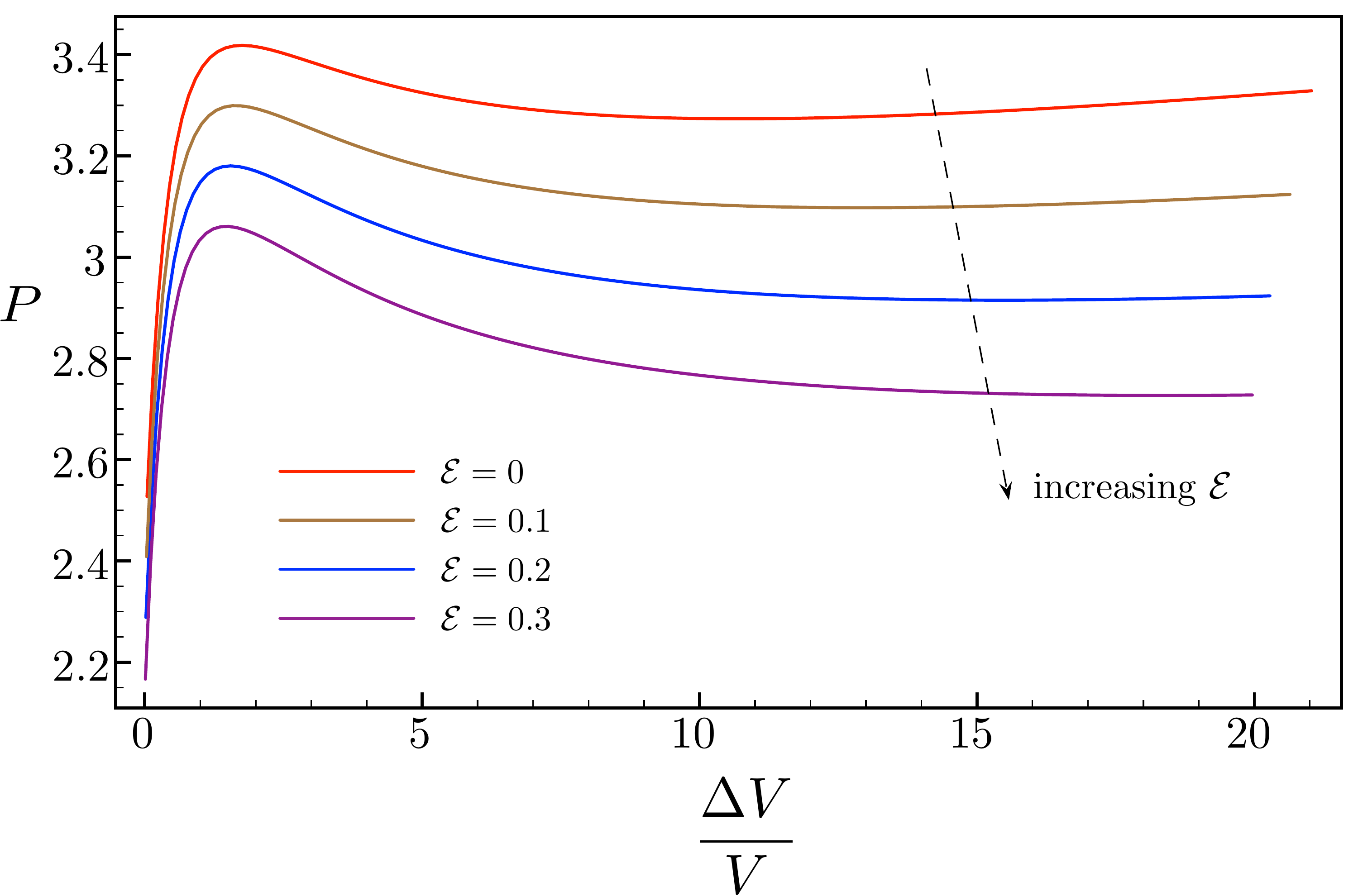}
  \caption{Plot of pressure against volume change for a toroidal membrane with geometry parameter $\gamma = 0.6$ and material parameter $\alpha = 0.2$ for different electrical loads $\mcal{E}$.} 
	\label{fig:gamma06_alpha02_1st}
\end{figure}

We first consider the behaviour of an inflating electro-elastic toroidal membrane under different electrical loads. Figure~\ref{fig:gamma06_alpha02_1st} shows the plots of pressure against volume change for a toroidal membrane with aspect ratio $\gamma = 0.6$, material coefficient $\alpha = 0.2$ and an electrical load $\mcal{E}$ increasing from $0$ to $0.3$. {The profiles of these plots are similar. The solution without electrical load is verified using the results in~\cite{Reddy2017}. }
At small deformation, the membrane has a large stiffness against inflation. Thus the ratio of pressure to volume change is large at the beginning. The ratio stays the same under different electrical loads. As the electrical load increases the limit point also occurs at a similar volume change, but the limit point pressure is significantly decreased. After the limit point for different electrical loads, the pressures all decrease with increasing volume. After a finite volume change, the membrane regains stiffness and the pressures start to increase as the volume increases. The same trend can be seen in Figures~\ref{fig:wrinkles_comparations} and~\ref{fig:Bifurcations} for a toroidal membrane with $\gamma = 0.4$ and $\alpha = 0.1, 0.2 \text{ and } 0.3$. 

\begin{figure}
\centering
\begin{subfigure}[b]{0.39\linewidth}
\centering
	\includegraphics[width=\linewidth]{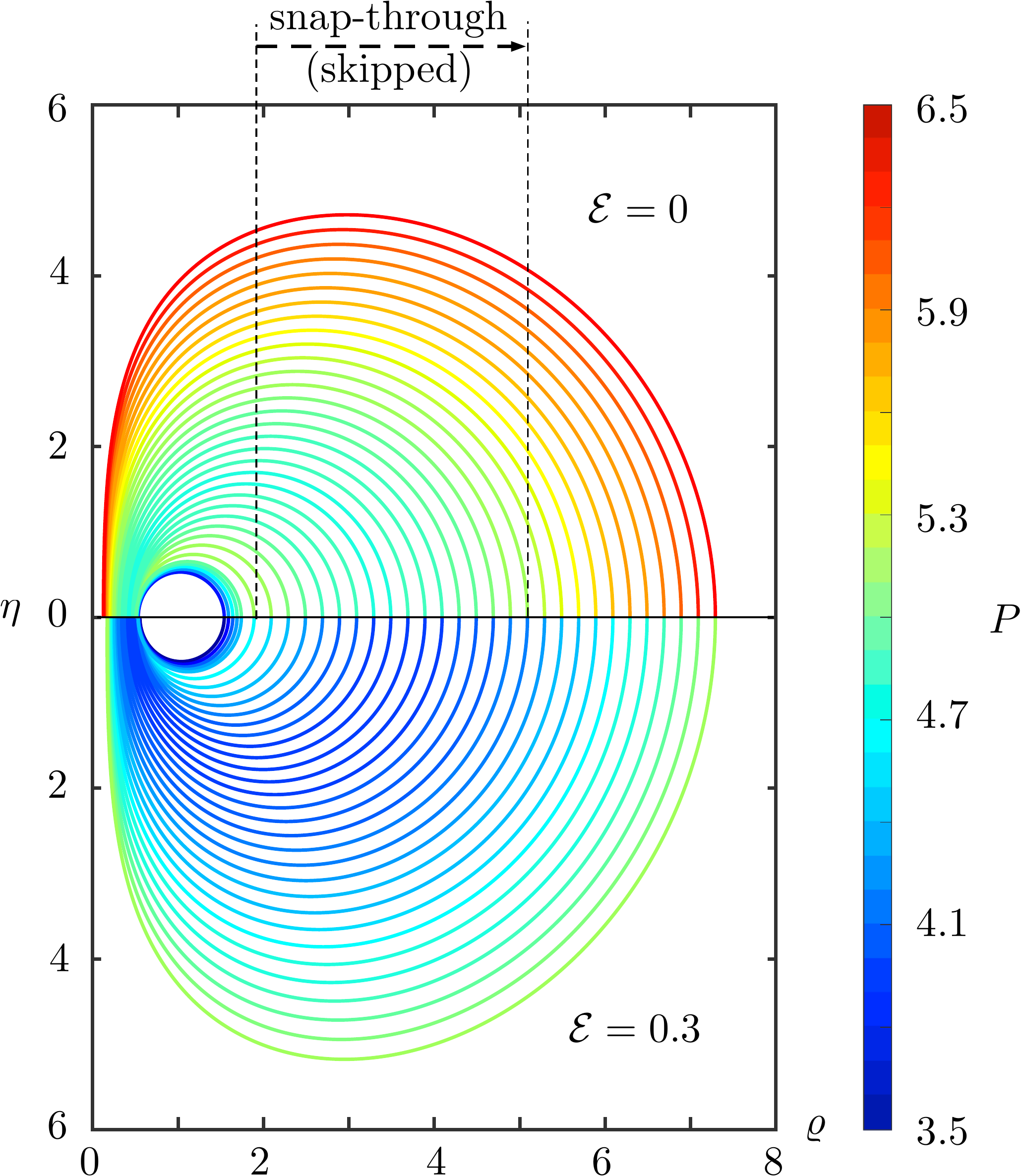}
	\caption{}
		\label{fig:profiles_pressure_1}
\end{subfigure}
\begin{subfigure}[b]{0.39\linewidth}
\centering
	\includegraphics[width=\linewidth]{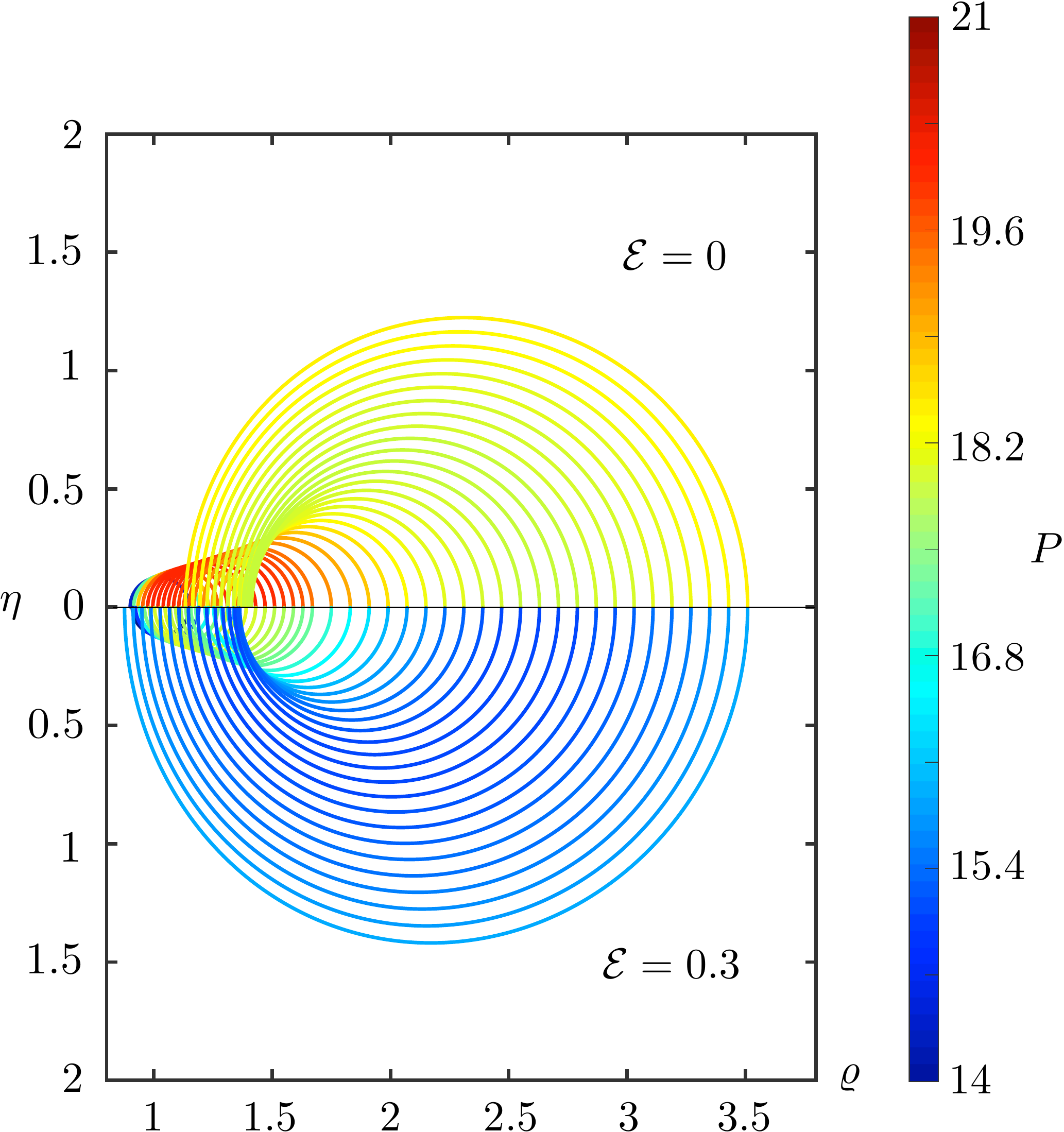}
	\caption{}
		\label{fig:profiles_pressure_2}
\end{subfigure}
	\caption{Cross-section profiles during inflation for toroidal membrane with (a) $\gamma = 0.4$ and (b) $\gamma = 0.1$. The material parameter $\alpha = 0.2$ for both cases. The upper halves represent the deformation under pure elastic loads. The lower halves represent the deformation with an electrical load $\mcal{E} = 0.3$.}  
	\label{fig:profiles_pressure}
\end{figure}

{ Figure~\ref{fig:profiles_pressure} compares and visualises the inflation of purely elastic membranes and eletroelastic membranes with $\mcal{E} = 0.3$. 
The principle solutions are solved for increasing $\vr$ at $\theta = 0$ and the deformed profiles for both cases are coloured by the pressure $P$. The electroelastic membrane can achieve the same volume change as the elastic membrane under considerably less pressure. } 
We also note the significant difference in inflation profiles of stout ($\gamma  = 0.4$) and slender ($\gamma = 0.1$) tori. 
For large $\gamma$ values shown in Figure \ref{fig:profiles_pressure_1}, the inflation (prescribed by position of outer end $\rho$ at $\theta = 0$) causes the inner end ($\theta  = \pi$) to move slightly inwards. However, for small $\gamma$ values the inner end ($\theta = \pi$) first moves outward and then inwards as shown in Figure \ref{fig:profiles_pressure_2}. The level of inward movement is significantly accentuated in the 
presence of the electric field. 
This dependence of $\rho(\pi)$ and $\rho(0)$ on the aspect ratio and material properties is of significance while designing actuator mechanisms from EAPs as detailed in~\cite{Venkata2020}.
\label{sec:principle_solution}

In an experimental setting, inflation of the membrane can be accomplished using either a pressure controlled or a volume controlled process.
%
 During the inflation of the membrane, one can control the pressure inside the membrane.  When the pressure reach the limit point in a pressure controlled experiment, there will be a sudden volume change of the membrane. This is known as a snap-through buckling of the hyperelastic membrane as described in Figures~\ref{fig:snap-through} and~\ref{fig:profiles_pressure_1}. { We also note that it would be possible for snap-through to occur prior to reaching the limit point if there are imperfections. The post bifurcation state can be analysed using a range of techniques including the Maxwell criterion~\cite{thompson2014quantified, thompson2015advances}. }

Figure~\ref{fig:profiles_pressure_2} shows the inflation process for a toroidal membrane with a smaller aspect ratio, $\gamma = 0.1$ and $\alpha = 0.2$.  The pressure to reach the limit point is much higher than a toroidal membrane with $\gamma = 0.4$. For such tori, the snap-through path is significantly longer and cannot be shown in Figure~\ref{fig:profiles_pressure_2}. 

The snap-through phenomenon can be eliminated if the torus is inflated in a volume-controlled experiment. Our calculations show that the snap-through phenomenon can also be eliminated in a mass-controlled experiment (assuming ideal gas law under constant temperature, mass is proportional to $PV$) since mass and volume increase monotonically for an electroelastic torus.

\subsection{Computation of wrinkling instability}
\begin{figure}
\centering
\begin{subfigure}[hb]{0.49\linewidth}
\centering
	\includegraphics[width=\linewidth]{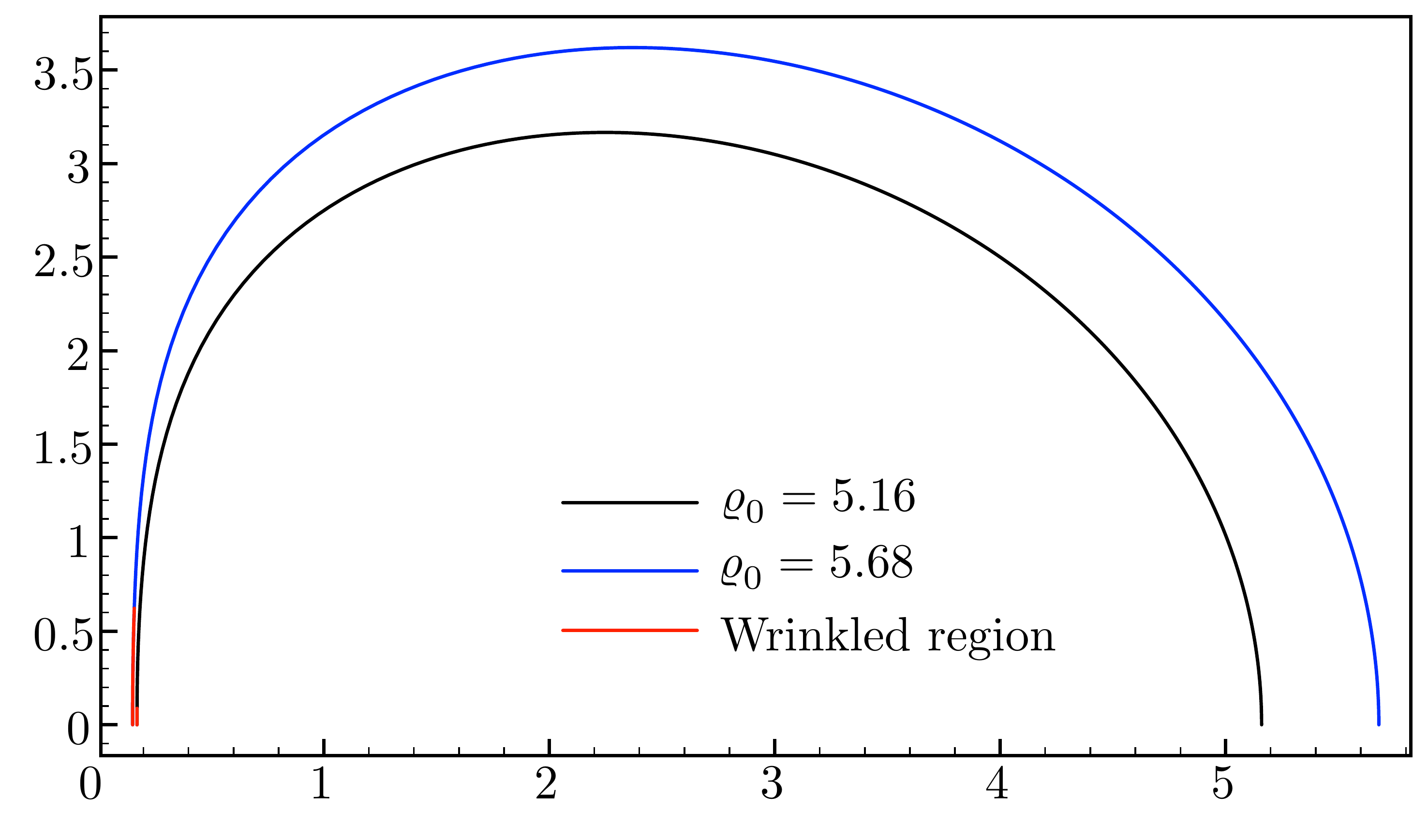}
	\caption{$\gamma = 0.4$.}
	\label{fig:wrinkling_profile_gamma04}
\end{subfigure}
\begin{subfigure}[hb]{0.49\linewidth}
\centering
	\includegraphics[width=\linewidth]{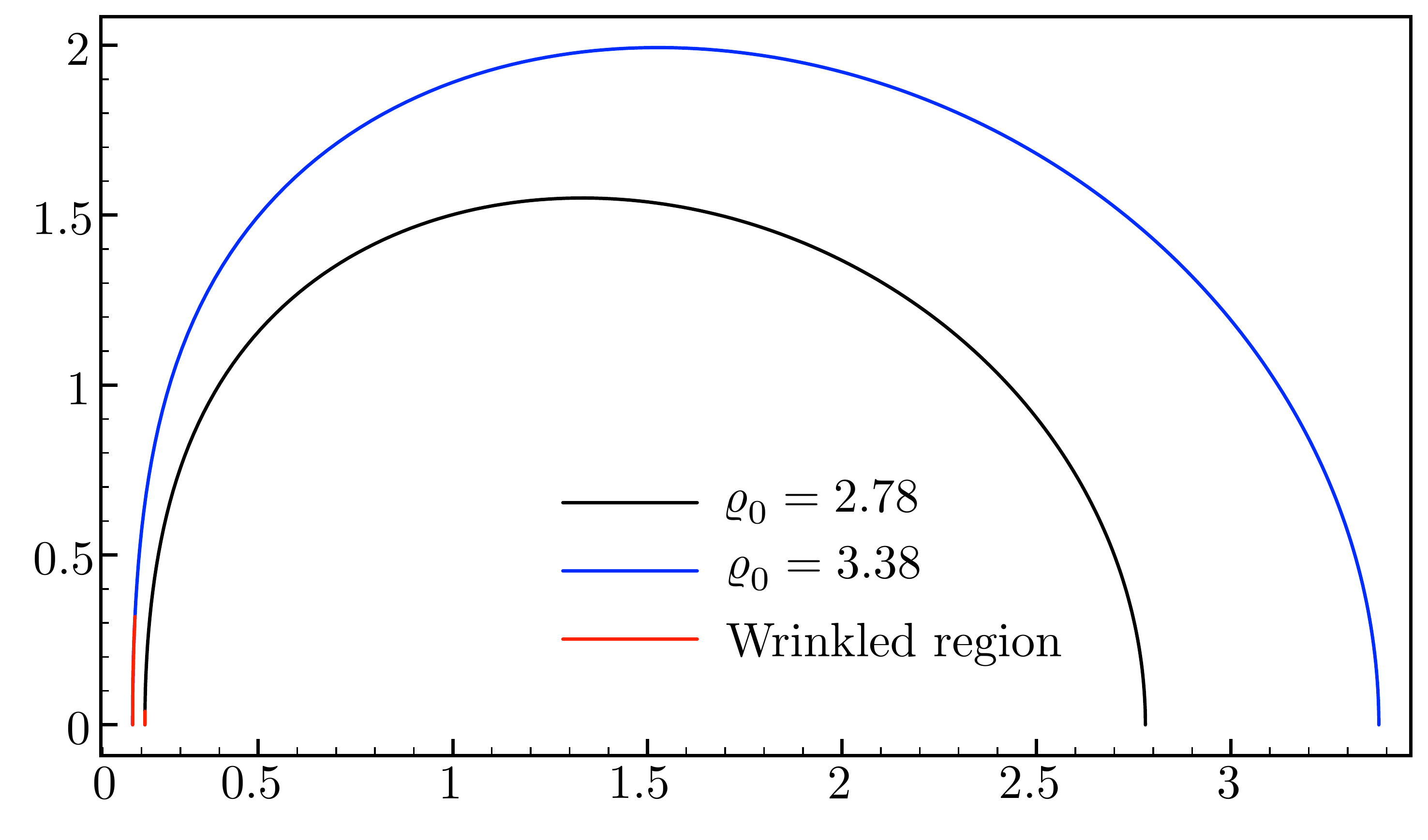}
	\caption{$\gamma = 0.6$.}
	\label{fig:wrinkling_profile_gamma06}
\end{subfigure}
	\caption{Membrane profiles showing wrinkling regions. The black profile shows the wrinkling starting to form and the blue profile indicate the wrinkling region is { $\theta = ({29}/{30})\pi$} to $\pi$. Both cases have the same material property $\alpha = 0.3,$  Electrical load is $\mcal{E} = 0.1$. A toroidal membrane with larger aspect ratio wrinkles with less inflation.}
	\label{fig:wrinkling_profiles}
\end{figure}

Wrinkling occurs during inflation when the in-plane stress in any direction becomes zero. 
The two in-plane principal stress components are calculated using the principal solutions obtained from~\eqref{eqn:stress_components}. 
For the toroidal membrane, wrinkling first occurs at the inner region where $\theta \approx \pi$.  Figures~\ref{fig:wrinkling_profile_gamma04} and~\ref{fig:wrinkling_profile_gamma06} are two examples of membranes with negative stresses calculated from the principal solutions. Figure~\ref{fig:wrinkling_profile_gamma04} shows a half cross section of the membrane with aspect ratio $\gamma = 0.4$, material parameter $\alpha = 0.3$, and the electrical load $\mcal{E} = 0.1$. 
Wrinkles start to form when { $\vr \approx 5.16$} at $\theta = 0$. 
When { $\vr \approx 5.68$} at $\theta = 0$, the wrinkling region is between { $\theta = ({29}/{30})\pi$} to $\pi$. 
Keeping the material parameter and electrical load the same, a torus with large aspect ratio $\gamma = 0.6$
starts to wrinkles when {$\vr \approx 2.78$} at $\theta = 0$.  When {$\vr \approx 3.38$}, the wrinkling region is between {$\theta = ({29}/{30})\pi$} to $\pi$. 

\label{sec:wrinkling_solution}

\begin{figure}
\centering
\begin{subfigure}[b]{\linewidth}
\centering
	\includegraphics[width=0.7\linewidth]{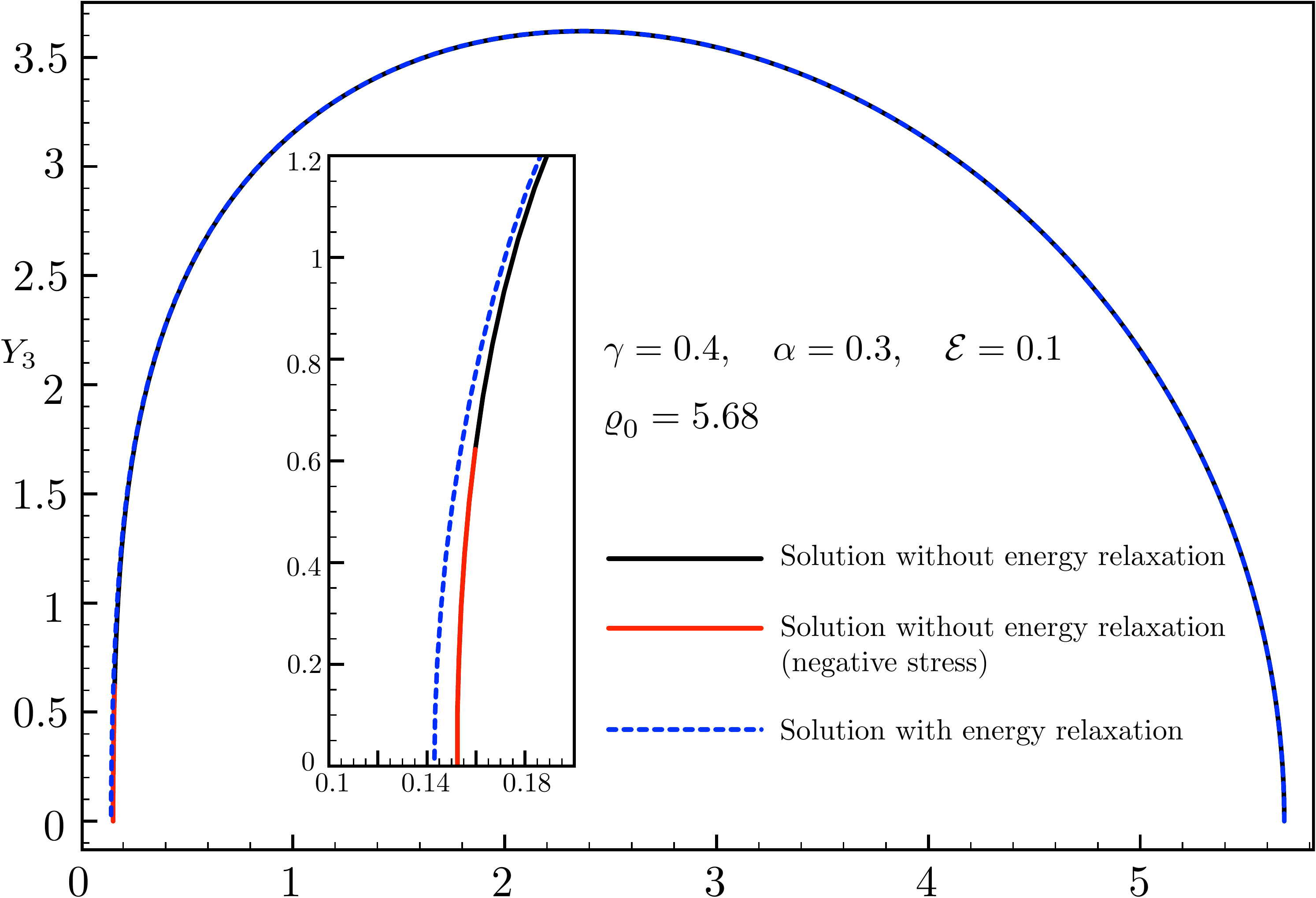}
	\caption{}
	\label{fig:wrinkles_compareg04a03e01}
\end{subfigure}
\begin{subfigure}[b]{\linewidth}
\centering
	\includegraphics[width=0.7\linewidth]{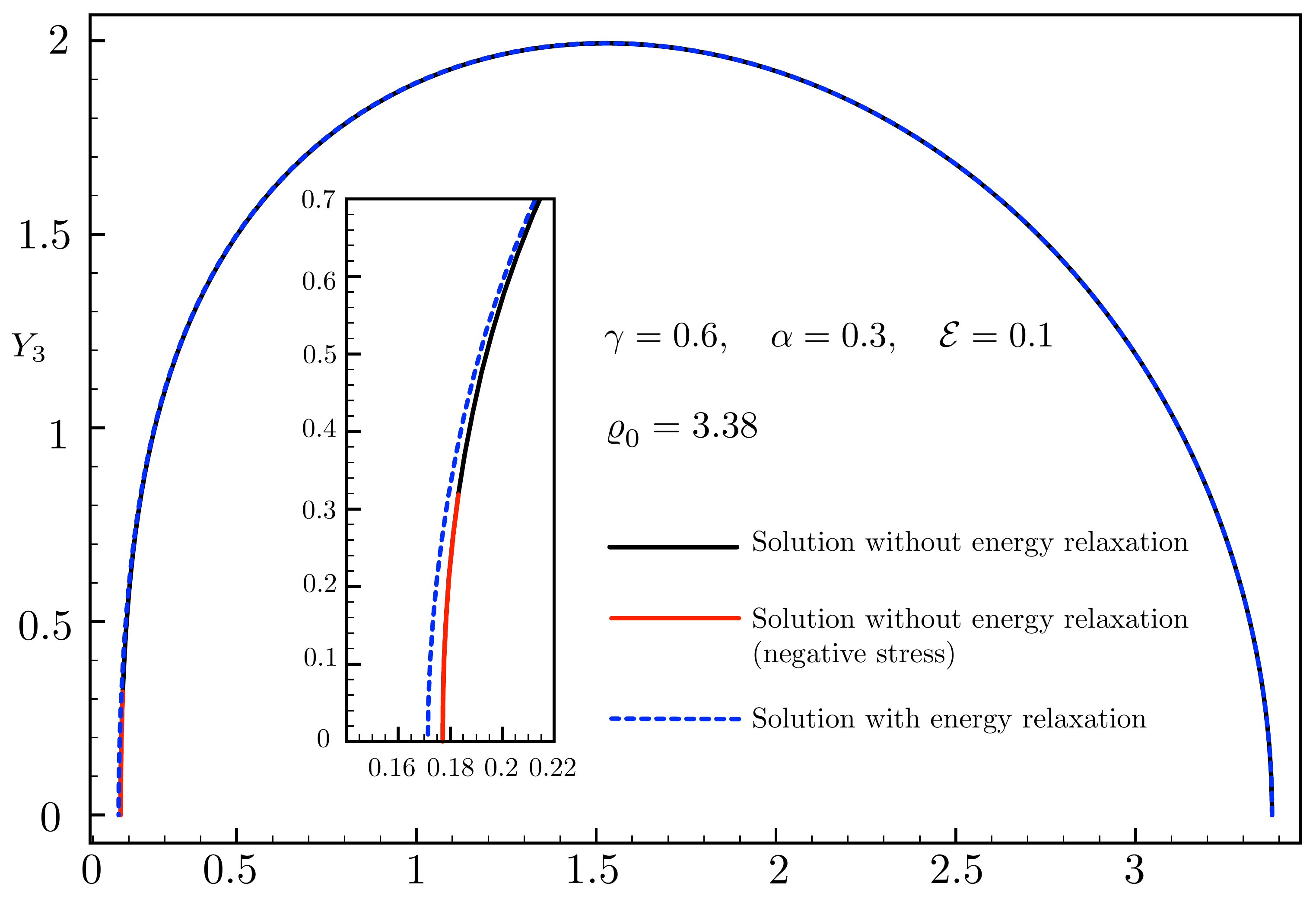}
	\caption{}
	\label{fig:wrinkles_compareg06a03e01}
\end{subfigure}
	\caption{ Comparison of the membrane profiles computed using the principal solution and the relaxed energy solution. }
	\label{fig:wrinkles_comparations}
\end{figure}

\begin{figure}
\centering
\begin{subfigure}[hb]{\linewidth}
\centering
	\includegraphics[width=0.6\linewidth]{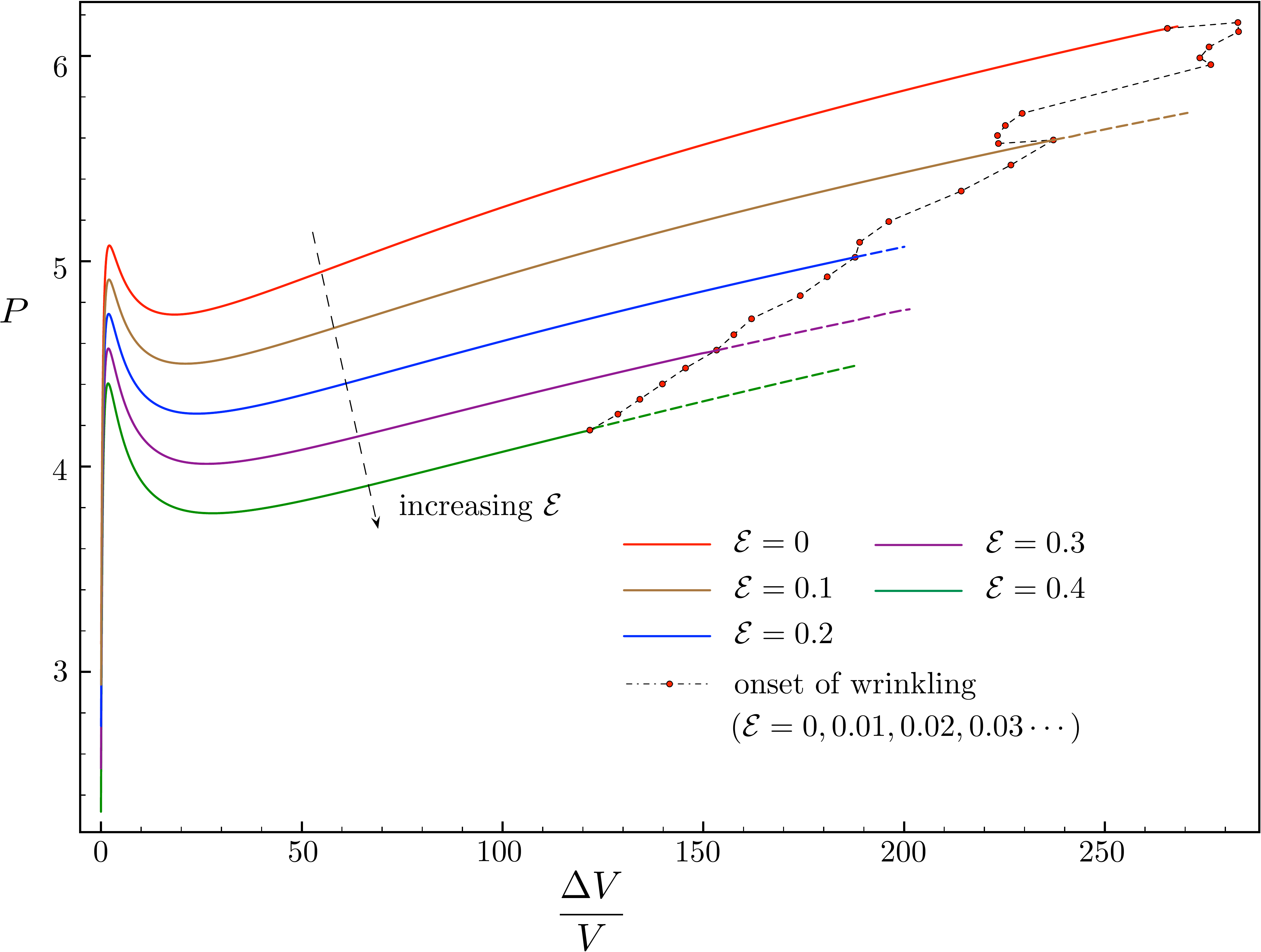}
	\caption{$\gamma = 0.4,\quad \alpha = 0.2$}
	\label{fig:gamma04_alpha02_wrinkle}
\end{subfigure}
\begin{subfigure}[hb]{\linewidth}
\centering
	\includegraphics[width=0.6\linewidth]{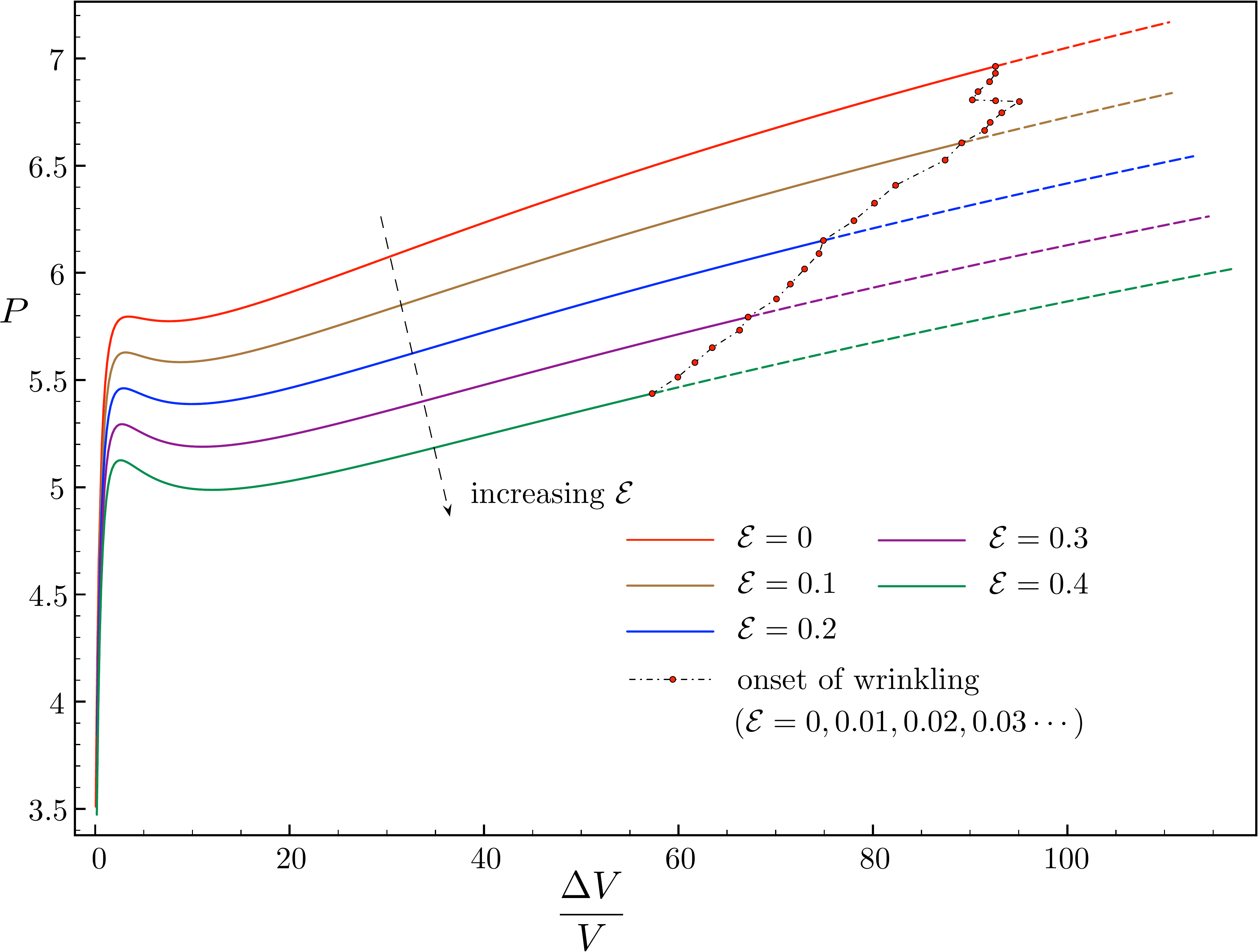}
	\caption{$\gamma = 0.4,\quad \alpha = 0.3$}
	\label{fig:gamma04_alpha03_wrinkle}
\end{subfigure}
	\caption{Plots of the pressure against the volume change with different electrical loads. Red dots indicate where the first wrinkles occur. {The dashed portions of the loading curves denote the unstable region post-wrinkling.} }
	\label{fig:wrinkles_comparations}
\end{figure}

The relaxed form of the energy density~\eqref{eqn:RE_energy_density} is adopted when negative stresses are detected. The modified ODE system~\eqref{eqn:RE_ODEs} is used for computing the profile of the wrinkled membrane. Figures~\ref{fig:wrinkles_compareg04a03e01} and~\ref{fig:wrinkles_compareg06a03e01} compare the {membrane profiles of the principal} solution with the {solution obtained using the relaxed energy density function for tori with two different aspect ratios ($\gamma = 0.4$ and $\gamma = 0.6$).
}
Figures~\ref{fig:gamma04_alpha02_wrinkle} and~\ref{fig:gamma04_alpha03_wrinkle} compare the plots of the pressure against the volume change with different electrical loads. Two toroidal membranes with same aspect ratio $\gamma = 0.4$ but different material parameters, $\alpha = 0.2$ and $0.3$, are investigated.  The red dots indicate the first appearance of wrinkling during inflation.
{The loading curve post the occurrence of wrinkling is dashed to demonstrate the unstable region.}
As the electrical loads increase, the wrinkles occur even at low values of pressure and inflation. {However, we note that the volume change for the first appearance of wrinkling does not decrease monotonically with increasing $\mcal{E}$. Significant variable changes in the onset points of wrinkling are observed in both cases when $\mcal{E}<0.1$. }

\subsection{Loss of symmetry}
\label{sec: loss of symmetry}
\begin{figure}
\centering
\begin{subfigure}[hb]{\linewidth}
\centering
	\includegraphics[width=0.52\linewidth]{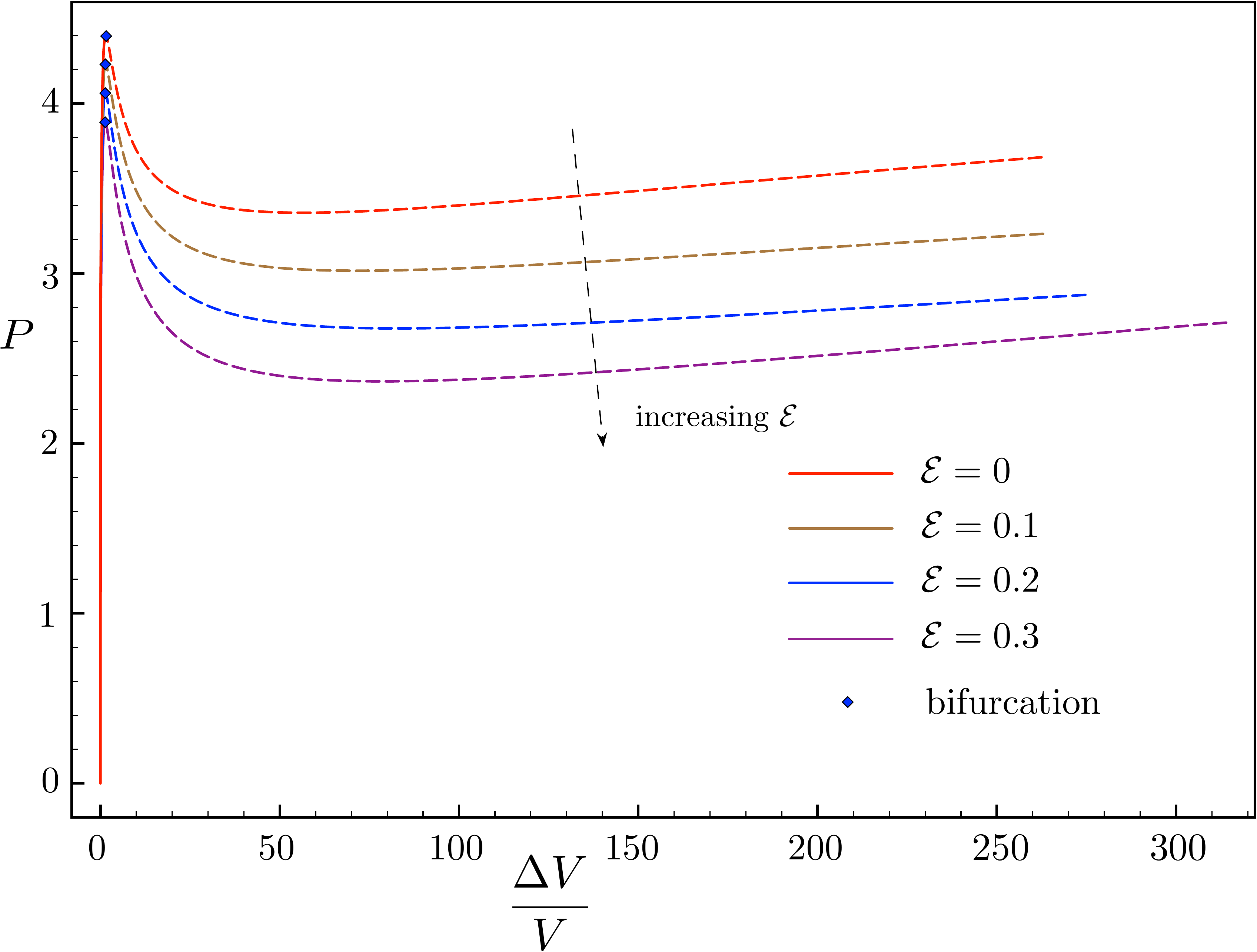}
	\caption{$\gamma = 0.4,\quad \alpha = 0.1$}
	\label{fig:gamma04_alpha01_epsilon}
\end{subfigure}
\begin{subfigure}[hb]{\linewidth}
\centering
	\includegraphics[width=0.52\linewidth]{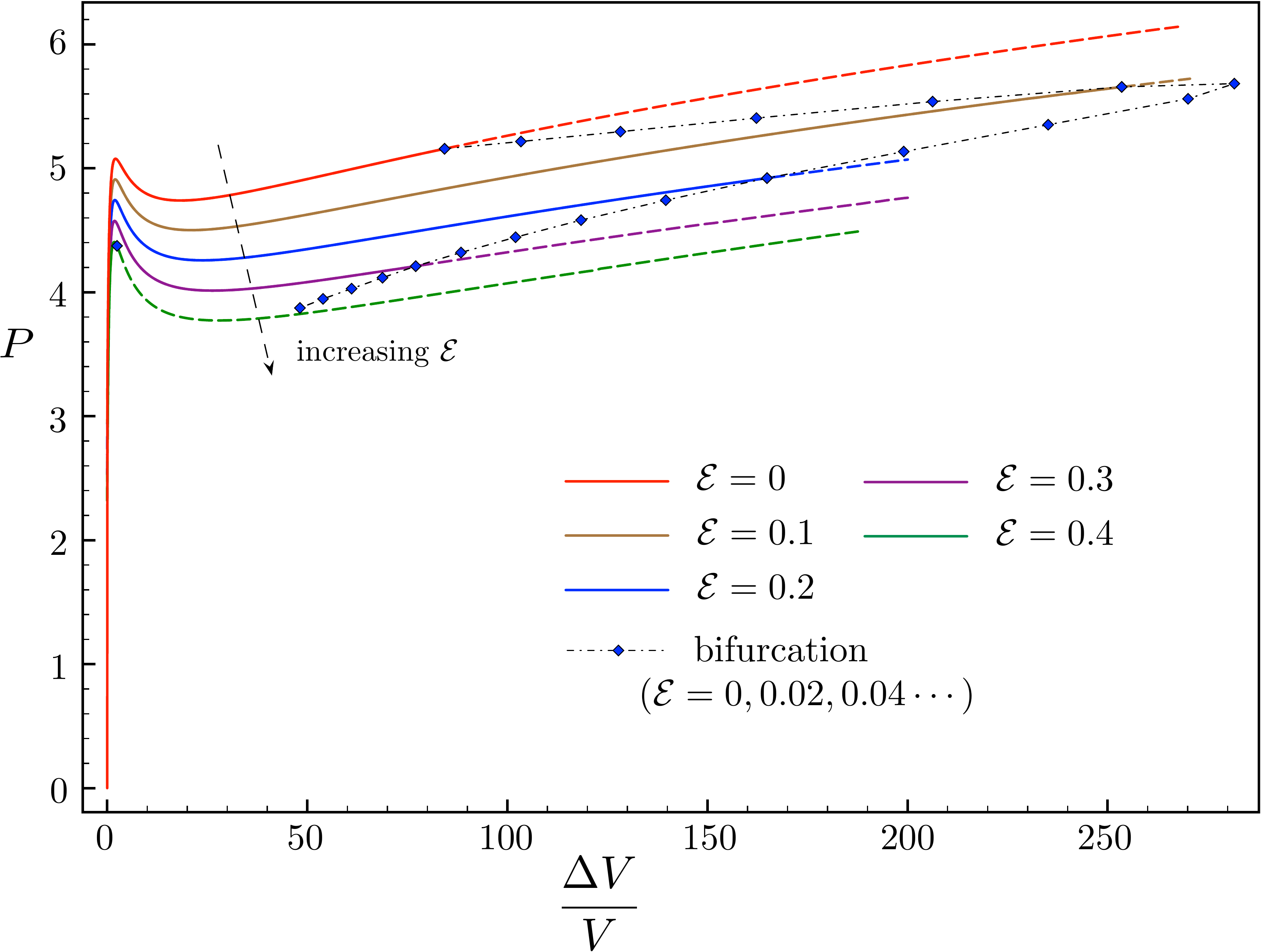}
	\caption{$\gamma = 0.4,\quad \alpha = 0.2$}
	\label{fig:gamma04_alpha02_epsilon}
\end{subfigure}
\begin{subfigure}[hb]{\linewidth}
\centering
	\includegraphics[width=0.52\linewidth]{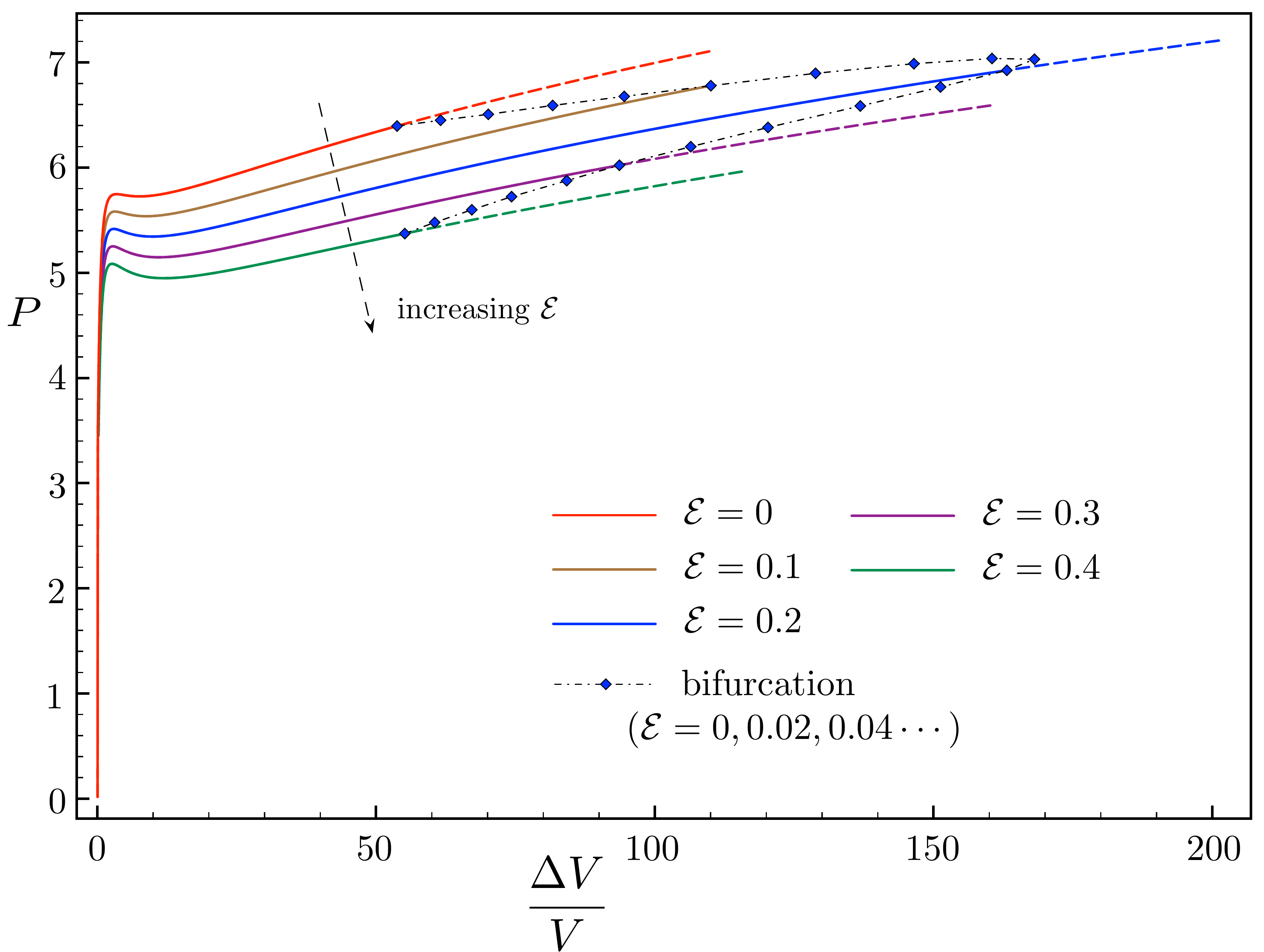}
	\caption{$\gamma = 0.4,\quad \alpha = 0.3$}
	\label{fig:gamma04_alpha03_epsilon}
\end{subfigure}
	\caption{ Plots of the pressure against the volume change with different electrical loads. Blue diamonds indicate {the onset of bifurcation that causes loss of symmetry. The dashed portions of the loading curves denote the unstable region after loss of symmetry}}
	\label{fig:Bifurcations}
\end{figure}

The loss of symmetry along the $\phi$ direction during inflation is examined. This bifurcation occurs when the second variation of the energy density function is zero as computed by checking the determinant of matrix $\mbf{M}$ in Equation~\eqref{eqn:matrix_system}. 
Three toroidal membranes with the same aspect ratio $\gamma = 0.4$ but different material parameter $\alpha = 0.1, 0.2 \text{ and } 0.3$ are compared. Figure~\ref{fig:gamma04_alpha01_epsilon} shows the plot of pressure against volume change for increasing electrical load $\mcal{E} = 0, 0.1, 0.2 \text{ and }0.3$ and the material parameter $\alpha = 0.1$. The bifurcation occurs close to the limit points which is similar to the neo-Hookean material investigated in~\cite{Venkata2020}.  
Figures~\ref{fig:gamma04_alpha02_epsilon} and~\ref{fig:gamma04_alpha03_epsilon} show the same plots for membranes with $\alpha = 0.2 \text{ and } 0.3$. 
In this case bifurcation occurs at the strain hardening stage. 
All the bifurcation points are found at buckling mode $m =1$. 
However, the electrical {load significantly influences the onset of bifurcation. }
When the electrical load remain small, it delays the appearance of bifurcation. 
After the electrical load exceeds a certain value, bifurcation happens with a smaller volume change. {For very large electrical loads, bifurcation occurs close to the limit point as shown for $\mcal{E}=0.4$ in Figure~\ref{fig:gamma04_alpha02_epsilon}.}

\begin{figure}
\centering
\begin{subfigure}[b]{0.49\linewidth}
\centering
	\includegraphics[width=\linewidth]{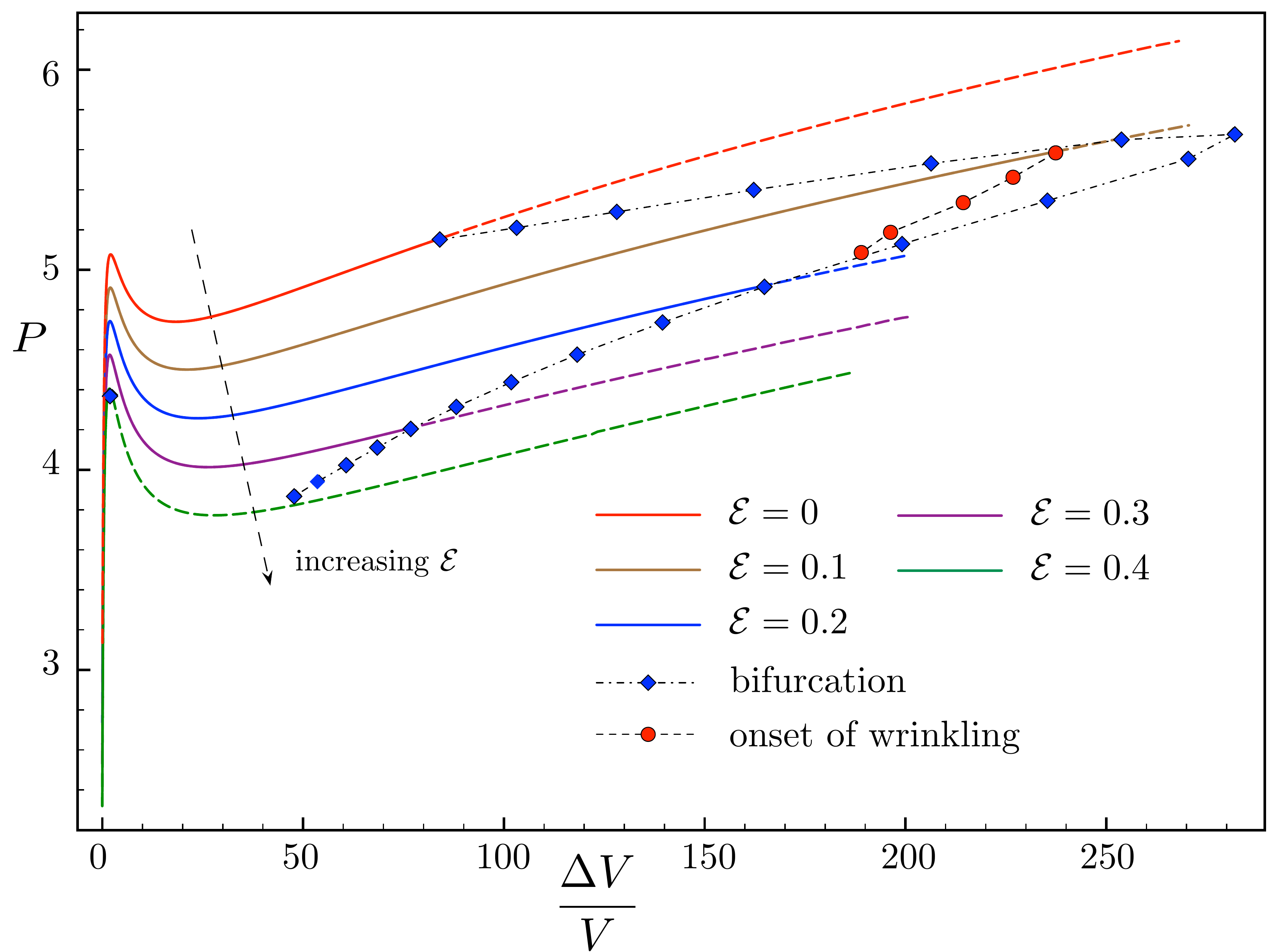}
	\caption{$\gamma = 0.4,\quad \alpha = 0.2$}
	\label{fig:gamma04_alpha02_two}
\end{subfigure}
\begin{subfigure}[b]{0.49\linewidth}
\centering
	\includegraphics[width=\linewidth]{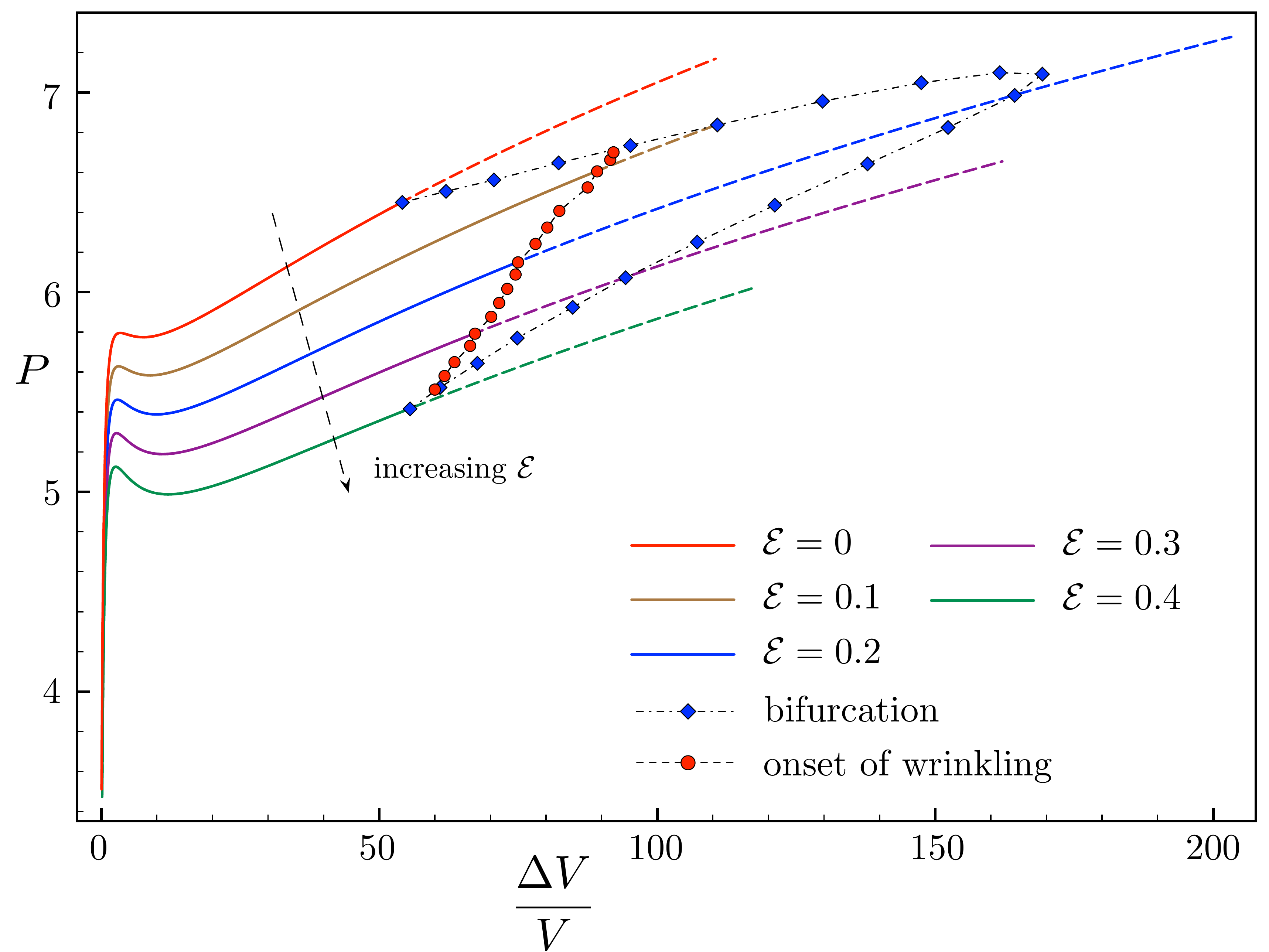}
	\caption{$\gamma = 0.4,\quad \alpha = 0.3$}
	\label{fig:gamma04_alpha03_two}
\end{subfigure}
\begin{subfigure}[b]{0.49\linewidth}
\centering
	\includegraphics[width=\linewidth]{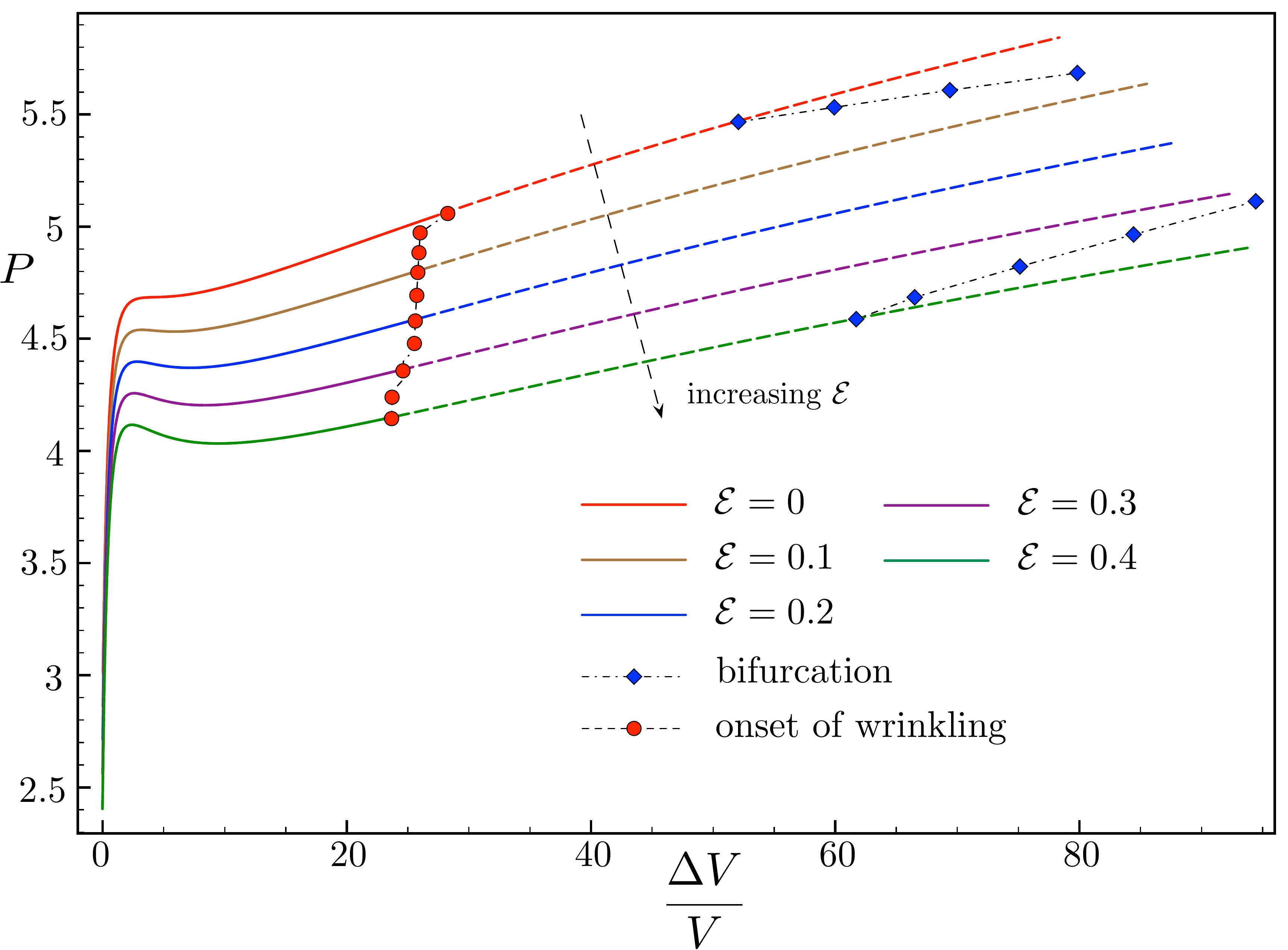}
	\caption{$\gamma = 0.5,\quad \alpha = 0.3$}
	\label{fig:gamma05_alpha03_two}
\end{subfigure}
	\caption{Plots of the pressure against the volume change with different electrical loads. Red dots indicate where the first wrinkle occurs. Blue diamonds indicate where the bifurcation points. { As the electric load increases, the bifurcation occurs at a larger volume change. However, after a certain value of $\mcal{E}$, the bifurcation occurs earlier. } }
	\label{fig:phase_changes}
\end{figure}

{Finally, we investigate the combined effect of both wrinkling and loss of symmetry. 
We compare the response of membranes with same geometry ($\gamma = 0.4$) and different material parameter ($\alpha = 0.2, 0.3$) in Figures~\ref{fig:gamma04_alpha02_two} and~\ref{fig:gamma04_alpha03_two}, and membranes with same material parameter ($\alpha = 0.3$) and different geometry ($\gamma = 0.4, 0.5$) in Figures~\ref{fig:gamma04_alpha03_two} and~\ref{fig:gamma05_alpha03_two}.
Wrinkling is the dominant instability mode for membranes with a large aspect ratio $\gamma = 0.5$ irrespective as to the electrical load.
For membranes with a lower aspect ratio $\gamma = 0.4$, the combined effect of the material parameter $\alpha$ and the electrical load $\mcal{E}$ dictates the instability mode.
Typically loss of symmetry occurs before wrinkling for very small and very large electrical loads, and wrinkling is preferred for moderate $\mcal{E}$ values.
A larger value of the material parameter $\alpha$ can increase the zone in which wrinkling is preferred.
}

\begin{figure}
 \centering
    \includegraphics[width=0.6\linewidth]{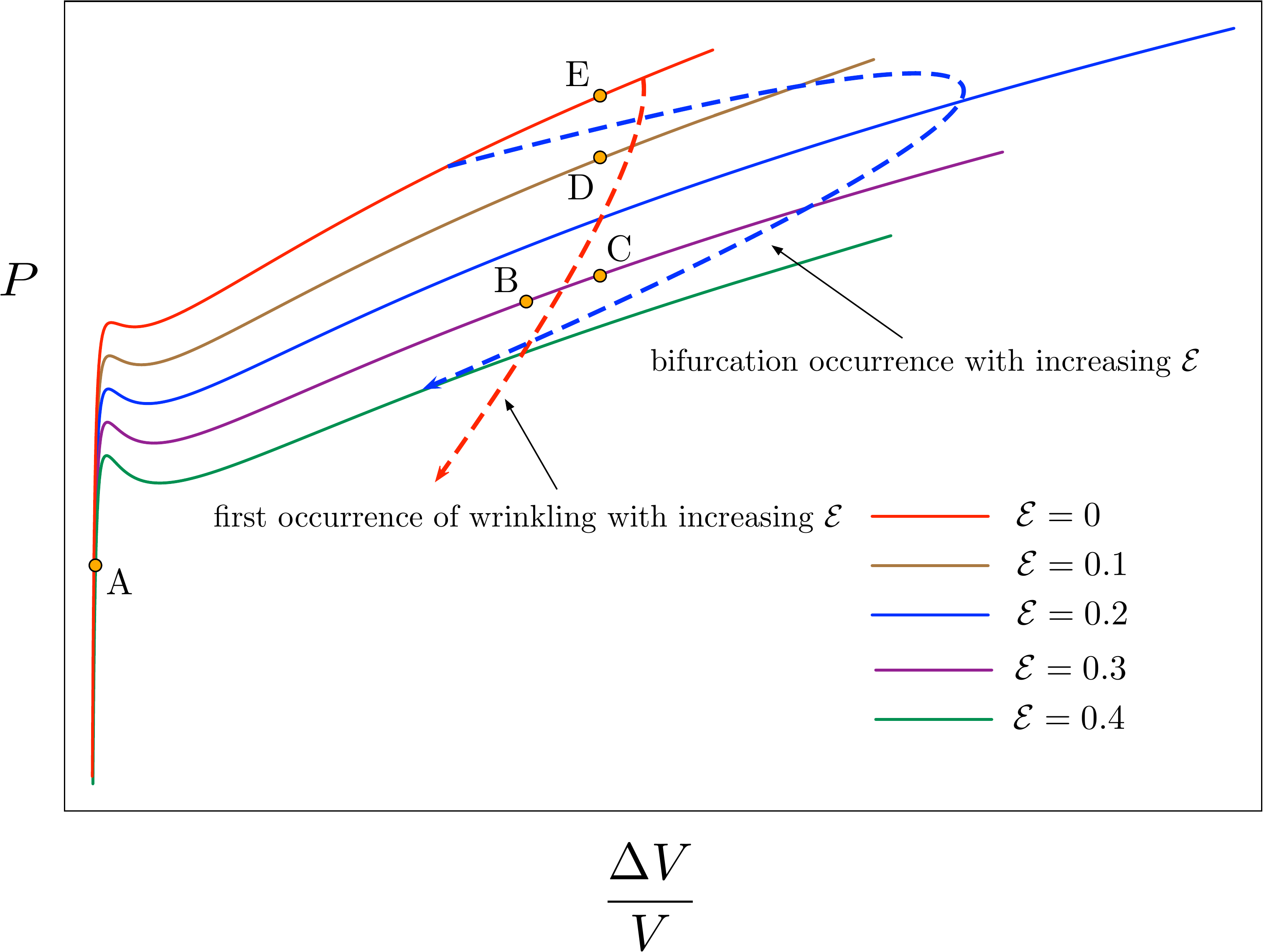}
    \caption{{Instability interaction diagram. One can  induce (or eliminate) wrinkles and loss of symmetry in a reversible manner by controlling the applied electric load $\mcal{E}$ and the volume of fluid in the membrane; for example by traversing the path A$\rightarrow$B$\rightarrow$C$\rightarrow$D$\rightarrow$E. } }
    \label{fig: sample application}
\end{figure}

This interesting interaction between the various modes of instabilities can be used as an actuation mechanism.
{For example, consider the idealised phase-space diagram depicted in Figure~\ref{fig: sample application} based on the response in Figure~\ref{fig:phase_changes}.}
An initial loading can be performed in the presence of electric field along the path {A$\rightarrow$B$\rightarrow$C.}
Moving from B to C induces wrinkles in the torus.
Upon reducing the intensity of the electric load while keeping the volume constant, one can remove wrinkles by moving to the state D.
Upon further reducing the electric load while keeping the volume constant, one can move to the state E where the torus is no longer symmetric.

\section{Conclusions}
\label{sec: conclusions}
A study of the mechanics of inflation of a toroidal membrane under large coupled electromechanical loading has been presented.
A numerical scheme has been developed to solve the highly nonlinear coupled ODEs to capture the rapidly changing solution for small inflation values together with the aid of an arc-length method.
The classical limit point instability for inflated membranes is recovered and the limit point pressure can be significantly reduced upon the application of a potential difference across the thickness of the torus.
Wrinkling instability has been modelled using an extension of the tension field theory to electroelasticity by employing a relaxed energy approach.
The torus loses its rotational symmetry at large values of inflation and this instability has been computed using a second variation based analysis.
This critical bifurcation point varies nonlinearly with the electrical load - it is delayed for moderate electric loads and is then favoured for large electric loads.

This interaction between the various instability modes can be exploited for the development of actuation mechanisms. One such example has been demonstrated in Figure~\ref{fig: sample application}. 
The phase space of this problem can be traversed by controlling the electric load and either one of the pressure or the volume thereby providing a great deal of flexibility in engineering design. Such features will be exploited in future works.

\section*{Acknowledgements}
This work was supported by the UK Engineering and Physical Sciences Research Council grant EP/R008531/1 for the Glasgow Computational Engineering Centre.\\
\indent Basant Lal Sharma acknowledges the partial support of MATRICS grant number MTR/2017/000013 from the Science and Engineering Research Board.\\
\indent Paul Steinmann also gratefully acknowledges financial support for this work by the Deutsche Forschungsgemeinschaft under GRK2495/B.

\newpage
\bibliography{required_refs}

\begin{thebibliography}{83}
\expandafter\ifx\csname natexlab\endcsname\relax\def\natexlab#1{#1}\fi
\providecommand{\url}[1]{\texttt{#1}}
\providecommand{\href}[2]{#2}
\providecommand{\path}[1]{#1}
\providecommand{\DOIprefix}{doi:}
\providecommand{\ArXivprefix}{arXiv:}
\providecommand{\URLprefix}{URL: }
\providecommand{\Pubmedprefix}{pmid:}
\providecommand{\doi}[1]{\href{http://dx.doi.org/#1}{\path{#1}}}
\providecommand{\Pubmed}[1]{\href{pmid:#1}{\path{#1}}}
\providecommand{\bibinfo}[2]{#2}
\ifx\xfnm\relax \def\xfnm[#1]{\unskip,\space#1}\fi
\bibitem[{Adams et~al.(2018)Adams, Sridar, Thalman, Copenhaver, Elsaad and
  Polygerinos}]{adams2018water}
\bibinfo{author}{Adams, W.}, \bibinfo{author}{Sridar, S.},
  \bibinfo{author}{Thalman, C.M.}, \bibinfo{author}{Copenhaver, B.},
  \bibinfo{author}{Elsaad, H.}, \bibinfo{author}{Polygerinos, P.},
  \bibinfo{year}{2018}.
\newblock \bibinfo{title}{Water pipe robot utilizing soft inflatable
  actuators}, in: \bibinfo{booktitle}{2018 IEEE International Conference on
  Soft Robotics (RoboSoft)}, \bibinfo{organization}{IEEE}. pp.
  \bibinfo{pages}{321--326}.
\bibitem[{Ahmad et~al.(2020)Ahmad, Patra and Hossain}]{Ahmad2020}
\bibinfo{author}{Ahmad, D.}, \bibinfo{author}{Patra, K.},
  \bibinfo{author}{Hossain, M.}, \bibinfo{year}{2020}.
\newblock \bibinfo{title}{{Experimental study and phenomenological modelling of
  flaw sensitivity of two polymers used as dielectric elastomers}}.
\newblock \bibinfo{journal}{Continuum Mechanics and Thermodynamics}
  \bibinfo{volume}{32}, \bibinfo{pages}{489--500}.
\bibitem[{Araromi et~al.(2014)Araromi, Gavrilovich, Shintake, Rosset, Richard,
  Gass and Shea}]{araromi2014rollable}
\bibinfo{author}{Araromi, O.A.}, \bibinfo{author}{Gavrilovich, I.},
  \bibinfo{author}{Shintake, J.}, \bibinfo{author}{Rosset, S.},
  \bibinfo{author}{Richard, M.}, \bibinfo{author}{Gass, V.},
  \bibinfo{author}{Shea, H.R.}, \bibinfo{year}{2014}.
\newblock \bibinfo{title}{Rollable multisegment dielectric elastomer minimum
  energy structures for a deployable microsatellite gripper}.
\newblock \bibinfo{journal}{IEEE/ASME Transactions on mechatronics}
  \bibinfo{volume}{20}, \bibinfo{pages}{438--446}.
\bibitem[{Ask et~al.(2012)Ask, Menzel and Ristinmaa}]{Ask2012}
\bibinfo{author}{Ask, A.}, \bibinfo{author}{Menzel, A.},
  \bibinfo{author}{Ristinmaa, M.}, \bibinfo{year}{2012}.
\newblock \bibinfo{title}{{Electrostriction in electro-viscoelastic polymers}}.
\newblock \bibinfo{journal}{Mechanics of Materials} \bibinfo{volume}{50},
  \bibinfo{pages}{9--21}.
\bibitem[{Bar-Cohen et~al.(2001)}]{bar2001electroactive}
\bibinfo{author}{Bar-Cohen, Y.}, et~al., \bibinfo{year}{2001}.
\newblock \bibinfo{title}{Electroactive polymer actuators as artificial
  muscles}.
\newblock \bibinfo{journal}{SPIE, Washington} .
\bibitem[{Barsotti(2015)}]{10.1115/1.4031243}
\bibinfo{author}{Barsotti, R.}, \bibinfo{year}{2015}.
\newblock \bibinfo{title}{{Approximated Solutions for Axisymmetric Wrinkled
  Inflated Membranes}}.
\newblock \bibinfo{journal}{Journal of Applied Mechanics} \bibinfo{volume}{82}.
\bibitem[{Benedict et~al.(1979)Benedict, Wineman and
  Yang}]{benedict1979determination}
\bibinfo{author}{Benedict, R.}, \bibinfo{author}{Wineman, A.},
  \bibinfo{author}{Yang, W.H.}, \bibinfo{year}{1979}.
\newblock \bibinfo{title}{The determination of limiting pressure in
  simultaneous elongation and inflation of nonlinear elastic tubes}.
\newblock \bibinfo{journal}{International Journal of Solids and Structures}
  \bibinfo{volume}{15}, \bibinfo{pages}{241--249}.
\bibitem[{Budiansky(1974)}]{budiansky1974theory}
\bibinfo{author}{Budiansky, B.}, \bibinfo{year}{1974}.
\newblock \bibinfo{title}{Theory of buckling and post-buckling behavior of
  elastic structures}, in: \bibinfo{booktitle}{Advances in applied mechanics}.
  \bibinfo{publisher}{Elsevier}. volume~\bibinfo{volume}{14}, pp.
  \bibinfo{pages}{1--65}.
\bibitem[{Bustamante et~al.(2009)Bustamante, Dorfmann and
  Ogden}]{bustamante2009nonlinear}
\bibinfo{author}{Bustamante, R.}, \bibinfo{author}{Dorfmann, A.},
  \bibinfo{author}{Ogden, R.W.}, \bibinfo{year}{2009}.
\newblock \bibinfo{title}{Nonlinear electroelastostatics: a variational
  framework}.
\newblock \bibinfo{journal}{Zeitschrift f{\"u}r angewandte Mathematik und
  Physik} \bibinfo{volume}{60}, \bibinfo{pages}{154--177}.
\bibitem[{Carroll(1987)}]{carroll1987pressure}
\bibinfo{author}{Carroll, M.}, \bibinfo{year}{1987}.
\newblock \bibinfo{title}{Pressure maximum behavior in inflation of
  incompressible elastic hollow spheres and cylinders}.
\newblock \bibinfo{journal}{Quarterly of applied mathematics}
  \bibinfo{volume}{45}, \bibinfo{pages}{141--154}.
\bibitem[{Chaudhuri and Dasgupta(2014)}]{Chaudhuri2014}
\bibinfo{author}{Chaudhuri, A.}, \bibinfo{author}{Dasgupta, A.},
  \bibinfo{year}{2014}.
\newblock \bibinfo{title}{{On the static and dynamic analysis of inflated
  hyperelastic circular membranes}}.
\newblock \bibinfo{journal}{Journal of the Mechanics and Physics of Solids}
  \bibinfo{volume}{64}, \bibinfo{pages}{302--315}.
\bibitem[{Crandall et~al.(1972)Crandall, Dahl and Lardner}]{Crandall1972}
\bibinfo{author}{Crandall, S.H.}, \bibinfo{author}{Dahl, N.C.},
  \bibinfo{author}{Lardner, T.J.}, \bibinfo{year}{1972}.
\newblock \bibinfo{title}{{An Introduction to the Mechanics of Solids}}.
\newblock \bibinfo{publisher}{McGraw-Hill}.
\bibitem[{De~Melo et~al.(2018)De~Melo, Pereira and Morais}]{de2018simulation}
\bibinfo{author}{De~Melo, F.J.}, \bibinfo{author}{Pereira, A.B.},
  \bibinfo{author}{Morais, A.B.}, \bibinfo{year}{2018}.
\newblock \bibinfo{title}{The simulation of an automotive air spring suspension
  using a pseudo-dynamic procedure}.
\newblock \bibinfo{journal}{Applied Sciences} \bibinfo{volume}{8},
  \bibinfo{pages}{1049}.
\bibitem[{De~Tommasi et~al.(2010)De~Tommasi, Puglisi, Saccomandi and
  Zurlo}]{de2010pull}
\bibinfo{author}{De~Tommasi, D.}, \bibinfo{author}{Puglisi, G.},
  \bibinfo{author}{Saccomandi, G.}, \bibinfo{author}{Zurlo, G.},
  \bibinfo{year}{2010}.
\newblock \bibinfo{title}{Pull-in and wrinkling instabilities of electroactive
  dielectric actuators}.
\newblock \bibinfo{journal}{Journal of Physics D: Applied Physics}
  \bibinfo{volume}{43}, \bibinfo{pages}{325501}.
\bibitem[{{De Tommasi} et~al.(2011){De Tommasi}, Puglisi and
  Zurlo}]{DeTommasi2011}
\bibinfo{author}{{De Tommasi}, D.}, \bibinfo{author}{Puglisi, G.},
  \bibinfo{author}{Zurlo, G.}, \bibinfo{year}{2011}.
\newblock \bibinfo{title}{{Compression-induced failure of electroactive
  polymeric thin films}}.
\newblock \bibinfo{journal}{Applied Physics Letters} \bibinfo{volume}{98}.
\bibitem[{D'Errico(2020)}]{fminsearch}
\bibinfo{author}{D'Errico, J.}, \bibinfo{year}{2020}.
\newblock \bibinfo{title}{fminsearchbnd, fminsearchcon}.
\newblock \URLprefix
  \url{https://www.mathworks.com/matlabcentral/fileexchange/8277-fminsearchbnd-fminsearchcon}.
\bibitem[{Di~Biasio et~al.(2014)Di~Biasio, Ambrosone and
  Cametti}]{di2014dielectric}
\bibinfo{author}{Di~Biasio, A.}, \bibinfo{author}{Ambrosone, L.},
  \bibinfo{author}{Cametti, C.}, \bibinfo{year}{2014}.
\newblock \bibinfo{title}{Dielectric response of shelled toroidal particles
  carrying localized surface charge distributions. the effect of concentric and
  confocal shells}.
\newblock \bibinfo{journal}{Bioelectrochemistry} \bibinfo{volume}{98},
  \bibinfo{pages}{76--86}.
\bibitem[{Dorfmann and Ogden(2005)}]{dorfmann2005nonlinear}
\bibinfo{author}{Dorfmann, A.}, \bibinfo{author}{Ogden, R.},
  \bibinfo{year}{2005}.
\newblock \bibinfo{title}{Nonlinear electroelasticity}.
\newblock \bibinfo{journal}{Acta Mechanica} \bibinfo{volume}{174},
  \bibinfo{pages}{167--183}.
\bibitem[{Dorfmann and Ogden(2006)}]{dorfmann2006nonlinear}
\bibinfo{author}{Dorfmann, A.}, \bibinfo{author}{Ogden, R.},
  \bibinfo{year}{2006}.
\newblock \bibinfo{title}{Nonlinear electroelastic deformations}.
\newblock \bibinfo{journal}{Journal of Elasticity} \bibinfo{volume}{82},
  \bibinfo{pages}{99--127}.
\bibitem[{Dorfmann and Ogden(2014a)}]{dorfmann2014instabilities}
\bibinfo{author}{Dorfmann, L.}, \bibinfo{author}{Ogden, R.W.},
  \bibinfo{year}{2014}a.
\newblock \bibinfo{title}{Instabilities of an electroelastic plate}.
\newblock \bibinfo{journal}{International Journal of Engineering Science}
  \bibinfo{volume}{77}, \bibinfo{pages}{79--101}.
\bibitem[{Dorfmann and Ogden(2014b)}]{dorfmann2014nonlinear}
\bibinfo{author}{Dorfmann, L.}, \bibinfo{author}{Ogden, R.W.},
  \bibinfo{year}{2014}b.
\newblock \bibinfo{title}{Nonlinear response of an electroelastic spherical
  shell}.
\newblock \bibinfo{journal}{International Journal of Engineering Science}
  \bibinfo{volume}{85}, \bibinfo{pages}{163--174}.
\bibitem[{Dorfmann and Ogden(2014c)}]{Dorfmann2014b}
\bibinfo{author}{Dorfmann, L.}, \bibinfo{author}{Ogden, R.W.},
  \bibinfo{year}{2014}c.
\newblock \bibinfo{title}{{Nonlinear theory of electroelastic and
  magnetoelastic interactions}}.
\newblock \bibinfo{publisher}{Springer}.
\bibitem[{Dorfmann and Ogden(2017)}]{dorfmann2017nonlinear}
\bibinfo{author}{Dorfmann, L.}, \bibinfo{author}{Ogden, R.W.},
  \bibinfo{year}{2017}.
\newblock \bibinfo{title}{Nonlinear electroelasticity: material properties,
  continuum theory and applications}.
\newblock \bibinfo{journal}{Proceedings of the Royal Society A: Mathematical,
  Physical and Engineering Sciences} \bibinfo{volume}{473},
  \bibinfo{pages}{20170311}.
\bibitem[{Dorfmann and Ogden(2019)}]{Dorfmann2017c}
\bibinfo{author}{Dorfmann, L.}, \bibinfo{author}{Ogden, R.W.},
  \bibinfo{year}{2019}.
\newblock \bibinfo{title}{{Instabilities of soft dielectrics}}.
\newblock \bibinfo{journal}{Philosophical Transactions of the Royal Society A}
  \bibinfo{volume}{377}, \bibinfo{pages}{20180077}.
\bibitem[{Eringen(1963)}]{eringen1963foundations}
\bibinfo{author}{Eringen, A.C.}, \bibinfo{year}{1963}.
\newblock \bibinfo{title}{On the foundations of electroelastostatics}.
\newblock \bibinfo{journal}{International Journal of Engineering Science}
  \bibinfo{volume}{1}, \bibinfo{pages}{127--153}.
\bibitem[{Greaney et~al.(2019)Greaney, Meere and Zurlo}]{Greaney2019}
\bibinfo{author}{Greaney, P.}, \bibinfo{author}{Meere, M.},
  \bibinfo{author}{Zurlo, G.}, \bibinfo{year}{2019}.
\newblock \bibinfo{title}{{The out-of-plane behaviour of dielectric membranes:
  Description of wrinkling and pull-in instabilities}}.
\newblock \bibinfo{journal}{Journal of the Mechanics and Physics of Solids}
  \bibinfo{volume}{122}, \bibinfo{pages}{84--97}.
\bibitem[{Hingorani and O'Donnell(1998)}]{hingorani1998toroidal}
\bibinfo{author}{Hingorani, M.M.}, \bibinfo{author}{O'Donnell, M.},
  \bibinfo{year}{1998}.
\newblock \bibinfo{title}{Toroidal proteins: running rings around dna}.
\newblock \bibinfo{journal}{Current Biology} \bibinfo{volume}{8},
  \bibinfo{pages}{R83--R86}.
\bibitem[{Hossain et~al.(2014)Hossain, Vu and Steinmann}]{Hossain2014}
\bibinfo{author}{Hossain, M.}, \bibinfo{author}{Vu, D.K.},
  \bibinfo{author}{Steinmann, P.}, \bibinfo{year}{2014}.
\newblock \bibinfo{title}{{A comprehensive characterization of the
  electro-mechanically coupled properties of VHB 4910 polymer}}.
\newblock \bibinfo{journal}{Archive of Applied Mechanics} \bibinfo{volume}{85},
  \bibinfo{pages}{523--537}.
\bibitem[{Khayat et~al.(1992)Khayat, Derdorri and
  Garc{\'\i}a-Rej{\'o}n}]{khayat1992inflation}
\bibinfo{author}{Khayat, R.E.}, \bibinfo{author}{Derdorri, A.},
  \bibinfo{author}{Garc{\'\i}a-Rej{\'o}n, A.}, \bibinfo{year}{1992}.
\newblock \bibinfo{title}{Inflation of an elastic cylindrical membrane:
  non-linear deformation and instability}.
\newblock \bibinfo{journal}{International journal of solids and structures}
  \bibinfo{volume}{29}, \bibinfo{pages}{69--87}.
\bibitem[{Kofod et~al.(2006)Kofod, Paajanen and Bauer}]{kofod2006self}
\bibinfo{author}{Kofod, G.}, \bibinfo{author}{Paajanen, M.},
  \bibinfo{author}{Bauer, S.}, \bibinfo{year}{2006}.
\newblock \bibinfo{title}{Self-organized minimum-energy structures for
  dielectric elastomer actuators}.
\newblock \bibinfo{journal}{Applied Physics A} \bibinfo{volume}{85},
  \bibinfo{pages}{141--143}.
\bibitem[{Kofod et~al.(2007)Kofod, Wirges, Paajanen and
  Bauer}]{kofod2007energy}
\bibinfo{author}{Kofod, G.}, \bibinfo{author}{Wirges, W.},
  \bibinfo{author}{Paajanen, M.}, \bibinfo{author}{Bauer, S.},
  \bibinfo{year}{2007}.
\newblock \bibinfo{title}{Energy minimization for self-organized structure
  formation and actuation}.
\newblock \bibinfo{journal}{Applied Physics Letters} \bibinfo{volume}{90},
  \bibinfo{pages}{081916}.
\bibitem[{Koiter(1970)}]{koiter1970stability}
\bibinfo{author}{Koiter, W.T.}, \bibinfo{year}{1970}.
\newblock \bibinfo{title}{The stability of elastic equilibrium}.
\newblock \bibinfo{type}{Technical Report}. Stanford Univ Ca Dept of
  Aeronautics and Astronautics.
\bibitem[{Kollosche et~al.(2012)Kollosche, Zhu, Suo and
  Kofod}]{kollosche2012complex}
\bibinfo{author}{Kollosche, M.}, \bibinfo{author}{Zhu, J.},
  \bibinfo{author}{Suo, Z.}, \bibinfo{author}{Kofod, G.}, \bibinfo{year}{2012}.
\newblock \bibinfo{title}{Complex interplay of nonlinear processes in
  dielectric elastomers}.
\newblock \bibinfo{journal}{Physical Review E} \bibinfo{volume}{85},
  \bibinfo{pages}{051801}.
\bibitem[{Lau et~al.(2017)Lau, Heng, Ahmed and Shrestha}]{lau2017dielectric}
\bibinfo{author}{Lau, G.K.}, \bibinfo{author}{Heng, K.R.},
  \bibinfo{author}{Ahmed, A.S.}, \bibinfo{author}{Shrestha, M.},
  \bibinfo{year}{2017}.
\newblock \bibinfo{title}{Dielectric elastomer fingers for versatile grasping
  and nimble pinching}.
\newblock \bibinfo{journal}{Applied Physics Letters} \bibinfo{volume}{110},
  \bibinfo{pages}{182906}.
\bibitem[{Leo-Macias et~al.(2011)Leo-Macias, Katz, Wei, Alimova, Katz, Rice,
  Diaz-Avalos, Hu, Stokes and Gottlieb}]{leo2011toroidal}
\bibinfo{author}{Leo-Macias, A.}, \bibinfo{author}{Katz, G.},
  \bibinfo{author}{Wei, H.}, \bibinfo{author}{Alimova, A.},
  \bibinfo{author}{Katz, A.}, \bibinfo{author}{Rice, W.J.},
  \bibinfo{author}{Diaz-Avalos, R.}, \bibinfo{author}{Hu, G.B.},
  \bibinfo{author}{Stokes, D.L.}, \bibinfo{author}{Gottlieb, P.},
  \bibinfo{year}{2011}.
\newblock \bibinfo{title}{Toroidal surface complexes of bacteriophage $\phi$12
  are responsible for host-cell attachment}.
\newblock \bibinfo{journal}{Virology} \bibinfo{volume}{414},
  \bibinfo{pages}{103--109}.
\bibitem[{Li et~al.(2013)Li, Keplinger, Baumgartner, Bauer, Yang and
  Suo}]{li2013giant}
\bibinfo{author}{Li, T.}, \bibinfo{author}{Keplinger, C.},
  \bibinfo{author}{Baumgartner, R.}, \bibinfo{author}{Bauer, S.},
  \bibinfo{author}{Yang, W.}, \bibinfo{author}{Suo, Z.}, \bibinfo{year}{2013}.
\newblock \bibinfo{title}{Giant voltage-induced deformation in dielectric
  elastomers near the verge of snap-through instability}.
\newblock \bibinfo{journal}{Journal of the Mechanics and Physics of Solids}
  \bibinfo{volume}{61}, \bibinfo{pages}{611--628}.
\bibitem[{Li and Steigmann(1995a)}]{Li1995a}
\bibinfo{author}{Li, X.}, \bibinfo{author}{Steigmann, D.J.},
  \bibinfo{year}{1995}a.
\newblock \bibinfo{title}{{Finite deformation of a pressurized toroidal
  membrane}}.
\newblock \bibinfo{journal}{International Journal of Non-Linear Mechanics}
  \bibinfo{volume}{30}, \bibinfo{pages}{583--595}.
\bibitem[{Li and Steigmann(1995b)}]{Li1995b}
\bibinfo{author}{Li, X.}, \bibinfo{author}{Steigmann, D.J.},
  \bibinfo{year}{1995}b.
\newblock \bibinfo{title}{{Point loads on a hemispherical elastic membrane}}.
\newblock \bibinfo{journal}{International Journal of Non-Linear Mechanics}
  \bibinfo{volume}{30}, \bibinfo{pages}{569--581}.
\bibitem[{Liu(2013)}]{Liu2013a}
\bibinfo{author}{Liu, L.}, \bibinfo{year}{2013}.
\newblock \bibinfo{title}{{On energy formulations of electrostatics for
  continuum media}}.
\newblock \bibinfo{journal}{Journal of the Mechanics and Physics of Solids}
  \bibinfo{volume}{61}, \bibinfo{pages}{968--990}.
\bibitem[{Liu(2020)}]{Liu2020github}
\bibinfo{author}{Liu, Z.}, \bibinfo{year}{2020}.
\newblock \bibinfo{title}{Electroelastic torodial membrane}.
\newblock \URLprefix \url{https://doi.org/10.5281/zenodo.3768563}.
\bibitem[{Lu et~al.(2020)Lu, Ma and Wang}]{lu2020mechanics}
\bibinfo{author}{Lu, T.}, \bibinfo{author}{Ma, C.}, \bibinfo{author}{Wang, T.},
  \bibinfo{year}{2020}.
\newblock \bibinfo{title}{Mechanics of dielectric elastomer structures: {A}
  review}.
\newblock \bibinfo{journal}{Extreme Mechanics Letters} ,
  \bibinfo{pages}{100752}.
\bibitem[{Mao et~al.(2018)Mao, Wu, Fu, Liu and Qu}]{Mao2018}
\bibinfo{author}{Mao, G.}, \bibinfo{author}{Wu, L.}, \bibinfo{author}{Fu, Y.},
  \bibinfo{author}{Liu, J.}, \bibinfo{author}{Qu, S.}, \bibinfo{year}{2018}.
\newblock \bibinfo{title}{{Voltage-controlled radial wrinkles of a trumpet-like
  dielectric elastomer structure}}.
\newblock \bibinfo{journal}{AIP Advances} \bibinfo{volume}{8}.
\bibitem[{MATLAB(2018)}]{MATLAB:2018b}
\bibinfo{author}{MATLAB}, \bibinfo{year}{2018}.
\newblock \bibinfo{title}{version 9.5.0 (R2018b)}.
\newblock \bibinfo{publisher}{The MathWorks Inc.}, \bibinfo{address}{Natick,
  Massachusetts}.
\bibitem[{McMeeking and Landis(2005)}]{mcmeeking2005electrostatic}
\bibinfo{author}{McMeeking, R.M.}, \bibinfo{author}{Landis, C.M.},
  \bibinfo{year}{2005}.
\newblock \bibinfo{title}{Electrostatic forces and stored energy for deformable
  dielectric materials}.
\newblock \bibinfo{journal}{Journal of Applied Mechanics} \bibinfo{volume}{72},
  \bibinfo{pages}{581--590}.
\bibitem[{Mehnert et~al.(2019)Mehnert, Hossain and
  Steinmann}]{MEHNERT2019103797}
\bibinfo{author}{Mehnert, M.}, \bibinfo{author}{Hossain, M.},
  \bibinfo{author}{Steinmann, P.}, \bibinfo{year}{2019}.
\newblock \bibinfo{title}{Experimental and numerical investigations of the
  electro-viscoelastic behavior of vhb 4905tm}.
\newblock \bibinfo{journal}{European Journal of Mechanics - A/Solids}
  \bibinfo{volume}{77}, \bibinfo{pages}{103797}.
\bibitem[{Melnikov and Ogden(2018)}]{Melnikov2018}
\bibinfo{author}{Melnikov, A.}, \bibinfo{author}{Ogden, R.W.},
  \bibinfo{year}{2018}.
\newblock \bibinfo{title}{{Bifurcation of finitely deformed thick-walled
  electroelastic cylindrical tubes subject to a radial electric field}}.
\newblock \bibinfo{journal}{Zeitschrift fur Angewandte Mathematik und Physik}
  \bibinfo{volume}{69}, \bibinfo{pages}{1--27}.
\bibitem[{Michel et~al.(2008)Michel, Bormann, Jordi and Fink}]{Michel2008}
\bibinfo{author}{Michel, S.}, \bibinfo{author}{Bormann, A.},
  \bibinfo{author}{Jordi, C.}, \bibinfo{author}{Fink, E.},
  \bibinfo{year}{2008}.
\newblock \bibinfo{title}{{Feasibility studies for a bionic propulsion system
  of a blimp based on dielectric elastomers}}.
\newblock \bibinfo{journal}{Proceedings of SPIE - EAPAD}
  \bibinfo{volume}{4332}, \bibinfo{pages}{1--15}.
\bibitem[{Miehe et~al.(2015)Miehe, Vallicotti and
  Z{\"a}h}]{miehe2015computational}
\bibinfo{author}{Miehe, C.}, \bibinfo{author}{Vallicotti, D.},
  \bibinfo{author}{Z{\"a}h, D.}, \bibinfo{year}{2015}.
\newblock \bibinfo{title}{Computational structural and material stability
  analysis in finite electro-elasto-statics of electro-active materials}.
\newblock \bibinfo{journal}{International Journal for Numerical Methods in
  Engineering} \bibinfo{volume}{102}, \bibinfo{pages}{1605--1637}.
\bibitem[{Mooney(1940)}]{mooney1940theory}
\bibinfo{author}{Mooney, M.}, \bibinfo{year}{1940}.
\newblock \bibinfo{title}{A theory of large elastic deformation}.
\newblock \bibinfo{journal}{Journal of applied physics} \bibinfo{volume}{11},
  \bibinfo{pages}{582--592}.
\bibitem[{Moretti et~al.(2019)Moretti, Papini, Daniele, Forehand, Ingram,
  Vertechy and Fontana}]{Moretti2019}
\bibinfo{author}{Moretti, G.}, \bibinfo{author}{Papini, G.P.R.},
  \bibinfo{author}{Daniele, L.}, \bibinfo{author}{Forehand, D.},
  \bibinfo{author}{Ingram, D.}, \bibinfo{author}{Vertechy, R.},
  \bibinfo{author}{Fontana, M.}, \bibinfo{year}{2019}.
\newblock \bibinfo{title}{{Modelling and field testing of a wave energy
  converter based on dielectric elastomer generators}}.
\newblock \bibinfo{journal}{Proceedings of the Royal Society A: Mathematical,
  Physical and Engineering Sciences} \bibinfo{volume}{475},
  \bibinfo{pages}{20180566}.
\bibitem[{Morrison et~al.(1962)Morrison, Riley and
  Zancanaro}]{morrison1962multiple}
\bibinfo{author}{Morrison, D.D.}, \bibinfo{author}{Riley, J.D.},
  \bibinfo{author}{Zancanaro, J.F.}, \bibinfo{year}{1962}.
\newblock \bibinfo{title}{Multiple shooting method for two-point boundary value
  problems}.
\newblock \bibinfo{journal}{Communications of the ACM} \bibinfo{volume}{5},
  \bibinfo{pages}{613--614}.
\bibitem[{M{\"u}ller and Struchtrup(2002)}]{muller2002inflating}
\bibinfo{author}{M{\"u}ller, I.}, \bibinfo{author}{Struchtrup, H.},
  \bibinfo{year}{2002}.
\newblock \bibinfo{title}{Inflating a rubber balloon}.
\newblock \bibinfo{journal}{Mathematics and Mechanics of Solids}
  \bibinfo{volume}{7}, \bibinfo{pages}{569--577}.
\bibitem[{Nayyar et~al.(2011)Nayyar, Ravi-Chandar and
  Huang}]{nayyar2011stretch}
\bibinfo{author}{Nayyar, V.}, \bibinfo{author}{Ravi-Chandar, K.},
  \bibinfo{author}{Huang, R.}, \bibinfo{year}{2011}.
\newblock \bibinfo{title}{Stretch-induced stress patterns and wrinkles in
  hyperelastic thin sheets}.
\newblock \bibinfo{journal}{International journal of solids and structures}
  \bibinfo{volume}{48}, \bibinfo{pages}{3471--3483}.
\bibitem[{O'Halloran et~al.(2008)O'Halloran, O'Malley and
  McHugh}]{O'Halloran2008}
\bibinfo{author}{O'Halloran, A.}, \bibinfo{author}{O'Malley, F.},
  \bibinfo{author}{McHugh, P.}, \bibinfo{year}{2008}.
\newblock \bibinfo{title}{{A review on dielectric elastomer actuators,
  technology, applications, and challenges}}.
\newblock \bibinfo{journal}{Journal of Applied Physics} \bibinfo{volume}{104},
  \bibinfo{pages}{71101--71110}.
\bibitem[{Ozsecen et~al.(2010)Ozsecen, Sivak and Mavroidis}]{Ozsecen2010}
\bibinfo{author}{Ozsecen, M.Y.}, \bibinfo{author}{Sivak, M.},
  \bibinfo{author}{Mavroidis, C.}, \bibinfo{year}{2010}.
\newblock \bibinfo{title}{{Haptic interfaces using dielectric electroactive
  polymers}}, in: \bibinfo{editor}{Tomizuka, M.}, \bibinfo{editor}{Yun, C.B.},
  \bibinfo{editor}{Giurgiutiu, V.}, \bibinfo{editor}{Lynch, J.P.} (Eds.),
  \bibinfo{booktitle}{Proceedings of SPIE - Sensors and Smart Structures
  Technologies for Civil, Mechanical, and Aerospace Systems}, p.
  \bibinfo{pages}{7647}.
\bibitem[{Pelrine et~al.(2000)Pelrine, Kornbluh, Pei and
  Joseph}]{pelrine2000high}
\bibinfo{author}{Pelrine, R.}, \bibinfo{author}{Kornbluh, R.},
  \bibinfo{author}{Pei, Q.}, \bibinfo{author}{Joseph, J.},
  \bibinfo{year}{2000}.
\newblock \bibinfo{title}{High-speed electrically actuated elastomers with
  strain greater than 100\%}.
\newblock \bibinfo{journal}{Science} \bibinfo{volume}{287},
  \bibinfo{pages}{836--839}.
\bibitem[{Pipkin(1986)}]{Pipkin1986}
\bibinfo{author}{Pipkin, A.C.}, \bibinfo{year}{1986}.
\newblock \bibinfo{title}{{The Relaxed Energy Density for Isotropic Elastic
  Membranes}}.
\newblock \bibinfo{journal}{IMA Journal of Applied Mathematics}
  \bibinfo{volume}{36}, \bibinfo{pages}{85--99}.
\bibitem[{Purnell et~al.(2018)Purnell, Butawan and Ramsey}]{purnell2018bio}
\bibinfo{author}{Purnell, M.C.}, \bibinfo{author}{Butawan, M.B.},
  \bibinfo{author}{Ramsey, R.D.}, \bibinfo{year}{2018}.
\newblock \bibinfo{title}{Bio-field array: a dielectrophoretic electromagnetic
  toroidal excitation to restore and maintain the golden ratio in human
  erythrocytes}.
\newblock \bibinfo{journal}{Physiological reports} \bibinfo{volume}{6},
  \bibinfo{pages}{e13722}.
\bibitem[{Reddy and Saxena(2017)}]{Reddy2017}
\bibinfo{author}{Reddy, N.H.}, \bibinfo{author}{Saxena, P.},
  \bibinfo{year}{2017}.
\newblock \bibinfo{title}{{Limit points in the free inflation of a
  magnetoelastic toroidal membrane}}.
\newblock \bibinfo{journal}{International Journal of Non-Linear Mechanics}
  \bibinfo{volume}{95}, \bibinfo{pages}{248--263}.
\bibitem[{Reddy and Saxena(2018)}]{reddy2018instabilities}
\bibinfo{author}{Reddy, N.H.}, \bibinfo{author}{Saxena, P.},
  \bibinfo{year}{2018}.
\newblock \bibinfo{title}{Instabilities in the axisymmetric magnetoelastic
  deformation of a cylindrical membrane}.
\newblock \bibinfo{journal}{International Journal of Solids and Structures}
  \bibinfo{volume}{136}, \bibinfo{pages}{203--219}.
\bibitem[{Rivlin(1948)}]{rivlin1948large}
\bibinfo{author}{Rivlin, R.}, \bibinfo{year}{1948}.
\newblock \bibinfo{title}{Large elastic deformations of isotropic materials. i.
  fundamental concepts}.
\newblock \bibinfo{journal}{Philosophical Transactions of the Royal Society of
  London. Series A, Mathematical and Physical Sciences} \bibinfo{volume}{240},
  \bibinfo{pages}{459--490}.
\bibitem[{Rudykh et~al.(2012)Rudykh, Bhattacharya and
  deBotton}]{rudykh2012snap}
\bibinfo{author}{Rudykh, S.}, \bibinfo{author}{Bhattacharya, K.},
  \bibinfo{author}{deBotton, G.}, \bibinfo{year}{2012}.
\newblock \bibinfo{title}{Snap-through actuation of thick-wall electroactive
  balloons}.
\newblock \bibinfo{journal}{International Journal of Non-Linear Mechanics}
  \bibinfo{volume}{47}, \bibinfo{pages}{206--209}.
\bibitem[{Saxena et~al.(2019)Saxena, Reddy and
  Pradhan}]{saxena2019magnetoelastic}
\bibinfo{author}{Saxena, P.}, \bibinfo{author}{Reddy, N.H.},
  \bibinfo{author}{Pradhan, S.P.}, \bibinfo{year}{2019}.
\newblock \bibinfo{title}{Magnetoelastic deformation of a circular membrane:
  wrinkling and limit point instabilities}.
\newblock \bibinfo{journal}{International Journal of Non-Linear Mechanics}
  \bibinfo{volume}{116}, \bibinfo{pages}{250--261}.
\bibitem[{Saxena and Sharma(2020)}]{saxena2019equilibrium}
\bibinfo{author}{Saxena, P.}, \bibinfo{author}{Sharma, B.L.},
  \bibinfo{year}{2020}.
\newblock \bibinfo{title}{On equilibrium equations and their perturbations
  using three different variational formulations of nonlinear
  electroelastostatics}.
\newblock \bibinfo{journal}{Mathematics and Mechanics of Solids} .
\bibitem[{Saxena et~al.(2014)Saxena, Vu and Steinmann}]{Saxena2014a}
\bibinfo{author}{Saxena, P.}, \bibinfo{author}{Vu, D.K.},
  \bibinfo{author}{Steinmann, P.}, \bibinfo{year}{2014}.
\newblock \bibinfo{title}{{On rate-dependent dissipation effects in
  electro-elasticity}}.
\newblock \bibinfo{journal}{International Journal of Non-Linear Mechanics}
  \bibinfo{volume}{62}, \bibinfo{pages}{1--11}.
\bibitem[{Shintake et~al.(2018)Shintake, Cacucciolo, Floreano and
  Shea}]{shintake2018soft}
\bibinfo{author}{Shintake, J.}, \bibinfo{author}{Cacucciolo, V.},
  \bibinfo{author}{Floreano, D.}, \bibinfo{author}{Shea, H.},
  \bibinfo{year}{2018}.
\newblock \bibinfo{title}{Soft robotic grippers}.
\newblock \bibinfo{journal}{Advanced Materials} \bibinfo{volume}{30},
  \bibinfo{pages}{1707035}.
\bibitem[{Steigmann(1990)}]{Steigmann1990}
\bibinfo{author}{Steigmann, D.J.}, \bibinfo{year}{1990}.
\newblock \bibinfo{title}{{Tension-Field Theory}}.
\newblock \bibinfo{journal}{Proceedings of the Royal Society A: Mathematical,
  Physical and Engineering Sciences} \bibinfo{volume}{429},
  \bibinfo{pages}{141--173}.
\bibitem[{Suo(2010)}]{suo2010theory}
\bibinfo{author}{Suo, Z.}, \bibinfo{year}{2010}.
\newblock \bibinfo{title}{Theory of dielectric elastomers}.
\newblock \bibinfo{journal}{Acta Mechanica Solida Sinica} \bibinfo{volume}{23},
  \bibinfo{pages}{549--578}.
\bibitem[{Swain and Gupta(2015)}]{Swain2015}
\bibinfo{author}{Swain, D.}, \bibinfo{author}{Gupta, A.}, \bibinfo{year}{2015}.
\newblock \bibinfo{title}{{Interfacial growth during closure of a cutaneous
  wound: Stress generation and wrinkle formation}}.
\newblock \bibinfo{journal}{Soft Matter} \bibinfo{volume}{11},
  \bibinfo{pages}{6499--6508}.
\bibitem[{Tamadapu et~al.(2013)Tamadapu, Dhavale and
  DasGupta}]{tamadapu2013geometrical}
\bibinfo{author}{Tamadapu, G.}, \bibinfo{author}{Dhavale, N.N.},
  \bibinfo{author}{DasGupta, A.}, \bibinfo{year}{2013}.
\newblock \bibinfo{title}{Geometrical feature of the scaling behavior of the
  limit-point pressure of inflated hyperelastic membranes}.
\newblock \bibinfo{journal}{Physical Review E} \bibinfo{volume}{88},
  \bibinfo{pages}{053201}.
\bibitem[{Thompson(2015)}]{thompson2015advances}
\bibinfo{author}{Thompson, J.M.T.}, \bibinfo{year}{2015}.
\newblock \bibinfo{title}{Advances in shell buckling: theory and experiments}.
\newblock \bibinfo{journal}{International Journal of Bifurcation and Chaos}
  \bibinfo{volume}{25}, \bibinfo{pages}{1530001}.
\bibitem[{Thompson and Van~der Heijden(2014)}]{thompson2014quantified}
\bibinfo{author}{Thompson, J.M.T.}, \bibinfo{author}{Van~der Heijden, G.},
  \bibinfo{year}{2014}.
\newblock \bibinfo{title}{Quantified" shock-sensitivity" above the {Maxwell}
  load}.
\newblock \bibinfo{journal}{International Journal of Bifurcation and Chaos}
  \bibinfo{volume}{24}, \bibinfo{pages}{1430009}.
\bibitem[{Tiersten(1978)}]{tiersten1978perturbation}
\bibinfo{author}{Tiersten, H.F.}, \bibinfo{year}{1978}.
\newblock \bibinfo{title}{Perturbation theory for linear electroelastic
  equations for small fields superposed on a bias}.
\newblock \bibinfo{journal}{The Journal of the Acoustical Society of America}
  \bibinfo{volume}{64}, \bibinfo{pages}{832--837}.
\bibitem[{Tiersten(1981)}]{tiersten1981electroelastic}
\bibinfo{author}{Tiersten, H.F.}, \bibinfo{year}{1981}.
\newblock \bibinfo{title}{Electroelastic interactions and the piezoelectric
  equations}.
\newblock \bibinfo{journal}{The Journal of the Acoustical Society of America}
  \bibinfo{volume}{70}, \bibinfo{pages}{1567--1576}.
\bibitem[{Toupin(1956)}]{toupin1956elastic}
\bibinfo{author}{Toupin, R.A.}, \bibinfo{year}{1956}.
\newblock \bibinfo{title}{The elastic dielectric}.
\newblock \bibinfo{journal}{Journal of Rational Mechanics and Analysis}
  \bibinfo{volume}{5}, \bibinfo{pages}{849--915}.
\bibitem[{Toupin(1963)}]{toupin1963dynamical}
\bibinfo{author}{Toupin, R.A.}, \bibinfo{year}{1963}.
\newblock \bibinfo{title}{A dynamical theory of elastic dielectrics}.
\newblock \bibinfo{journal}{International Journal of Engineering Science}
  \bibinfo{volume}{1}, \bibinfo{pages}{101--126}.
\bibitem[{Venkata and Saxena(2020)}]{Venkata2020}
\bibinfo{author}{Venkata, S.P.}, \bibinfo{author}{Saxena, P.},
  \bibinfo{year}{2020}.
\newblock \bibinfo{title}{{Instabilities in the free inflation of a nonlinear
  hyperelastic toroidal membrane}}.
\newblock \bibinfo{journal}{Journal of Mechanics of Materials and Structures}
  \bibinfo{volume}{14}, \bibinfo{pages}{473--496}.
\bibitem[{Vu et~al.(2007)Vu, Steinmann and Possart}]{vu2007numerical}
\bibinfo{author}{Vu, D.}, \bibinfo{author}{Steinmann, P.},
  \bibinfo{author}{Possart, G.}, \bibinfo{year}{2007}.
\newblock \bibinfo{title}{Numerical modelling of non-linear electroelasticity}.
\newblock \bibinfo{journal}{International Journal for Numerical Methods in
  Engineering} \bibinfo{volume}{70}, \bibinfo{pages}{685--704}.
\bibitem[{Wong and Pellegrino(2006)}]{Wong2006}
\bibinfo{author}{Wong, W.}, \bibinfo{author}{Pellegrino, S.},
  \bibinfo{year}{2006}.
\newblock \bibinfo{title}{{Wrinkled membranes II: analytical models}}.
\newblock \bibinfo{journal}{Journal of Mechanics of Materials and Structures}
  \bibinfo{volume}{1}, \bibinfo{pages}{27--61}.
\bibitem[{Xie et~al.(2016)Xie, Liu and Fu}]{Xie2016}
\bibinfo{author}{Xie, Y.X.}, \bibinfo{author}{Liu, J.C.}, \bibinfo{author}{Fu,
  Y.B.}, \bibinfo{year}{2016}.
\newblock \bibinfo{title}{{Bifurcation of a dielectric elastomer balloon under
  pressurized inflation and electric actuation}}.
\newblock \bibinfo{journal}{International Journal of Solids and Structures}
  \bibinfo{volume}{78-79}, \bibinfo{pages}{182--188}.
\bibitem[{Zang et~al.(2020)Zang, Liao, Lang, Zhao, Yuan and
  Feng}]{zang2020bionic}
\bibinfo{author}{Zang, H.}, \bibinfo{author}{Liao, B.}, \bibinfo{author}{Lang,
  X.}, \bibinfo{author}{Zhao, Z.L.}, \bibinfo{author}{Yuan, W.},
  \bibinfo{author}{Feng, X.Q.}, \bibinfo{year}{2020}.
\newblock \bibinfo{title}{Bionic torus as a self-adaptive soft grasper in
  robots}.
\newblock \bibinfo{journal}{Applied Physics Letters} \bibinfo{volume}{116},
  \bibinfo{pages}{023701}.
\bibitem[{Zhang et~al.(2016)Zhang, Sun, Chen, Liu, Li and Li}]{Zhang2016}
\bibinfo{author}{Zhang, C.}, \bibinfo{author}{Sun, W.}, \bibinfo{author}{Chen,
  H.}, \bibinfo{author}{Liu, L.}, \bibinfo{author}{Li, B.},
  \bibinfo{author}{Li, D.}, \bibinfo{year}{2016}.
\newblock \bibinfo{title}{{Electromechanical deformation of conical dielectric
  elastomer actuator with hydrogel electrodes}}.
\newblock \bibinfo{journal}{Journal of Applied Physics} \bibinfo{volume}{119}.
\bibitem[{Zhao and Suo(2007)}]{zhao2007method}
\bibinfo{author}{Zhao, X.}, \bibinfo{author}{Suo, Z.}, \bibinfo{year}{2007}.
\newblock \bibinfo{title}{Method to analyze electromechanical stability of
  dielectric elastomers}.
\newblock \bibinfo{journal}{Applied Physics Letters} \bibinfo{volume}{91},
  \bibinfo{pages}{061921}.

\end{thebibliography}
\appendix 
 \section{Pressure term in the variational formulation} \label{appendix: pressure term}
 The pressure term in the total potential energy functional
 \begin{equation}
  \int\limits_{V_0}^{V_0 + \Delta V} \wt{P} \mathrm{d}V
 \end{equation}
 should be such that upon taking its first variation, the virtual work $(\delta V)$ obtained is of the form [63]
 \begin{equation}
  \delta V = \int\limits_{\Gamma} \big[\wt{P} \mbf{n} \mathrm{d}s_0 \big] \cdot \delta \mbf{x}
 \end{equation}
 where $\Gamma$ is the domain comprising the deformed mid-surface of the membrane and $\delta \mbf{x}$ is a virtual displacement of a point on the mid-surface. The normal vector is given by
\begin{equation}
 \mbf{n} = \frac{1}{\sqrt{g}} \left[ \wtr \wte_\theta \cos \phi \mbf{E}_1 + \wtr \wte_\theta  \sin \phi \mbf{E}_2 - \wtr \wtr_\theta  \mbf{E}_3 \right] \quad \text{where} \quad \sqrt{g} = \wtr \left[\wtr_\theta ^2 + \wte_\theta ^2 \right]^{1/2}. \label{eqn: deformed normal vector}
\end{equation}
 From equations (3) and \eqref{eqn: deformed normal vector}, one obtains
 \begin{equation}
  \delta V = \int\limits_0^{2 \pi} \int\limits_0^{2 \pi} \wt{P} \left[ \wtr \wte_\theta \delta \wtr - \wtr \wtr_\theta \delta \wte \right] \mathrm{d}\theta \, \mathrm{d}\phi .
 \end{equation}
 It can be shown that this is the first variation of the functional following
 \begin{equation}
  \frac{1}{2} \int\limits_{0}^{2 \pi} \int\limits_{0}^{2 \pi} \wt{P} \wtr^2 \wte_\theta \, \mathrm{d}\theta \, \mathrm{d}\phi , \label{eqn: primitive of virtual work}
 \end{equation}
 after using  the condition $\delta \eta|_{\theta = 0} = \delta \eta|_{\theta = 2 \pi}$. 
 
 Further note that total volume of the deformed torus is given by
 \begin{equation}
  V (\wtr , \wte) = 2 \int\limits_{\wtr(\pi)}^{\wtr(0)} 2\pi \wtr \wte \, \mathrm{d}\wtr = -4 \pi \int\limits_0^\pi \wtr \wtr_\theta \wte \, \mathrm{d}\theta .
 \end{equation}
 In the domain $\theta \in [0, 2\pi]$, it can be shown that the function $\wtr$ is even while $\wtr_\theta$ and $\wte$ are odd with respect to the point $\theta = \pi$.
 Thus the product $\wtr \wtr_\theta \wte$ is an even function and the above integrals can be written as
 \begin{equation}
  V (\wtr , \wte) = - 2 \pi \int\limits_0^{2 \pi} \wtr \wtr_\theta \wte \, \mathrm{d}\theta = - \int\limits_0^{2 \pi} \int\limits_0^{2 \pi} \wtr \wtr_\theta \wte \, \mathrm{d}\theta \, \mathrm{d}\phi.
 \end{equation}
 Upon using the identity
 \begin{equation}
  \int\limits_{0}^{2 \pi} \left[ \wtr^2 \wte \right]_\theta \mathrm{d}\theta = 0 \quad \Rightarrow \int\limits_{0}^{2 \pi} \wtr^2 \wte_\theta \, \mathrm{d}\theta = - 2 \int\limits_{0}^{2 \pi} \wtr \wtr_\theta \wte \, \mathrm{d}\theta,
 \end{equation}
 we can write the above volume integral as
 \begin{equation}
  V (\wtr , \wte) = \frac{1}{2} \int\limits_0^{2 \pi} \int\limits_0^{2 \pi} \wtr^2 \wte_\theta \, \mathrm{d}\theta \, \mathrm{d}\phi.
 \end{equation}

 If the torus deforms from the reference configuration with volume $V_0$ to the current configuration with volume $V$ at constant pressure $\wt{P}$ then the total work done can be written as
 \begin{equation}
  \frac{1}{2} \int\limits_0^{2 \pi} \int\limits_0^{2 \pi} \wt{P} \wtr^2 \wte_\theta \, \mathrm{d}\theta \, \mathrm{d}\phi - \wt{P} V_0,
 \end{equation}
 which is the same expression as  \eqref{eqn: primitive of virtual work} barring the constant term.
 
 \section{Derivatives of the energy density function}\label{appendix: derivatives} The first derivatives of the right Cauchy--Green deformation tensor $\CGright$ as expressed in Equation~(8)  with respect to $\vr$, $\vr_\theta$, $\eta$ and $\eta_\theta$ are given by
\begin{align}
 \CGright_{\vr} &= \text{diag} \left(0,  \frac{2 \lambda_2}{1 + \gamma \cos \theta}, \frac{-2}{\lambda_1^2 \lambda_2^3[1 + \gamma \cos \theta]} \right), \nonumber \\
 \CGright_{\vr_\theta} &= \text{diag} \left( \frac{2 \vr_\theta}{\gamma^2}, 0, \frac{-2 \vr_\theta}{\lambda_1^4 \lambda_2^2 \gamma^2} \right), \nonumber\\ 
 \CGright_{\eta} &= \mbf{0}, \nonumber \\
  \CGright_{\eta_\theta} &= \text{diag} \left(  \frac{2 \eta_\theta}{\gamma^2}, 0, \frac{-2 \eta_\theta}{\lambda_1^4 \lambda_2^2 \gamma^2} \right). 
\end{align}
Then non-zero second derivatives of $\CGright$ are
\begin{align}
 \CGright_{\vr\vr} &= \text{diag} \left(0,  \frac{2}{[1 + \gamma \cos \theta]^2}, \frac{6}{\lambda_1^2 \lambda_2^4[1 + \gamma \cos \theta]^2} \right), \nonumber \\
 \CGright_{\vr\vr_\theta} &= \text{diag} \left( 0, 0, \frac{4\vr_\theta}{\gamma^2\lambda_1^4 \lambda_2^3 [1+\gamma\cos \theta]} \right), \nonumber \\
  \CGright_{\vr\eta_\theta} &= \text{diag} \left( 0, 0, \frac{4\eta_\theta}{\gamma^2\lambda_1^4 \lambda_2^3 [1+\gamma\cos \theta]} \right), \nonumber \\
  \CGright_{\vr_\theta \vr_\theta} &= \text{diag} \left( \frac{2}{\gamma^2}, 0, \frac{8\vr_\theta^2}{\gamma^4\lambda_1^6\lambda_2^2} - \frac{2}{\gamma^2 \lambda_1^4 \lambda_2^2} \right), \nonumber \\
   \CGright_{\eta_\theta \eta_\theta} &= \text{diag} \left( \frac{2}{\gamma^2}, 0, \frac{8\eta_\theta^2}{\gamma^4\lambda_1^6\lambda_2^2} - \frac{2}{\gamma^2 \lambda_1^4 \lambda_2^2} \right), \nonumber \\
     \CGright_{\vr_\theta \eta_\theta} &= \text{diag} \left( 0,0, \frac{8\vr_\theta\eta_\theta}{\gamma^4\lambda_1^6 \lambda_2^2}\right).
\end{align}
The full derivatives of some of the second derivatives with respect to $\theta$ are computed as
\begin{align}
\fd{  \CGright_{\vr\vr_\theta}}{ \theta} & = \text{diag} \left( 0,0, \frac{4}{\gamma^2} \left[ \frac{\vr_{\theta\theta}}{\lambda_1^4\lambda_2^3 [1+\gamma\cos\theta]}+\frac{\vr_\theta\gamma\sin\theta}{\lambda_1^4\lambda_2^3[1+\gamma\cos\theta]^2} - \frac{3\vr_\theta{\lambda_2}_\theta}{\lambda_1^4\lambda_2^4[1+\gamma\cos\theta]} - \frac{4\vr_\theta{\lambda_1}_\theta}{\lambda_1^5\lambda_2^3[1+\gamma\cos\theta]} \right] \right), \nonumber \\
\fd{  \CGright_{\vr\eta_\theta}}{ \theta} & = \text{diag} \left( 0,0, \frac{4}{\gamma^2} \left[ \frac{\eta_{\theta\theta}}{\lambda_1^4\lambda_2^3 [1+\gamma\cos\theta]}+\frac{\eta_\theta\gamma\sin\theta}{\lambda_1^4\lambda_2^3[1+\gamma\cos\theta]^2} - \frac{3\eta_\theta{\lambda_2}_\theta}{\lambda_1^4\lambda_2^4[1+\gamma\cos\theta]} - \frac{4\eta_\theta{\lambda_1}_\theta}{\lambda_1^5\lambda_2^3[1+\gamma\cos\theta]} \right] \right), \nonumber \\
 \fd{\CGright_{\vr_\theta \eta_\theta }}{\theta} &= \text{diag} \left( 0,0, \frac{8\vr_{\theta\theta}\eta_\theta}{\gamma^4\lambda_1^6 \lambda_2^2} + \frac{8\vr_\theta\eta_{\theta\theta}}{\gamma^4\lambda_1^6 \lambda_2^2} -  \frac{16\vr_\theta\eta_{\theta} {\lambda_2}_{\theta}}{\gamma^4\lambda_1^6 \lambda_2^3} - \frac{48\vr_\theta\eta_{\theta} {\lambda_1}_{\theta}}{\gamma^4\lambda_1^7 \lambda_2^2} \right), \nonumber \\
 \fd{\CGright_{\eta_\theta \eta_\theta }}{\theta} &= \text{diag} \left( 0,0, \frac{8}{\gamma^4}\left[ \frac{2\eta_\theta \eta_{\theta \theta}}{\lambda_1^6 \lambda_2^2} - \frac{6\eta_\theta ^2}{\lambda_1^7 \lambda_2^2}{\lambda_1}_{\theta} - \frac{2\eta_\theta ^2}{\lambda_1^6 \lambda_2^3}{\lambda_2}_{\theta}  \right] + \frac{2}{\gamma^2}\left[ \frac{4}{\lambda_1^5\lambda_2^2} {\lambda_1}_{\theta} + \frac{2}{\lambda_1^4\lambda_2^3} {\lambda_2}_{\theta} \right]\right).
\end{align}
Given the energy density function in equation~(29), the first derivatives of the energy density with respect to $\vr$ is computed as
\begin{align}
\Omega_{\vr} = &C_1 \pd{I_1}{\vr} + C_2 \pd{I_2}{\vr} + \beta [{\CGright}_{\vr} \disR] \cdot \disR \nonumber \\
= &C_1 \frac{2\lambda_2}{[1+ \gamma \cos\theta]} \Bigg[ \left[ 1+\alpha \lambda_1^2 \right] \left[ 1 - \frac{1}{\lambda_1^2 \lambda_2^4} \right] \Bigg]  - \frac{\Phi_0^2 \lambda_2 \lambda_1^2}{2 \beta  H^2 [1 + \gamma \cos \theta]}  .
\end{align}
\begin{subequations}
Similarly, other first derivatives of energy density function $\Omega$ are given by
\begin{align}
\Omega_{\disR} & = 2 \beta \CGright \disR , \nonumber \\
\Omega_{\vr_\theta} & = C_1 \frac{2 \varrho_\theta}{\gamma^2} \Bigg[ \left[ 1 + \alpha \lambda_2^2 \right] \left[ 1 - \frac{1}{\lambda_1^4 \lambda_2^2} \right] \bigg] - \frac{ \vr_\theta \Phi_0^2 \lambda_2^2}{ 2\beta H^2 \gamma^2 } , \nonumber  \\
\Omega_{\eta_\theta} & = C_1 \frac{2\eta_\theta}{\gamma^2} \Bigg[ \left[ 1 + \alpha \lambda_2^2 \right] \left[ 1 - \frac{1}{\lambda_1^4 \lambda_2^2} \right] \Bigg] - \frac{ \eta_\theta \Phi_0^2 \lambda_2^2}{2 \beta  H^2 \gamma^2 } , \nonumber \\
\Omega_{\eta} &= 0 .
\end{align}
\end{subequations}

The non-zero second derivatives of the energy density are given by
\begin{align}
\Omega_{\vr\vr} &=2 C_1  \frac{\left[ 1+\alpha \lambda_1^2 \right]}{[1+ \gamma \cos\theta]^2} \Bigg[ 1 + \frac{3}{\lambda_1^2\lambda_2^4} \Bigg]   + \frac{3\Phi_0^2 \lambda_1^2}{ 2 \beta  H^2 [1 + \gamma \cos \theta]^2} , \nonumber \\
\Omega_{\vr \vr_\theta} &=  C_1 \frac{4\lambda_2}{[1+ \gamma \cos\theta]} \frac{\vr_\theta}{\gamma^2}  \Bigg[ \alpha+\frac{1}{\lambda_1^4\lambda_2^4} \Bigg]+ \frac{\vr_\theta}{\gamma^2}  \frac{\Phi_0^2\lambda_2}{ \beta H^2 [1 + \gamma \cos \theta]} , \nonumber \\
\Omega_{\vr\eta_\theta} & =  C_1 \frac{4\lambda_2}{[1+ \gamma \cos\theta]} \frac{\eta_\theta}{\gamma^2}  \Bigg[ \alpha + \frac{1}{\lambda_1^4\lambda_2^4} \Bigg] + \frac{\eta_\theta}{\gamma^2}  \frac{\Phi_0^2\lambda_2}{ \beta H^2 [1 + \gamma \cos \theta]} , \nonumber \\
\Omega_{\vr_\theta \vr_\theta} & = \frac{2C_1}{\gamma^2}[1+\alpha \lambda_2^2] \left[ \left[ 1- \frac{1}{\lambda_1^4 \lambda_2^2}\right] + \frac{4 \vr_\theta^2}{\gamma^2 \lambda_1^6 \lambda_2^2} \right] + \frac{2\Phi_0^2 \vr_\theta^2\lambda_2^2}{\beta H^2\gamma^4\lambda_1^2} - \frac{\Phi_0^2\lambda_2^2}{2\beta H^2\gamma^2}, \nonumber \\
\Omega_{\vr_\theta \eta_\theta} & = 8C_1\frac{\vr_\theta \eta_\theta}{\gamma^4}\left[ 1+\alpha\lambda_2^2 \right]\frac{1}{\lambda_1^6\lambda_2^2}+\frac{\vr_\theta \eta_\theta}{\gamma^4}\frac{2\Phi_0^2 \lambda_2^2}{ \beta H^2 \lambda_1^2}, \nonumber  \\
\Omega_{\eta_\theta \eta_\theta} & = \frac{2C_1}{\gamma^2}[1+\alpha \lambda_2^2] \left[ \left[ 1- \frac{1}{\lambda_1^4 \lambda_2^2}\right] + \frac{4 \eta_\theta^2}{\gamma^2 \lambda_1^6 \lambda_2^2} \right] + \frac{2\Phi_0^2 \eta_\theta^2\lambda_2^2}{\beta H^2\gamma^4\lambda_1^2} - \frac{\Phi_0^2\lambda_2^2}{2\beta H^2\gamma^2}.
\end{align}
Full derivatives of some second derivatives of the energy density function with respect to $\theta$ are computed as
\begin{align}
\fd{\Omega_{\eta_\theta \eta_\theta}}{\theta} = & \frac{2C_1}{\gamma^2} \Bigg[ 2\alpha \lambda_2 {\lambda_2}_{\theta} \left[ \left[ 1- \frac{1}{\lambda_1^4 \lambda_2^2}\right] + \frac{4 \eta_\theta^2}{\gamma^2 \lambda_1^6 \lambda_2^2} \right] \nonumber  \\
&+  [1+\alpha \lambda_2^2] \left[ \frac{4{\lambda_1}_\theta}{\lambda_1^5\lambda_2^2} + \frac{2{\lambda_2}_\theta}{\lambda_1^4\lambda_2^3} + \frac{4}{\gamma^2} \left[ \frac{2\eta_\theta \eta_{\theta\theta}}{\lambda_1^6 \lambda_2^2} - \frac{6\eta_\theta^2{\lambda_1}_\theta}{\lambda_1^7 \lambda_2^2} - \frac{2\eta_\theta^2 {\lambda_2}_{\theta}}{\lambda_1^6 \lambda_2^3} \right]   \right] \Bigg] \nonumber \\
&+ \frac{\Phi_0^2}{\beta H^2}\left[\frac{2}{\gamma^4}\left[ \frac{2\eta_\theta \eta_{\theta \theta}\lambda_2^2}{\lambda_1^2} - \frac{6\eta_\theta ^2 \lambda_2^2}{\lambda_1^3}{\lambda_1}_{\theta} - \frac{2\eta_\theta ^2 \lambda_2}{\lambda_1^2}{\lambda_2}_{\theta}  \right] + \frac{1}{2\gamma^2}\left[ \frac{4\lambda_2^2}{\lambda_1} {\lambda_1}_{\theta} + {2\lambda_2} {\lambda_2}_{\theta} \right]\right] \nonumber \\
\fd{\Omega_{\vr\vr_\theta}}{ \theta} = &\frac{4C_1}{\gamma^2}\Bigg[ \left[\frac{\vr_\theta \gamma \sin\theta}{[1+\gamma \cos\theta]^2} + \frac{\vr_{\theta \theta}}{[1+\gamma \cos\theta]}\right]\left[ \alpha \lambda_2 + \frac{1}{\lambda_1^4 \lambda_2^3} \right] \nonumber \\
&+ \frac{\vr_\theta}{[1+\gamma\cos\theta]} \left[ \alpha {\lambda_2}_{\theta} - \frac{4}{\lambda_1^5 \lambda_2^3} {\lambda_1}_{\theta} - \frac{3}{\lambda_1^4\lambda_2^4}   {\lambda_2}_\theta \right] \Bigg] \nonumber \\
 &- \frac{\Phi_0^2}{\beta H^2\gamma^2} \left[ \frac{\vr_{\theta \theta}\lambda_2}{ [1+\gamma\cos\theta]}+\frac{\vr_{\theta}\gamma\sin\theta \lambda_2}{[1+\gamma\cos\theta]^2} - \frac{3\vr_\theta{\lambda_2}_{\theta}}{[1+\gamma\cos\theta]} - \frac{4\vr_\theta \lambda_2{\lambda_1}_{\theta}}{\lambda_1[1+\gamma\cos\theta]} \right] \nonumber \\
\fd{\Omega_{\vr\eta_\theta}}{ \theta} = &\frac{4C_1}{\gamma^2}\Bigg[ \left[\frac{\eta_\theta \gamma \sin\theta}{[1+\gamma \cos\theta]^2} + \frac{\eta_{\theta \theta}}{[1+\gamma \cos\theta]}\right]\left[ \alpha \lambda_2 + \frac{1}{\lambda_1^4 \lambda_2^3} \right] \nonumber \\
&+ \frac{\eta_\theta}{[1+\gamma\cos\theta]} \left[ \alpha {\lambda_2}_{\theta} - \frac{4}{\lambda_1^5 \lambda_2^3} {\lambda_1}_{\theta} - \frac{3}{\lambda_1^4\lambda_2^4}   {\lambda_2}_\theta \right] \Bigg] \nonumber \\
 &- \frac{\Phi_0^2}{\beta H^2\gamma^2} \left[ \frac{\eta_{\theta \theta}\lambda_2}{ [1+\gamma\cos\theta]}+\frac{\eta_{\theta}\gamma\sin\theta \lambda_2}{[1+\gamma\cos\theta]^2} - \frac{3\eta_\theta{\lambda_2}_{\theta}}{[1+\gamma\cos\theta]} - \frac{4\eta_\theta \lambda_2 {\lambda_1}_{\theta}}{\lambda_1[1+\gamma\cos\theta]} \right] \nonumber \\
\fd{\Omega_{\vr_\theta \eta_\theta}}{ \theta} =  &\frac{8C_1}{\gamma^4}\Bigg[\left[ \vr_{\theta \theta} \eta_{\theta} + \vr_{\theta } \eta_{\theta \theta} \right] \left[ 1+\alpha \lambda_2^2 \right]\frac{1}{\lambda_1^6\lambda_2^2} +  {\vr_\theta \eta_\theta}\left[ 2\alpha\lambda_2 {\lambda_2}_{\theta} \right]\frac{1}{\lambda_1^6\lambda_2^2} \nonumber \\
&- {\vr_\theta \eta_\theta}{\gamma^4}\left[ 1+\alpha\lambda_2^2 \right] \left[\frac{6{\lambda_1}_{\theta}}{\lambda_1^7\lambda_2^2} + \frac{2{\lambda_2}_{\theta}}{\lambda_1^6 \lambda_2^3}\right]\Bigg] \nonumber \\
& + \frac{\Phi_0^2}{\beta H^2} \left[ \frac{2\vr_{\theta\theta}\eta_\theta \lambda_2^2}{\gamma^4\lambda_1^2} + \frac{2\vr_\theta\eta_{\theta\theta} \lambda_2^2}{\gamma^4\lambda_1^2} -  \frac{4\vr_\theta\eta_{\theta} {\lambda_2}_{\theta} \lambda_2}{\gamma^4\lambda_1^2} - \frac{12\vr_\theta\eta_{\theta} {\lambda_1}_{\theta}\lambda_2^2}{\gamma^4\lambda_1^3}\right]
\end{align}
where
 \begin{align}
  \lambda_{1 \theta} = \frac{\vr_\theta \vr_{\theta \theta} + \eta_\theta \eta_{\theta \theta}}{\gamma \left[ \vr_\theta^2 + \eta_\theta^2 \right]^{1/2}} = \frac{\vr_\theta \vr_{\theta \theta} + \eta_\theta \eta_{\theta \theta}}{\gamma^2 \lambda_1} , \quad {\lambda_2}_\theta =  \frac{\vr_\theta}{1 + \gamma \cos \theta} + \frac{\vr \gamma \sin \theta}{[1 + \gamma \cos \theta]^2}.
 \end{align}
 
The second various derivatives of $\Omega$ with respect to $\disR$ are computed as
\begin{subequations}
 \begin{align}
  \Omega_{\disR \disR} &= 2 \beta \CGright, \nonumber \\
  \Omega_{\disR \vr} &= 2 \beta \CGright_{\vr} \disR = \frac{\Phi_0}{H}\CGright_{\vr} \CGright^{-1}\mathbf{N}, \nonumber \\
    \Omega_{\disR \vr_\theta} &= 2 \beta \CGright_{\vr_\theta} \disR = \frac{\Phi_0}{H}\CGright_{\vr_\theta} \CGright^{-1}\mathbf{N}, \nonumber \\
  \Omega_{\disR \eta_\theta} &= 2 \beta \CGright_{\eta_\theta} \disR = \frac{\Phi_0}{H}\CGright_{\eta_\theta} \CGright^{-1}\mathbf{N}, \nonumber \\
 \fd{ \Omega_{\disR \vr_\theta }}{\theta} &= 2 \beta \CGright_{\vr_\theta \theta} \disR =\frac{\Phi_0}{H}\fd{\CGright_{\vr_\theta}}{\theta} \CGright^{-1}\mathbf{N} , \nonumber \\
 \fd{ \Omega_{\disR \eta_\theta}}{\theta} &= 2 \beta \CGright_{\eta_\theta \theta} \disR = \frac{\Phi_0}{H} \fd{\CGright_{\eta_\theta}}{\theta} \CGright^{-1}\mathbf{N},
 \end{align}
\end{subequations}
where
\begin{subequations}
 \begin{align}
   \fd{ \CGright_{\vr_\theta}}{ \theta} &= \text{diag} \bigg( -\frac{2\vr_{\theta \theta}}{\gamma^2},0,-\frac{2}{\gamma^2}\left( \frac{\vr_{\theta \theta}}{\lambda_1^4\lambda_2^2} - \frac{4\vr_\theta}{\lambda_1^5\lambda_2^2}\lambda_{1\theta} -\frac{2\vr_\theta}{\lambda_1^4 \lambda_2^3}\lambda_{2\theta}\right)\bigg), \\
   \fd{\CGright_{\eta_\theta }}{\theta} &= \text{diag} \bigg( -\frac{2\eta_{\theta \theta}}{\gamma^2},0,-\frac{2}{\gamma^2}\left( \frac{\eta_{\theta \theta}}{\lambda_1^4\lambda_2^2} - \frac{4\eta_\theta}{\lambda_1^5\lambda_2^2}\lambda_{1\theta} -\frac{2\eta_\theta}{\lambda_1^4 \lambda_2^3}\lambda_{2\theta}\right)\bigg).
 \end{align}
\end{subequations}
Upon rewriting, we get
\begin{subequations}
\begin{align}
\Omega_{\disR \disR} &= 2\beta\text{diag}\left( \lambda_1^2, \lambda_2^2, \lambda_3^2 \right), \\
 \Omega_{\disR \vr} & = \frac{\Phi_0}{H}  \text{diag} \bigg( 0,\frac{2}{\lambda_2 (1+\gamma \cos(\theta))},-\frac{2}{\lambda_2 (1+\gamma \cos(\theta))} \bigg) \mathbf{N},\\
  \Omega_{\disR \vr_\theta} & = \frac{\Phi_0}{H}  \text{diag} \bigg(\frac{2\vr_\theta}{\lambda_1^2 \gamma^2},0,\frac{-2\vr_\theta}{\lambda_1^2\gamma^2}\bigg)\mathbf{N}, \\
  \Omega_{\disR \eta_\theta} & = \frac{\Phi_0}{H}  \text{diag} \bigg(\frac{2\eta_\theta}{\lambda_1^2 \gamma^2},0,\frac{-2\eta_\theta}{\lambda_1^2\gamma^2}\bigg)\mathbf{N}, \\
\fd{ \Omega_{\disR \vr_\theta}}{ \theta} &= \frac{\Phi_0}{H}  \text{diag} \bigg( -\frac{2\vr_{\theta \theta}}{\gamma^2\lambda_1^2},0,-\frac{2}{\gamma^2}\left( \frac{\vr_{\theta \theta}}{\lambda_1^2} - \frac{4\vr_\theta}{\lambda_1^3}\lambda_{1\theta} -\frac{2\vr_\theta}{\lambda_1^2 \lambda_2}\lambda_{2\theta}\right)\bigg) \mathbf{N}, \\
\fd{ \Omega_{\disR \eta_\theta}}{ \theta} &=  \frac{\Phi_0}{H}   \text{diag} \bigg( -\frac{2\eta_{\theta \theta}}{\gamma^2\lambda_1^2},0,-\frac{2}{\gamma^2}\left( \frac{\eta_{\theta \theta}}{\lambda_1^2} - \frac{4\eta_\theta}{\lambda_1^3}\lambda_{1\theta} -\frac{2\eta_\theta}{\lambda_1^2 \lambda_2}\lambda_{2\theta}\right)\bigg) \mathbf{N}.
\end{align}
\end{subequations}

\section{Second derivatives for computing the second variation of the potential energy}
\label{appendix: second_derivatives}
In Section~5, the loss of symmetry in the $\phi$ direction was considered. Bifurcation occurs when the second variation of the potential energy function becomes zero. However, the symmetric assumption in solutions is no longer valid and $\lambda_1$, $\lambda_2$ and $\lambda_3$ in the right Cauchy--Green deformation tensor are computed using the full expressions~(6) and~(7). The second derivatives of the energy density function with respect to $\vr$ is expressed as
\begin{align}
\Omega_{\vr \vr} = C_1[ {I_1}_{\vr \vr} + \alpha {I_2}_{\vr \vr}] + \beta [ \CGright_{\vr \vr} \disR]\cdot\disR,
\end{align}
where 
\begin{align}
{I_1}_{\vr \vr} &= {\lambda_1}^2_{\vr\vr} +{\lambda_2}^2_{\vr\vr}+{\lambda_3}^2_{\vr\vr}, \nonumber \\
{I_2}_{\vr \vr} &= {\lambda_1}^{-2}_{\vr\vr} +{\lambda_2}^{-2}_{\vr\vr}+{\lambda_3}^{-2}_{\vr\vr}.
\end{align}
On substituting the expressions~(6) and~(7) into the previous equations, one obtains
\begin{align}
 &{I_1}_{\vr \vr} = \frac{2}{[1+\gamma \cos \theta]^2} - \frac{2\gamma^2[1+\gamma \cos \theta]^2[\vr_\theta^2 + \eta_\theta^2]}{\left[[\vr_\theta \eta_\phi - \vr_\phi \eta_\theta]^2 + \vr^2 [\vr_\theta^2 + \eta_\theta^2]\right]^2} + \frac{8\gamma^2 [1 + \gamma \cos \theta]^2 \vr^2 [\vr_\theta^2 + \eta_\theta^2]^2}{\left[[\vr_\theta \eta_\phi - \vr_\phi \eta_\theta]^2 + \vr^2 [\vr_\theta^2 + \eta_\theta^2]\right]^3}\nonumber\\
  &{I_2}_{\vr \vr} = \frac{ 2 \gamma^2 }{[\vr_\theta \eta_\phi - \vr_\phi \eta_\theta]^2 + \vr^2[\vr^2_\theta + \eta^2_\theta]} - \frac{4\gamma^2 \vr^2[\vr_\theta^2+\eta_\theta^2]}{\left[[\vr_\theta \eta_\phi - \vr_\phi \eta_\theta]^2 + \vr^2[\vr^2_\theta + \eta^2_\theta]\right]^2} \nonumber\\
   &- \frac{2[\vr_\theta^2 + \eta_\theta^2]\left[\gamma^2[\vr_\phi^2 + \eta_\phi^2 + \vr^2]+[\vr_\theta^2 + \eta_\theta^2][1+\gamma \cos \theta]^2 \right]}{\left[[\vr_\theta \eta_\phi - \vr_\phi \eta_\theta]^2 + \vr^2[\vr^2_\theta + \eta^2_\theta]\right]^2} \nonumber\\
    &+ \frac{8\vr^2[\vr_\theta^2 + \eta_\theta^2]^2\left[ \gamma^2[\vr_\phi^2 + \eta_\phi^2 +\vr^2] + [\vr_\theta^2 + \eta_\theta^2] [1+\gamma \cos \theta]^2\right]}{\left[[\vr_\theta \eta_\phi - \vr_\phi \eta_\theta]^2 + \vr^2[\vr^2_\theta + \eta^2_\theta]\right]^3} +\frac{2[\vr_\theta^2 + \eta_\theta^2]}{\gamma^2[1+\gamma \cos \theta]^2},
\end{align}
and
\begin{align}
\beta [ \CGright_{\vr\vr} \disR]\cdot\disR = \frac{2\Phi_0^2\vr^2[\vr_\theta^2 + \eta_\theta^2]^2}{\beta H^2 \gamma^2[1+\gamma \cos \theta]^2\left[[\vr_\theta \eta_\phi - \vr_\phi \eta_\theta]^2 + \vr^2[\vr^2_\theta + \eta^2_\theta]\right]}- \frac{\Phi_0^2[\vr_\theta^2 + \eta_\theta^2]}{2\beta H^2\gamma^2[1+\gamma \cos \theta]^2}.
\end{align}
Similarly, the second derivative of $\Omega$ with respect to $\vr$ and $\vr_\theta$ can be computed as
\begin{align}
    \Omega_{\vr \vr_\theta} = C_1[ {I_1}_{\vr \vr_\theta} + \alpha {I_2}_{\vr \vr_\theta}] + \beta [ \CGright_{\vr \vr_\theta} \disR]\cdot\disR,
\end{align}
where
\begin{align}   
   & {I_1}_{\vr \vr_\theta} = \frac{4\gamma^2\vr[\vr_\theta^2 + \eta_\theta^2]\left[ 2\vr^2\vr_\theta + 2\eta_\phi[\eta_\phi \vr_\theta - \vr_\phi \eta_\theta] \right][1+\gamma \cos \theta]^2}{\left[[\vr_\theta \eta_\phi - \vr_\phi \eta_\theta]^2 + \vr^2[\vr^2_\theta + \eta^2_\theta]\right]^3} - \frac{4\gamma^2 \vr \vr_\theta [1+\gamma \cos \theta]^2}{\left[[\vr_\theta \eta_\phi - \vr_\phi \eta_\theta]^2 + \vr^2[\vr^2_\theta + \eta^2_\theta]\right]^2}, \\
    &{I_2}_{\vr \vr_\theta} = -\frac{4\gamma^2 \vr \left[ 2\vr^2\vr_\theta + 2\eta_\phi[\eta_\phi \vr_\theta - \vr_\phi \eta_\theta] \right] }{\left[[\vr_\theta \eta_\phi - \vr_\phi \eta_\theta]^2 + \vr^2[\vr^2_\theta + \eta^2_\theta]\right]^2} - \frac{4 \vr \vr_\theta [\vr_\theta^2 + \eta_\theta^2] [1+\gamma \cos \theta]^2 }{\left[[\vr_\theta \eta_\phi - \vr_\phi \eta_\theta]^2 + \vr^2[\vr^2_\theta + \eta^2_\theta]\right]^2}\nonumber\\
    &-\frac{4\vr\vr_\theta\left[[\vr^2_\theta + \eta^2_\theta][1+\gamma \cos \theta]^2 + [\vr^2_\phi + \eta^2_\phi + \vr^2]\gamma^2\right]}{\left[[\vr_\theta \eta_\phi - \vr_\phi \eta_\theta]^2 + \vr^2[\vr^2_\theta + \eta^2_\theta]\right]^2}\nonumber\\ 
    &+\frac{4 \vr [\vr_\theta^2 + \eta_\theta^2]\left[2\vr^2\vr_\theta + 2 \eta_\phi [\eta_\phi \vr_\theta - \vr_\phi \eta_\theta] \right]\left[[\vr^2_\theta + \eta^2_\theta][1+\gamma \cos \theta]^2 + [\vr^2_\phi + \eta^2_\phi + \vr^2]\gamma^2\right] }{\left[[\vr_\theta \eta_\phi - \vr_\phi \eta_\theta]^2 + \vr^2[\vr^2_\theta + \eta^2_\theta]\right]^3} + \frac{4\vr\vr_\theta}{\gamma^2[1+\gamma \cos \theta]^2},\\
   & \beta [ \CGright_{\vr \vr_\theta} \disR]\cdot\disR = \frac{\Phi_0^2 \vr [\vr_\theta^2 + \eta_\theta^2][2\vr^2 \vr_\theta + 2 \eta_\phi\left[\eta_\phi \vr_\theta - \vr_\phi \eta_\theta]\right]}{\beta H^2 \left[[\vr_\theta \eta_\phi - \vr_\phi \eta_\theta]^2 + \vr^2[\vr^2_\theta + \eta^2_\theta]\right]\gamma^2[1 + \gamma\cos\theta]^2} - \frac{\Phi_0^2 \vr \vr_\theta}{\beta H^2 \gamma^2[1+\gamma\cos\theta]^2}.
\end{align}
The second derivative of $\Omega$ with respect to $\vr$ and $\eta_\theta$ can be computed as
\begin{align}
  &  \Omega_{\vr \eta_\theta} = C_1[ {I_1}_{\vr \eta_\theta} + \alpha {I_2}_{\vr \eta_\theta}] + \beta [ \CGright_{\vr \eta_\theta} \disR]\cdot\disR ,
\end{align}
where
\begin{align}   
   & {I_1}_{\vr \eta_\theta} = \frac{4\gamma^2\vr[\vr_\theta^2 + \eta_\theta^2]\left[ 2\vr^2\eta_\theta - 2\vr_\phi[\eta_\phi \vr_\theta - \vr_\phi \eta_\theta] \right][1+\gamma \cos \theta]^2}{\left[[\vr_\theta \eta_\phi - \vr_\phi \eta_\theta]^2 + \vr^2[\vr^2_\theta + \eta^2_\theta]\right]^3} - \frac{4\gamma^2 \vr \eta_\theta [1+\gamma \cos \theta]^2}{\left[[\vr_\theta \eta_\phi - \vr_\phi \eta_\theta]^2 + \vr^2[\vr^2_\theta + \eta^2_\theta]\right]^2}, \\
    &{I_2}_{\vr \eta_\theta} = -\frac{4\gamma^2 \vr \left[ 2\vr^2\eta_\theta - 2\vr_\phi[\eta_\phi \vr_\theta - \vr_\phi \eta_\theta] \right] }{\left[[\vr_\theta \eta_\phi - \vr_\phi \eta_\theta]^2 + \vr^2[\vr^2_\theta + \eta^2_\theta]\right]^2} - \frac{4 \vr \eta_\theta [\vr_\theta^2 + \eta_\theta^2] [1+\gamma \cos \theta]^2 }{\left[[\vr_\theta \eta_\phi - \vr_\phi \eta_\theta]^2 + \vr^2[\vr^2_\theta + \eta^2_\theta]\right]^2} \nonumber \\
    &-\frac{4\vr\eta_\theta\left[[\vr^2_\theta + \eta^2_\theta][1+\gamma \cos \theta]^2 + [\vr^2_\phi + \eta^2_\phi + \vr^2]\gamma^2\right]}{\left[[\vr_\theta \eta_\phi - \vr_\phi \eta_\theta]^2 + \vr^2[\vr^2_\theta + \eta^2_\theta]\right]^2} \nonumber \\ 
    &+\frac{4 \vr [\vr_\theta^2 + \eta_\theta^2]\left[2\vr^2\eta_\theta - 2 \vr_\phi [\eta_\phi \vr_\theta - \vr_\phi \eta_\theta] \right]\left[[\vr^2_\theta + \eta^2_\theta][1+\gamma \cos \theta]^2 + [\vr^2_\phi + \eta^2_\phi + \vr^2]\gamma^2\right] }{\left[[\vr_\theta \eta_\phi - \vr_\phi \eta_\theta]^2 + \vr^2[\vr^2_\theta + \eta^2_\theta]\right]^3} \nonumber \\
    &+ \frac{4\vr\eta_\theta}{\gamma^2[1+\gamma \cos \theta]^2}, \\
    &\beta [ \CGright_{\vr \eta_\theta} \disR]\cdot\disR = \frac{\Phi_0^2 \vr [\vr_\theta^2 + \eta_\theta^2][2\vr^2 \eta_\theta - 2 \vr_\phi\left[\eta_\phi \vr_\theta - \vr_\phi \eta_\theta]\right]}{\beta H^2 \left[[\vr_\theta \eta_\phi - \vr_\phi \eta_\theta]^2 + \vr^2[\vr^2_\theta + \eta^2_\theta]\right]\gamma^2[1 + \gamma\cos\theta]^2} - \frac{\Phi_0^2 \vr \eta_\theta}{\beta H^2 \gamma^2[1+\gamma\cos\theta]^2}
\end{align}
The second derivative of $\Omega$ with respect to $\vr$ and $\vr_\phi$ can be computed as
\begin{align}
    \Omega_{\vr \vr_\phi} = C_1[ {I_1}_{\vr \vr_\phi} + \alpha {I_2}_{\vr \vr_\phi}] + \beta [ \CGright_{\vr \vr_\phi} \disR]\cdot\disR ,
\end{align}
where
\begin{align}   
  &  {I_1}_{\vr \vr_\phi} = -\frac{8\vr\eta_\theta[\eta_\phi\vr_\theta - \vr_\phi \eta_\theta][\vr_\theta^2 + \eta_\theta^2]\gamma^2[1+\gamma \cos\theta]^2}{\left[[\vr_\theta \eta_\phi - \vr_\phi \eta_\theta]^2 + \vr^2[\vr^2_\theta + \eta^2_\theta]\right]^3}, \\
   & {I_2}_{\vr \vr_\phi} = \frac{4 \vr \gamma^2 \eta_\theta[\vr_\theta \eta_\phi - \vr_\phi \eta_\theta]}{\left[[\vr_\theta \eta_\phi - \vr_\phi \eta_\theta]^2 + \vr^2[\vr^2_\theta + \eta^2_\theta] \right]^2} -\frac{4\vr_\phi \vr [\vr_\theta^2+ \eta_\theta^2]\gamma^2}{\left[[\vr_\theta \eta_\phi - \vr_\phi \eta_\theta]^2 + \vr^2[\vr^2_\theta + \eta^2_\theta] \right]^2} \nonumber \\
    &- \frac{8\vr\eta_\theta[\vr_\theta \eta_\phi - \vr_\phi \eta_\theta][\vr_\theta^2 + \eta_\theta^2]\left[[\vr^2_\theta + \eta^2_\theta][1+\gamma \cos \theta]^2 + [\vr^2_\phi + \eta^2_\phi + \vr^2]\gamma^2\right]}{\left[[\vr_\theta \eta_\phi - \vr_\phi \eta_\theta]^2 + \vr^2[\vr^2_\theta + \eta^2_\theta] \right]^3} , \\
    &\beta [ \CGright_{\vr \vr_\phi} \disR]\cdot\disR = -\frac{2\Phi_0^2\vr\eta_\theta[\eta_\phi\vr_\theta - \vr_\phi \eta_\theta][\vr_\theta^2 + \eta_\theta^2]}{\beta H^2 \left[[\vr_\theta \eta_\phi - \vr_\phi \eta_\theta]^2 + \vr^2[\vr^2_\theta + \eta^2_\theta]\right]\gamma^2[1+\gamma \cos\theta]^2}.
\end{align}
The second derivative of $\Omega$ with respect to $\vr$ and $\eta_\phi$ is given by
\begin{align}
   \Omega_{\vr \eta_\phi} = C_1[ {I_1}_{\vr \eta_\phi} + \alpha {I_2}_{\vr \eta_\phi}] + \beta [ \CGright_{\vr \eta_\phi} \disR]\cdot\disR ,
\end{align}
where
\begin{align}   
    & {I_1}_{\vr \eta_\phi} = \frac{8\vr\vr_\theta[\eta_\phi\vr_\theta - \vr_\phi \eta_\theta][\vr_\theta^2 + \eta_\theta^2]\gamma^2[1+\gamma \cos\theta]^2}{\left[[\vr_\theta \eta_\phi - \vr_\phi \eta_\theta]^2 + \vr^2[\vr^2_\theta + \eta^2_\theta]\right]^3} , \\
    & {I_2}_{\vr \eta_\phi} = -\frac{4 \vr \gamma^2 \vr_\theta[\vr_\theta \eta_\phi - \vr_\phi \eta_\theta]}{\left[[\vr_\theta \eta_\phi - \vr_\phi \eta_\theta]^2 + \vr^2[\vr^2_\theta + \eta^2_\theta] \right]^2} -\frac{4\eta_\phi \vr [\vr_\theta^2+ \eta_\theta^2]\gamma^2}{\left[[\vr_\theta \eta_\phi - \vr_\phi \eta_\theta]^2 + \vr^2[\vr^2_\theta + \eta^2_\theta] \right]^2} \nonumber \\
    &+ \frac{8\vr\vr_\theta[\vr_\theta \eta_\phi - \vr_\phi \eta_\theta][\vr_\theta^2 + \eta_\theta^2]\left[[\vr^2_\theta + \eta^2_\theta][1+\gamma \cos \theta] + [\vr^2_\phi + \eta^2_\phi + \vr^2]\gamma^2\right]}{\left[[\vr_\theta \eta_\phi - \vr_\phi \eta_\theta]^2 + \vr^2[\vr^2_\theta + \eta^2_\theta] \right]^3}, \\
   &  \beta [ \CGright_{\vr \vr_\phi} \disR]\cdot\disR = \frac{2\Phi_0^2\vr\vr_\theta[\eta_\phi\vr_\theta - \vr_\phi \eta_\theta][\vr_\theta^2 + \eta_\theta^2]}{\beta H^2 \left[[\vr_\theta \eta_\phi - \vr_\phi \eta_\theta]^2 + \vr^2[\vr^2_\theta + \eta^2_\theta]\right]\gamma^2[1+\gamma \cos\theta]^2}
\end{align}
The second derivative of $\Omega$ with respect to $\vr_\theta$ is given by
\begin{align}
    &\Omega_{\vr_\theta \vr_\theta} = C_1[ {I_1}_{\vr_\theta \vr_\theta} + \alpha {I_2}_{\vr_\theta \vr_\theta} ]+ \beta [ \CGright_{\vr_\theta \vr_\theta} \disR]\cdot\disR,
\end{align}
where
\begin{align}   
    & {I_1}_{\vr_\theta \vr_\theta}= \frac{2}{\gamma^2} + \frac{2\gamma^2[1+\gamma \cos \theta]^2[2\eta_\phi^2 \vr_\theta - 2\eta_\phi \vr_\phi \eta_\theta + 2\vr^2\vr_\theta]^2 }{\left[[\vr_\theta \eta_\phi - \vr_\phi \eta_\theta]^2 + \vr^2[\vr^2_\theta + \eta^2_\theta]\right]^3} - \frac{\gamma^2[1+\gamma \cos \theta]^2[2\eta_\phi^2  + 2\vr^2] }{\left[[\vr_\theta \eta_\phi - \vr_\phi \eta_\theta]^2 + \vr^2[\vr^2_\theta + \eta^2_\theta]\right]^2}, \\
    & {I_2}_{\vr_\theta \vr_\theta} = \frac{2\vr^2 + 2\eta_\phi^2}{\gamma^2[1+\gamma \cos \theta]^2} - \frac{4\vr_\theta[1+\gamma \cos \theta]^2[2\vr^2\vr_\theta +2\eta_\phi^2\vr_\theta -2\eta_\theta \vr_\phi \eta_\phi] }{\left[[\vr_\theta \eta_\phi - \vr_\phi \eta_\theta]^2 + \vr^2[\vr^2_\theta + \eta^2_\theta]\right]^2} + \frac{2[1+\gamma \cos \theta]^2 }{[\vr_\theta \eta_\phi - \vr_\phi \eta_\theta]^2 + \vr^2[\vr^2_\theta + \eta^2_\theta]}\nonumber\\
    &+ \frac{2\left[[\vr^2_\theta + \eta^2_\theta][1+\gamma \cos \theta]^2 + [\vr^2_\phi + \eta^2_\phi + \vr^2]\gamma^2 \right] [2\eta_\phi^2 \vr_\theta - 2\eta_\phi \vr_\phi \eta_\theta + 2\vr^2\vr_\theta]^2}{\left[ [\vr_\theta \eta_\phi - \vr_\phi \eta_\theta]^2 + \vr^2[\vr^2_\theta + \eta^2_\theta] \right]^3}\nonumber\\
    &- \frac{\left[[\vr^2_\theta + \eta^2_\theta][1+\gamma \cos \theta]^2 + [\vr^2_\phi + \eta^2_\phi + \vr^2]\gamma^2 \right] [2\eta_\phi^2 + 2\vr^2]}{\left[ [\vr_\theta \eta_\phi - \vr_\phi \eta_\theta]^2 + \vr^2[\vr^2_\theta + \eta^2_\theta] \right]^2},\\
    &\beta [ \CGright_{\vr_\theta \vr_\theta} \disR]\cdot\disR = \frac{2\Phi_0^2[\eta_\phi^2 \vr_\theta - \eta_\phi \vr_\phi \eta_\theta + \vr^2\vr_\theta]^2 }{\beta H^2 \left[[\vr_\theta \eta_\phi - \vr_\phi \eta_\theta]^2 + \vr^2[\vr^2_\theta + \eta^2_\theta]\right]\gamma^2[1+\gamma \cos \theta]^2} - \frac{\Phi_0^2[\eta_\phi^2  + \vr^2] }{2\beta H^2 \gamma^2[1+\gamma \cos \theta]^2}.
\end{align}
The second derivative of $\Omega$ with respect to $\vr_\theta$ and $\eta_\theta$ is given by
\begin{align}
    &\Omega_{\vr_\theta \eta_\theta} = C_1[ {I_1}_{\vr_\theta \eta_\theta} + \alpha {I_2}_{\vr_\theta \eta_\theta} ]+ \beta [ \CGright_{\vr_\theta \eta_\theta} \disR]\cdot\disR,
\end{align}
where
\begin{align}   
    &{I_1}_{\vr_\theta \eta_\theta} = \frac{2\gamma^2[1+\gamma \cos \theta]^2[2\vr_\phi^2 \eta_\theta - 2\eta_\phi \vr_\phi \vr_\theta + 2\vr^2\eta_\theta][2\vr^2\vr_\theta + 2\vr_\theta\eta_\phi^2 - 2\vr_\phi\eta_\phi \eta_\theta] }{\left[[\vr_\theta \eta_\phi - \vr_\phi \eta_\theta]^2 + \vr^2[\vr^2_\theta + \eta^2_\theta]\right]^3} ,\\
    &- \frac{2\vr_\phi\eta_\phi\gamma^2[1+\gamma \cos \theta]^2}{\left[[\vr_\theta \eta_\phi - \vr_\phi \eta_\theta]^2 + \vr^2[\vr^2_\theta + \eta^2_\theta]\right]^2} \nonumber \\
    &{I_2}_{\vr_\theta \eta_\theta} = -\frac{2\vr_\phi\eta_\phi}{\gamma^2[1+\gamma \cos \theta]^2} - \frac{2\vr_\theta[1+\gamma \cos \theta]^2[2\vr^2\eta_\theta -2\vr_\phi\eta_\phi\vr_\theta + 2 \vr_\phi^2 \eta_\theta] }{\left[[\vr_\theta \eta_\phi - \vr_\phi \eta_\theta]^2 + \vr^2[\vr^2_\theta + \eta^2_\theta]\right]^2}\nonumber \\
    &-\frac{2\eta_\theta[1+\gamma \cos\theta]^2[2\vr^2\vr_\theta+2\eta_\phi^2\vr_\theta-2\eta_\phi\vr_\phi\eta_\theta]}{\left[[\vr_\theta \eta_\phi - \vr_\phi \eta_\theta]^2 + \vr^2[\vr^2_\theta + \eta^2_\theta]\right]^2} + \frac{2\vr_\phi \eta_\phi\left[[\vr^2_\theta + \eta^2_\theta][1+\gamma \cos \theta]^2 + [\vr^2_\phi + \eta^2_\phi + \vr^2]\gamma^2 \right] }{\left[[\vr_\theta \eta_\phi - \vr_\phi \eta_\theta]^2 + \vr^2[\vr^2_\theta + \eta^2_\theta]\right]^2}\nonumber \\
    &+\frac{2[2\vr^2\eta_\theta -2\vr_\phi\eta_\phi\vr_\theta + 2 \vr_\phi^2 \eta_\theta][2\vr^2\vr_\theta+2\eta_\phi^2\vr_\theta-2\eta_\phi\vr_\phi\eta_\theta]\left[[\vr^2_\theta + \eta^2_\theta][1+\gamma \cos \theta]^2 + [\vr^2_\phi + \eta^2_\phi + \vr^2]\gamma^2 \right]}{\left[[\vr_\theta \eta_\phi - \vr_\phi \eta_\theta]^2 + \vr^2[\vr^2_\theta + \eta^2_\theta]\right]^3} , \\
    &\beta [ \CGright_{\vr_\theta \eta_\theta} \disR]\cdot\disR = \frac{2\Phi_0^2[\vr_\phi^2 \eta_\theta - \eta_\phi \vr_\phi \vr_\theta + \vr^2\eta_\theta][\vr^2\vr_\theta + \vr_\theta\eta_\phi^2 - \vr_\phi\eta_\phi \eta_\theta] }{\beta H^2 \left[[\vr_\theta \eta_\phi - \vr_\phi \eta_\theta]^2 + \vr^2[\vr^2_\theta + \eta^2_\theta]\right] \gamma^2[1+\gamma \cos \theta]^2 } - \frac{\Phi_0^2\vr_\phi\eta_\phi}{2\beta H^2 \gamma^2[1+\gamma \cos \theta]^2} .
\end{align}
The second derivative of $\Omega$ with respect to $\eta_\theta$ and $\vr_\phi$ is computed as
\begin{align}
   & \Omega_{\eta_\theta \vr_\phi} = C_1[ {I_1}_{\eta_\theta \vr_\phi} + \alpha {I_2}_{\eta_\theta \vr_\phi} ]+ \beta [ \CGright_{\eta_\theta \vr_\phi} \disR]\cdot\disR,
\end{align}
where
\begin{align}   
  &  {I_1}_{\eta_\theta \vr_\phi} = -\frac{4\gamma^2[1+\gamma \cos \theta]^2 [\vr_\theta\eta_\theta\eta_\phi - \eta_\theta^2\vr_\phi] [2\vr_\phi^2 \eta_\theta - 2\eta_\phi \vr_\phi \vr_\theta + 2\vr^2\eta_\theta]}{\left[[\vr_\theta \eta_\phi - \vr_\phi \eta_\theta]^2 + \vr^2[\vr^2_\theta + \eta^2_\theta]\right]^3} \nonumber\\
  &- \frac{[2\vr_\phi\eta_\theta - 2[\vr_\theta\eta_\phi - \vr_\phi \eta_\theta]]\gamma^2[1+\gamma \cos \theta]^2}{\left[[\vr_\theta \eta_\phi - \vr_\phi \eta_\theta]^2 + \vr^2[\vr^2_\theta + \eta^2_\theta]\right]^2} , \\
    & {I_2}_{\eta_\theta \vr_\phi} = \frac{2\eta_\theta \vr_\phi - 2[\eta_\phi \vr_\theta - \vr_\phi \eta_\theta]}{\gamma^2[1+\gamma \cos \theta]^2} + \frac{4\eta_\theta^2 [\eta_\phi\vr_\theta - \vr_\phi \eta_\theta] [1+\gamma \cos \theta]^2 }{\left[[\vr_\theta \eta_\phi - \vr_\phi \eta_\theta]^2 + \vr^2[\vr^2_\theta + \eta^2_\theta]\right]^2} \nonumber\\
   & - \frac{2\vr_\phi\gamma^2\left[2\vr^2\eta_\theta - 2\vr_\phi[\eta_\phi\vr_\theta - \vr_\phi \eta_\theta]\right]}{\left[[\vr_\theta \eta_\phi - \vr_\phi \eta_\theta]^2 + \vr^2[\vr^2_\theta + \eta^2_\theta]\right]^2} \nonumber \\
    &- \frac{\left[2\vr_\phi\eta_\theta - 2[\eta_\phi\vr_\theta - \vr_\phi\eta_\theta] \right] \left[[\vr^2_\theta + \eta^2_\theta][1+\gamma \cos \theta]^2 + [\vr^2_\phi + \eta^2_\phi + \vr^2]\gamma^2 \right] }{\left[ [\vr_\theta \eta_\phi - \vr_\phi \eta_\theta]^2 + \vr^2[\vr^2_\theta + \eta^2_\theta] \right]^2} \nonumber\\
    &- \frac{4\eta_\theta[\eta_\phi\vr_\theta - \vr_\phi\eta_\theta]\left[ 2\vr^2\vr_\theta - 2\vr_\phi[\eta_\phi\vr_\theta - \vr_\phi \eta_\theta] \right]\left[[\vr^2_\theta + \eta^2_\theta][1+\gamma \cos \theta]^2 + [\vr^2_\phi + \eta^2_\phi + \vr^2]\gamma^2 \right]}{\left[ [\vr_\theta \eta_\phi - \vr_\phi \eta_\theta]^2 + \vr^2[\vr^2_\theta + \eta^2_\theta] \right]^3} ,\\
    &\beta [ \CGright_{\eta_\theta \vr_\phi} \disR]\cdot\disR = -\frac{\Phi_0^2[\vr_\theta\eta_\theta\eta_\phi - \eta_\theta^2\vr_\phi] [2\vr_\phi^2 \eta_\theta - 2\eta_\phi \vr_\phi \vr_\theta + 2\vr^2\eta_\theta]}{\beta H^2 \left[[\vr_\theta \eta_\phi - \vr_\phi \eta_\theta]^2 + \vr^2[\vr^2_\theta + \eta^2_\theta]\right]\gamma^2[1+\gamma \cos \theta]^2} - \frac{\Phi_0^2[\vr_\phi\eta_\theta - \vr_\theta\eta_\phi + \vr_\phi \eta_\theta]}{2\beta H^2\gamma^2[1+\gamma \cos \theta]^2}
\end{align}
The second derivative of $\Omega$ with respect to $\eta_\theta$ is computed as
\begin{align}
    &\Omega_{\eta_\theta \eta_\theta} = C_1[ {I_1}_{\eta_\theta \eta_\theta} + \alpha {I_2}_{\eta_\theta \eta_\theta} ]+ \beta [ \CGright_{\eta_\theta \eta_\theta} \disR]\cdot\disR,
\end{align}
where
\begin{align}   
    & {I_1}_{\eta_\theta \eta_\theta} = \frac{2}{\gamma^2} + \frac{2\gamma^2[1+\gamma \cos \theta]^2[2\vr_\phi^2 \eta_\theta - 2\eta_\phi \vr_\phi \vr_\theta + 2\vr^2\eta_\theta]^2 }{\left[[\vr_\theta \eta_\phi - \vr_\phi \eta_\theta]^2 + \vr^2[\vr^2_\theta + \eta^2_\theta]\right]^3} - \frac{\gamma^2[1+\gamma \cos \theta]^2[2\vr_\phi^2  + 2\vr^2] }{\left[[\vr_\theta \eta_\phi - \vr_\phi \eta_\theta]^2 + \vr^2[\vr^2_\theta + \eta^2_\theta]\right]^2} , \\
    & {I_2}_{\eta_\theta \eta_\theta} = \frac{2\vr^2 + 2\vr_\phi^2}{\gamma^2[1+\gamma \cos \theta]^2} - \frac{4\eta_\theta[1+\gamma \cos \theta]^2[2\vr^2\eta_\theta + 2\vr_\phi^2\eta_\theta -2\vr_\theta \vr_\phi \eta_\phi] }{\left[[\vr_\theta \eta_\phi - \vr_\phi \eta_\theta]^2 + \vr^2[\vr^2_\theta + \eta^2_\theta]\right]^2} + \frac{2[1+\gamma \cos \theta]^2 }{[\vr_\theta \eta_\phi - \vr_\phi \eta_\theta]^2 + \vr^2[\vr^2_\theta + \eta^2_\theta]}\nonumber \\
    &+ \frac{2\left[[\vr^2_\theta + \eta^2_\theta][1+\gamma \cos \theta]^2 + [\vr^2_\phi + \eta^2_\phi + \vr^2]\gamma^2 \right] [2\vr_\phi^2 \eta_\theta - 2\eta_\phi \vr_\phi \vr_\theta + 2\vr^2\eta_\theta]^2}{\left[ [\vr_\theta \eta_\phi - \vr_\phi \eta_\theta]^2 + \vr^2[\vr^2_\theta + \eta^2_\theta] \right]^3}\nonumber\\
    &- \frac{\left[[\vr^2_\theta + \eta^2_\theta][1+\gamma \cos \theta]^2 + [\vr^2_\phi + \eta^2_\phi + \vr^2]\gamma^2 \right] [2\vr_\phi^2 + 2\vr^2]}{\left[ [\vr_\theta \eta_\phi - \vr_\phi \eta_\theta]^2 + \vr^2[\vr^2_\theta + \eta^2_\theta] \right]^2}, \\
    & \beta [ \CGright_{\eta_\theta \eta_\theta} \disR]\cdot\disR = \frac{2\Phi_0^2[\vr_\phi^2 \eta_\theta - \eta_\phi \vr_\phi \vr_\theta + \vr^2\eta_\theta]^2 }{\beta H^2 \left[[\vr_\theta \eta_\phi - \vr_\phi \eta_\theta]^2 + \vr^2[\vr^2_\theta + \eta^2_\theta]\right]\gamma^2[1+\gamma \cos \theta]^2} - \frac{\Phi_0^2[\vr_\phi^2  + \vr^2] }{2\beta H^2 \gamma^2[1+\gamma \cos \theta]^2}
\end{align}
The second derivative of $\Omega$ with respect to $\eta_\theta$ and $\eta_\phi$ is given by
\begin{align}
   & \Omega_{\eta_\theta \eta_\phi} = C_1[ {I_1}_{\eta_\theta \eta_\phi} + \alpha {I_2}_{\eta_\theta \eta_\phi} ]+ \beta [ \CGright_{\eta_\theta \eta_\phi} \disR]\cdot\disR ,
\end{align}
where
\begin{align}   
   & {I_1}_{\eta_\theta \eta_\phi} = \frac{4 \vr_\theta[\eta_\phi\vr_\theta -\vr_\phi\eta_\theta] \gamma^2[1+\gamma \cos \theta]^2[2\vr_\phi^2 \eta_\theta - 2\vr_\phi \eta_\phi \vr_\theta + 2\vr^2\eta_\theta] }{\left[[\vr_\theta \eta_\phi - \vr_\phi \eta_\theta]^2 + \vr^2[\vr^2_\theta + \eta^2_\theta]\right]^3} + \frac{ 2\vr_\phi\vr_\theta \gamma^2[1+\gamma \cos \theta]^2}{\left[[\vr_\theta \eta_\phi - \vr_\phi \eta_\theta]^2 + \vr^2[\vr^2_\theta + \eta^2_\theta]\right]^2}, \\
    &{I_2}_{\eta_\theta \eta_\phi} = -\frac{2\vr_\phi \vr_\theta}{\gamma^2[1+\gamma \cos \theta]^2} -\frac{4\vr_\theta\eta_\theta[\eta_\phi \vr_\theta - \vr_\phi \eta_\theta][1+\gamma \cos \theta]^2}{\left[ [\vr_\theta \eta_\phi - \vr_\phi \eta_\theta]^2 + \vr^2[\vr^2_\theta + \eta^2_\theta] \right]^2} - \frac{2\eta_\phi\left[2\vr^2\eta_\theta - 2\vr_\phi[\eta_\phi\vr_\theta - \vr_\phi\eta_\theta]\right]\gamma^2 }{\left[ [\vr_\theta \eta_\phi - \vr_\phi \eta_\theta]^2 + \vr^2[\vr^2_\theta + \eta^2_\theta] \right]^2} \nonumber \\
     &+ \frac{2\vr_\phi\vr_\theta\left[[\vr^2_\theta + \eta^2_\theta][1+\gamma \cos \theta]^2 + [\vr^2_\phi + \eta^2_\phi + \vr^2]\gamma^2 \right]}{\left[ [\vr_\theta \eta_\phi - \vr_\phi \eta_\theta]^2 + \vr^2[\vr^2_\theta + \eta^2_\theta] \right]^2} \nonumber\\
     &+ \frac{4\vr_\theta[\eta_\phi\vr_\theta - \vr_\phi\eta_\theta]\left[2\vr^2\eta_\theta - 2\vr_\phi[\eta_\phi\vr_\theta - \vr_\phi\eta_\theta]\right]\left[[\vr^2_\theta + \eta^2_\theta][1+\gamma \cos \theta]^2 + [\vr^2_\phi + \eta^2_\phi + \vr^2]\gamma^2 \right]}{\left[ [\vr_\theta \eta_\phi - \vr_\phi \eta_\theta]^2 + \vr^2[\vr^2_\theta + \eta^2_\theta] \right]^3},\\
    &\beta [ \CGright_{\eta_\theta \eta_\phi} \disR]\cdot\disR = \frac{\Phi_0^2 \vr_\theta [\eta_\phi\vr_\theta -\vr_\phi\eta_\theta][2\vr_\phi^2 \eta_\theta - 2\vr_\phi \eta_\phi \vr_\theta + 2\vr^2\eta_\theta] }{\beta H^2 \left[[\vr_\theta \eta_\phi - \vr_\phi \eta_\theta]^2 + \vr^2[\vr^2_\theta + \eta^2_\theta]\right]\gamma^2[1+\gamma \cos \theta]^2} -\frac{\Phi_0^2 \vr_\phi \vr_\theta }{2\beta H^2 \gamma^2[1+\gamma \cos \theta]^2} .
\end{align}
The second derivative of $\Omega$ with respect to $\vr_\phi$ is given by
\begin{align}
    & \Omega_{\vr_\phi \vr_\phi} = C_1[ {I_1}_{\vr_\phi \vr_\phi} + \alpha {I_2}_{\vr_\phi \vr_\phi}] + \beta [ \CGright_{\vr_\phi \vr_\phi} \disR]\cdot\disR ,
\end{align}
where
\begin{align}   
   &  {I_1}_{\vr_\phi \vr_\phi} = \frac{2}{[1+\gamma \cos \theta]^2} - \frac{2\eta_\theta^2\gamma^2[1+\gamma \cos \theta]^2}{\left[ [\vr_\theta \eta_\phi - \vr_\phi \eta_\theta]^2 + \vr^2[\vr^2_\theta + \eta^2_\theta] \right]^2} + \frac{8\eta_\theta^2[\vr_\theta \eta_\phi - \vr_\phi \eta_\theta]^2\gamma^2[1+\gamma \cos \theta]^2 }{\left[ [\vr_\theta \eta_\phi - \vr_\phi \eta_\theta]^2 + \vr^2[\vr^2_\theta + \eta^2_\theta] \right]^3} , \\
   & {I_2}_{\vr_\phi \vr_\phi} = \frac{2\gamma^2}{[\vr_\theta \eta_\phi - \vr_\phi \eta_\theta]^2 + \vr^2[\vr^2_\theta + \eta^2_\theta]} + \frac{8\vr_\phi\eta_\theta[\vr_\theta \eta_\phi - \vr_\phi \eta_\theta]\gamma^2}{\left[[\vr_\theta \eta_\phi - \vr_\phi \eta_\theta]^2 + \vr^2[\vr^2_\theta + \eta^2_\theta]\right]^2} + \frac{2\eta_\theta^2}{\gamma^2[1+\gamma\cos \theta]^2}\nonumber\\
    &- \frac{2\eta_\theta^2\left[ [\vr^2_\theta + \eta^2_\theta][1+\gamma \cos \theta]^2 + [\vr^2_\phi + \eta^2_\phi + \vr^2]\gamma^2 \right]}{\left[[\vr_\theta \eta_\phi - \vr_\phi \eta_\theta]^2 + \vr^2[\vr^2_\theta + \eta^2_\theta] \right]^2} \nonumber \\
    & + \frac{8\eta_\theta^2[\vr_\theta \eta_\phi - \vr_\phi \eta_\theta]^2\left[ [\vr^2_\theta + \eta^2_\theta][1+\gamma \cos \theta]^2 + [\vr^2_\phi + \eta^2_\phi + \vr^2]\gamma^2 \right]}{\left[[\vr_\theta \eta_\phi - \vr_\phi \eta_\theta]^2 + \vr^2[\vr^2_\theta + \eta^2_\theta] \right]^3} , \\
    &\beta [ \CGright_{\vr_\phi \vr_\phi} \disR]\cdot\disR = \frac{\Phi_0^2\eta_\theta^2}{2\beta H^2 \gamma^2[1+\gamma \cos \theta]^2} + \frac{2\Phi_0^2\eta_\theta^2[\vr_\theta \eta_\phi - \vr_\phi \eta_\theta]^2}{\beta H^2 \left[ [\vr_\theta \eta_\phi - \vr_\phi \eta_\theta]^2 + \vr^2[\vr^2_\theta + \eta^2_\theta] \right]\gamma^2[1+\gamma \cos \theta]^2 }.
\end{align}
The second derivative of $\Omega$ with respect to $\eta_\phi$ is given by
\begin{align}
    &\Omega_{\eta_\phi \eta_\phi} = C_1[ {I_1}_{\eta_\phi \eta_\phi} + \alpha {I_2}_{\eta_\phi \eta_\phi}] + \beta [ \CGright_{\eta_\phi \eta_\phi} \disR]\cdot\disR ,
\end{align}
where
\begin{align}   
    & {I_1}_{\eta_\phi \eta_\phi} = \frac{2}{[1+\gamma \cos \theta]^2} - \frac{2\vr_\theta^2\gamma^2[1+\gamma \cos \theta]^2}{\left[ [\vr_\theta \eta_\phi - \vr_\phi \eta_\theta]^2 + \vr^2[\vr^2_\theta + \eta^2_\theta] \right]^2} + \frac{8\vr_\theta^2[\vr_\theta \eta_\phi - \vr_\phi \eta_\theta]^2\gamma^2[1+\gamma \cos \theta]^2 }{\left[ [\vr_\theta \eta_\phi - \vr_\phi \eta_\theta]^2 + \vr^2[\vr^2_\theta + \eta^2_\theta] \right]^3}, \\
   & {I_2}_{\eta_\phi \eta_\phi} = \frac{2\gamma^2}{[\vr_\theta \eta_\phi - \vr_\phi \eta_\theta]^2 + \vr^2[\vr^2_\theta + \eta^2_\theta]} - \frac{8\eta_\phi\vr_\theta[\vr_\theta \eta_\phi - \vr_\phi \eta_\theta]\gamma^2}{\left[[\vr_\theta \eta_\phi - \vr_\phi \eta_\theta]^2 + \vr^2[\vr^2_\theta + \eta^2_\theta]\right]^2} + \frac{2\vr_\theta^2}{\gamma^2[1+\gamma\cos \theta]^2}\nonumber \\
    &- \frac{2\vr_\theta^2\left[ [\vr^2_\theta + \eta^2_\theta][1+\gamma \cos \theta]^2 + [\vr^2_\phi + \eta^2_\phi + \vr^2]\gamma^2 \right]}{\left[[\vr_\theta \eta_\phi - \vr_\phi \eta_\theta]^2 + \vr^2[\vr^2_\theta + \eta^2_\theta] \right]^2}\nonumber \\
    & + \frac{8\vr_\theta^2[\vr_\theta \eta_\phi - \vr_\phi \eta_\theta]^2\left[ [\vr^2_\theta + \eta^2_\theta][1+\gamma \cos \theta]^2 + [\vr^2_\phi + \eta^2_\phi + \vr^2]\gamma^2 \right]}{\left[[\vr_\theta \eta_\phi - \vr_\phi \eta_\theta]^2 + \vr^2[\vr^2_\theta + \eta^2_\theta] \right]^3} , \\
   & \beta [ \CGright_{\eta_\phi \eta_\phi} \disR]\cdot\disR = \frac{\Phi_0^2\vr_\theta^2}{2\beta H^2 \gamma^2[1+\gamma \cos \theta]^2} + \frac{2\Phi_0^2\vr_\theta^2[\vr_\theta \eta_\phi - \vr_\phi \eta_\theta]^2}{\beta H^2 \left[ [\vr_\theta \eta_\phi - \vr_\phi \eta_\theta]^2 + \vr^2[\vr^2_\theta + \eta^2_\theta] \right]\gamma^2[1+\gamma \cos \theta]^2 }.
\end{align}
The second derivative of $\Omega$ with respect to $\vr_\phi$ and $\eta_\phi$ is given by
\begin{align}
  &  \Omega_{\vr_\phi \eta_\phi} = C_1[ {I_1}_{\vr_\phi \eta_\phi} + \alpha {I_2}_{\vr_\phi \eta_\phi}] + \beta [ \CGright_{\vr_\phi \eta_\phi} \disR]\cdot\disR ,
\end{align}
where
\begin{align}   
   &  {I_1}_{\vr_\phi \eta_\phi} = \frac{2\vr_\theta \eta_\theta\gamma^2[1+\gamma\cos\theta]^2}{\left[ [\vr_\theta \eta_\phi - \vr_\phi \eta_\theta]^2 + \vr^2[\vr^2_\theta + \eta^2_\theta] \right]^2} -  \frac{8\vr_\theta \eta_\theta[\vr_\theta \eta_\phi - \vr_\phi \eta_\theta]^2 \gamma^2[1+\gamma\cos\theta]^2}{\left[ [\vr_\theta \eta_\phi - \vr_\phi \eta_\theta]^2 + \vr^2[\vr^2_\theta + \eta^2_\theta] \right]^3} , \\
    & {I_2}_{\vr_\phi \eta_\phi}  =-\frac{2\vr_\theta \eta_\theta}{\gamma^2[1+\gamma\cos\theta]^2} - \frac{4\vr_\phi\vr_\theta[\vr_\theta \eta_\phi - \vr_\phi \eta_\theta]\gamma^2}{\left[ [\vr_\theta \eta_\phi - \vr_\phi \eta_\theta]^2 + \vr^2[\vr^2_\theta + \eta^2_\theta] \right]^2} + \frac{4\eta_\phi\eta_\theta[\vr_\theta \eta_\phi - \vr_\phi \eta_\theta]\gamma^2}{\left[ [\vr_\theta \eta_\phi - \vr_\phi \eta_\theta]^2 + \vr^2[\vr^2_\theta + \eta^2_\theta] \right]^2} \nonumber \\
    & +\frac{8\vr_\theta\eta_\theta[\vr_\theta \eta_\phi - \vr_\phi \eta_\theta]^2\left[ [\vr^2_\theta + \eta^2_\theta][1+\gamma \cos \theta]^2 + [\vr^2_\phi + \eta^2_\phi + \vr^2]\gamma^2 \right]}{\left[[\vr_\theta \eta_\phi - \vr_\phi \eta_\theta]^2 + \vr^2[\vr^2_\theta + \eta^2_\theta] \right]^3} \nonumber \\
    &+ \frac{2\vr_\theta\eta_\theta\left[ [\vr^2_\theta + \eta^2_\theta][1+\gamma \cos \theta]^2 + [\vr^2_\phi + \eta^2_\phi + \vr^2]\gamma^2 \right]}{\left[[\vr_\theta \eta_\phi - \vr_\phi \eta_\theta]^2 + \vr^2[\vr^2_\theta + \eta^2_\theta] \right]^2} , \\
    & \beta [ \CGright_{\vr_\phi \eta_\phi} \disR]\cdot\disR = \frac{\Phi_0^2\vr_\theta \eta_\theta}{2\beta H^2\gamma^2[1+\gamma\cos\theta]^2} -  \frac{2\Phi_0^2\vr_\theta \eta_\theta[\vr_\theta \eta_\phi - \vr_\phi \eta_\theta]^2}{\beta H^2 \left[ [\vr_\theta \eta_\phi - \vr_\phi \eta_\theta]^2 + \vr^2[\vr^2_\theta + \eta^2_\theta] \right]\gamma^2[1+\gamma\cos\theta]^2} .
\end{align}
Finally, the full derivative of $\Omega_{\vr \vr_\phi}$ with respect to $\phi$ is expressed as
\begin{align}
    &\fd{\Omega_{\vr \vr_\phi}}{\phi} = C_1[ \fd{{I_1}_{\vr \vr_\phi}}{\phi} + \alpha \fd{{I_2}_{\vr \vr_\phi}}{\phi}] + \beta [ \fd{\CGright_{\vr \vr_\phi}}{\phi} \disR]\cdot\disR ,
\end{align}
where
\begin{align}
    &\fd{{I_1}_{\vr \vr_\phi}}{\phi} = -\frac{8\vr\eta_\theta[\eta_{\phi\phi}\vr_\theta - \vr_{\phi\phi} \eta_\theta][\vr_\theta^2 + \eta_\theta^2]\gamma^2[1+\gamma \cos\theta]^2}{\left[[\vr_\theta \eta_\phi - \vr_\phi \eta_\theta]^2 + \vr^2[\vr^2_\theta + \eta^2_\theta]\right]^3} 
    -\frac{8\vr_{\phi}\eta_\theta[\eta_\phi\vr_\theta - \vr_\phi \eta_\theta][\vr_\theta^2 + \eta_\theta^2]\gamma^2[1+\gamma \cos\theta]^2}{\left[[\vr_\theta \eta_\phi - \vr_\phi \eta_\theta]^2 + \vr^2[\vr^2_\theta + \eta^2_\theta]\right]^3}\nonumber \\
    &+ \frac{24\vr\eta_\theta[\eta_\phi\vr_\theta - \vr_\phi \eta_\theta][\vr_\theta^2 + \eta_\theta^2]\gamma^2[1+\gamma \cos\theta]^2}{\left[[\vr_\theta \eta_\phi - \vr_\phi \eta_\theta]^2 + \vr^2[\vr^2_\theta + \eta^2_\theta]\right]^4}\left[ 2[\eta_\phi\vr_\theta - \vr_\phi \eta_\theta][\eta_{\phi\phi}\vr_\theta - \vr_{\phi\phi}\eta_\theta] + 2\vr_\phi \vr[\vr^2_\theta + \eta^2_\theta] \right], \nonumber \\
    &\fd{{I_2}_{\vr \vr_\phi}}{\phi} = \frac{4 \vr_{\phi} \gamma^2 \eta_\theta[\vr_\theta \eta_\phi - \vr_\phi \eta_\theta]}{\left[[\vr_\theta \eta_\phi - \vr_\phi \eta_\theta]^2 + \vr^2[\vr^2_\theta + \eta^2_\theta] \right]^2} + \frac{4 \vr \gamma^2 \eta_\theta[\vr_\theta \eta_{\phi\phi} - \vr_{\phi\phi} \eta_\theta]}{\left[[\vr_\theta \eta_{\phi} - \vr_{\phi} \eta_\theta]^2 + \vr^2[\vr^2_\theta + \eta^2_\theta] \right]^2} \nonumber \\
    &- \frac{8 \vr \gamma^2 \eta_\theta[\vr_\theta \eta_\phi - \vr_\phi \eta_\theta]}{\left[[\vr_\theta \eta_\phi - \vr_\phi \eta_\theta]^2 + \vr^2[\vr^2_\theta + \eta^2_\theta] \right]^3} \left[ 2[\vr_\theta \eta_\phi - \vr_\phi \eta_\theta][\vr_\theta \eta_{\phi\phi} - \vr_{\phi\phi} \eta_\theta] + 2\vr\vr_\phi[\vr^2_\theta + \eta^2_\theta] \right] \nonumber \\
    &- \frac{4\vr_{\phi\phi} \vr [\vr_\theta^2+ \eta_\theta^2]\gamma^2}{\left[[\vr_\theta \eta_\phi - \vr_\phi \eta_\theta]^2 + \vr^2[\vr^2_\theta + \eta^2_\theta] \right]^2}- \frac{4\vr_\phi \vr_\phi [\vr_\theta^2+ \eta_\theta^2]\gamma^2}{\left[[\vr_\theta \eta_\phi - \vr_\phi \eta_\theta]^2 + \vr^2[\vr^2_\theta + \eta^2_\theta] \right]^2} \nonumber \\
    &+ \frac{8\vr_\phi \vr [\vr_\theta^2+ \eta_\theta^2]\gamma^2}{\left[[\vr_\theta \eta_\phi - \vr_\phi \eta_\theta]^2 + \vr^2[\vr^2_\theta + \eta^2_\theta] \right]^3}\left[ 2[\vr_\theta \eta_\phi - \vr_\phi \eta_\theta][\vr_\theta \eta_{\phi\phi} - \vr_{\phi\phi} \eta_\theta] + 2\vr\vr_\phi[\vr^2_\theta + \eta^2_\theta] \right] \nonumber \\
     &- \frac{8\vr_\phi \eta_\theta[\vr_\theta \eta_\phi - \vr_\phi \eta_\theta][\vr_\theta^2 + \eta_\theta^2]\left[[\vr^2_\theta + \eta^2_\theta][1+\gamma \cos \theta]^2 + [\vr^2_\phi + \eta^2_\phi + \vr^2]\gamma^2\right]}{\left[[\vr_\theta \eta_\phi - \vr_\phi \eta_\theta]^2 + \vr^2[\vr^2_\theta + \eta^2_\theta] \right]^3} \nonumber \\
     &- \frac{8\vr\eta_\theta[\vr_\theta \eta_{\phi\phi} - \vr_{\phi\phi} \eta_\theta][\vr_\theta^2 + \eta_\theta^2]\left[[\vr^2_\theta + \eta^2_\theta][1+\gamma \cos \theta]^2 + [\vr^2_\phi + \eta^2_\phi + \vr^2]\gamma^2\right]}{\left[[\vr_\theta \eta_\phi - \vr_\phi \eta_\theta]^2 + \vr^2[\vr^2_\theta + \eta^2_\theta] \right]^3} \nonumber \\
      &- \frac{8\vr\eta_\theta[\vr_\theta \eta_\phi - \vr_\phi \eta_\theta][\vr_\theta^2 + \eta_\theta^2]\left [2\vr_\phi\vr_{\phi\phi} + 2\eta_\phi \eta_{\phi\phi} + 2\vr\vr_\phi]\gamma^2\right]}{\left[[\vr_\theta \eta_\phi - \vr_\phi \eta_\theta]^2 + \vr^2[\vr^2_\theta + \eta^2_\theta] \right]^3}\nonumber \\
      &+ \frac{24\vr\eta_\theta[\vr_\theta \eta_\phi - \vr_\phi \eta_\theta][\vr_\theta^2 + \eta_\theta^2]\left[[\vr^2_\theta + \eta^2_\theta][1+\gamma \cos \theta]^2 + [\vr^2_\phi + \eta^2_\phi + \vr^2]\gamma^2\right]}{\left[[\vr_\theta \eta_\phi - \vr_\phi \eta_\theta]^2 + \vr^2[\vr^2_\theta + \eta^2_\theta] \right]^4} \nonumber\\
      &\cdot\left[ 2[\vr_\theta \eta_\phi - \vr_\phi \eta_\theta][\vr_\theta \eta_{\phi\phi} - \vr_{\phi\phi} \eta_\theta] + 2\vr\vr_\phi[\vr^2_\theta + \eta^2_\theta] \right] \nonumber \\
    &+ \frac{24\vr\eta_\theta[\eta_\phi\vr_\theta - \vr_\phi \eta_\theta][\vr_\theta^2 + \eta_\theta^2]\gamma^2[1+\gamma \cos\theta]^2}{\left[[\vr_\theta \eta_\phi - \vr_\phi \eta_\theta]^2 + \vr^2[\vr^2_\theta + \eta^2_\theta]\right]^4}\left[ 2[\eta_\phi\vr_\theta - \vr_\phi \eta_\theta][\eta_{\phi\phi}\vr_\theta - \vr_{\phi\phi}\eta_\theta] + 2\vr_\phi \vr[\vr^2_\theta + \eta^2_\theta] \right], \nonumber \\
    &\beta [ \fd{\CGright_{\vr \vr_\phi}}{\phi} \disR]\cdot\disR =  -\frac{2\Phi_0^2\vr\eta_\theta[\eta_{\phi\phi}\vr_\theta - \vr_{\phi\phi} \eta_\theta][\vr_\theta^2 + \eta_\theta^2]}{\beta H^2 \left[[\vr_\theta \eta_\phi - \vr_\phi \eta_\theta]^2 + \vr^2[\vr^2_\theta + \eta^2_\theta]\right]\gamma^2[1+\gamma \cos\theta]^2} \nonumber\\
    &-\frac{2\Phi_0^2\vr_{\phi}\eta_\theta[\eta_\phi\vr_\theta - \vr_\phi \eta_\theta][\vr_\theta^2 + \eta_\theta^2]}{\beta H^2 \left[[\vr_\theta \eta_\phi - \vr_\phi \eta_\theta]^2 + \vr^2[\vr^2_\theta + \eta^2_\theta]\gamma^2[1+\gamma \cos\theta]^2\right]} \nonumber \\
    &+ \frac{6\Phi_0^2\vr\eta_\theta[\eta_\phi\vr_\theta - \vr_\phi \eta_\theta][\vr_\theta^2 + \eta_\theta^2]}{\beta H^2 \left[[\vr_\theta \eta_\phi - \vr_\phi \eta_\theta]^2 + \vr^2[\vr^2_\theta + \eta^2_\theta]\right]^2\gamma^2[1+\gamma \cos\theta]^2}\left[ 2[\eta_\phi\vr_\theta - \vr_\phi \eta_\theta][\eta_{\phi\phi}\vr_\theta - \vr_{\phi\phi}\eta_\theta] + 2\vr_\phi \vr[\vr^2_\theta + \eta^2_\theta] \right]. 
\end{align}
\section{Reformulation of the coupled ODEs}
\label{appendix: coefficients}
The derivatives of the stretches in equations (9) with respect to $\theta$ are expressed as
 \begin{align}
  {\lambda_1}_\theta = \frac{\vr_\theta \vr_{\theta\theta} + \eta_\theta \eta_{\theta\theta}}{\gamma \left[ \vr_\theta^2 + \eta_\theta^2 \right]^{1/2}} = \frac{\vr_\theta \vr_{\theta\theta}  + \eta_\theta \eta_{\theta\theta} }{\gamma^2 \lambda_1} , \quad {\lambda_2}_\theta =  \frac{\vr_\theta}{1 + \gamma \cos \theta} + \frac{\vr \gamma \sin \theta}{[1 + \gamma \cos \theta]^2}.
 \end{align}
 
 The equations (31a) and (31b) can be rewritten as
\begin{align}
 & - \gamma \sin \theta \bigg[  \frac{2 \varrho_\theta}{\gamma^2}  \left[ 1 + \alpha \lambda_2^2 \right] \left[ 1 - \frac{1}{\lambda_1^4 \lambda_2^2} \right] -  \frac{\vr_\theta \mcal{E}\lambda_2^2}{2 \gamma^2} \bigg] \nonumber \\
&+   [1 + \gamma \cos \theta] \Bigg[  \frac{2 \varrho_{\theta\theta}}{\gamma^2}  \left[ 1 + \alpha \lambda_2^2 \right] \left[ 1 - \frac{1}{\lambda_1^4 \lambda_2^2} \right] + \frac{2 \varrho_\theta}{\gamma^2}  \left[ 2 \alpha \lambda_2 {\lambda_2}_\theta \right] \left[ 1 - \frac{1}{\lambda_1^4 \lambda_2^2} \right] \nonumber \\
&+ \frac{2 \varrho_\theta}{\gamma^2}  \left[ 1 + \alpha \lambda_2^2 \right] \left[   \frac{2{\lambda_2}_\theta}{\lambda_2^3 \lambda_1^4} + \frac{4{\lambda_1}_\theta}{\lambda_2^2 \lambda_1^5} \right]  - \frac{ \mcal{E}\lambda_2^2}{2 \gamma^2}\rho_{\theta \theta} - \frac{2\rho_\theta \mcal{E}}{ \gamma^2}\lambda_2 {\lambda_2}_\theta \Bigg] \nonumber \\
 &-   2\lambda_2  \left[ 1+\alpha \lambda_1^2 \right] \left[ 1 - \frac{1}{\lambda_1^2 \lambda_2^4} \right]  + \frac{\mcal{E}\lambda_2 \lambda_1^2}{2  }   + \frac{P \vr \eta_\theta}{\gamma} = 0,
\end{align}
 \begin{align}
 &   - \gamma \sin \theta  \Bigg[ \frac{2\eta_\theta}{\gamma^2} \left[ 1 + \alpha \lambda_2^2 \right] \left[ 1 - \frac{1}{\lambda_1^4 \lambda_2^2} \right] - \frac{\mcal{E} \eta_\theta \lambda_2^2 }{2  \gamma^2} \Bigg] \nonumber \\
  &  +  [1 + \gamma \cos \theta] \Bigg[  \frac{2 \eta_{\theta \theta}}{\gamma^2}  \left[ 1 + \alpha \lambda_2^2 \right] \left[ 1 - \frac{1}{\lambda_1^4 \lambda_2^2} \right] + \frac{2 \eta_\theta}{\gamma^2}  \left[ 2 \alpha \lambda_2 {\lambda_2}_\theta \right] \left[ 1 - \frac{1}{\lambda_1^4 \lambda_2^2} \right] \nonumber \\
  &  + \frac{2 \eta_\theta}{\gamma^2}  \left[ 1 + \alpha \lambda_2^2 \right] \left[   \frac{2{\lambda_2}_\theta}{\lambda_2^3 \lambda_1^4} + \frac{4{\lambda_1}_\theta}{\lambda_2^2 \lambda_1^5} \right]  - \frac{ \mcal{E}\lambda_2^2}{2 \gamma^2}\eta_{\theta \theta} - \frac{2\eta_\theta \mcal{E}}{ \gamma^2}\lambda_2 {\lambda_2}_\theta \Bigg] - \frac{P \varrho {\varrho}_\theta}{\gamma} = 0 .
 \end{align}
From the first equation, the coefficient of $\vr_{\theta \theta}$ is
 \begin{align}
 A_1 = [1 + \gamma \cos \theta]  \Bigg[ \frac{2 }{\gamma^2}  \left[ 1 + \alpha \lambda_2^2 \right] \left[ 1 - \frac{1}{\lambda_1^4 \lambda_2^2} \right] + \frac{8 \varrho_\theta^2}{\gamma^4 \lambda_1^6 \lambda_2^2}  \left[ 1 + \alpha \lambda_2^2 \right]  - \frac{ \mcal{E}\lambda_2^2}{2 \gamma^2} \Bigg].
 \end{align}
The coefficient of $\eta_{\theta \theta}$ is given by
  \begin{align}
 A_2 = [1 + \gamma \cos \theta]  \frac{8 \varrho_\theta \eta_\theta}{\gamma^4 \lambda_1^6 \lambda_2^2}  \left[ 1 + \alpha \lambda_2^2 \right],
  \end{align}
and the remaining term is
\begin{align}
  A_3 = &  - \gamma \sin \theta  \Bigg[ \frac{2\vr_\theta}{\gamma^2} \left[ 1 + \alpha \lambda_2^2 \right] \left[ 1 - \frac{1}{\lambda_1^4 \lambda_2^2} \right] - \frac{\mcal{E} \vr_\theta \lambda_2^2 }{2  \gamma^2} \Bigg] \nonumber \\  
   &+ [1 + \gamma \cos \theta] \Bigg[ \frac{2 \varrho_\theta}{\gamma^2}  \left[ 2 \alpha \lambda_2 {\lambda_2}_\theta \right] \left[ 1 - \frac{1}{\lambda_1^4 \lambda_2^2} \right] + \frac{4 \vr_\theta {\lambda_2}_\theta}{\gamma^2 \lambda_2^3 \lambda_1^4} [1 + \alpha \lambda_2^2]- \frac{2\rho_\theta \mcal{E}}{ \gamma^2}\lambda_2 {\lambda_2}_\theta \Bigg] \nonumber \\
   &-   2\lambda_2  \left[ 1+\alpha \lambda_1^2 \right] \left[ 1 - \frac{1}{\lambda_1^2 \lambda_2^4} \right]  + \frac{\mcal{E}\lambda_2 \lambda_1^2}{2  }   + \frac{P \vr \eta_\theta}{\gamma}
  \end{align}
From the second equation, the coefficient of $\vr''$ is
{
\begin{align}
B_1  = [1 + \gamma \cos \theta] \frac{8 \varrho_\theta \eta_\theta}{\gamma^4 \lambda_1^6 \lambda_2^2}  \left[ 1 + \alpha \lambda_2^2 \right]
\end{align}
}
and the coefficient of $\eta''$ is
 \begin{align}
B_2 =  [1 + \gamma \cos \theta]  \Bigg[ \frac{2 }{\gamma^2}  \left[ 1 + \alpha \lambda_2^2 \right] \left[ 1 - \frac{1}{\lambda_1^2 \lambda_2^4} \right] + \frac{8 \eta_\theta^2}{\gamma^4 \lambda_1^6 \lambda_2^2}  \left[ 1 + \alpha \lambda_2^2 \right]  - \frac{ \mcal{E}\lambda_2^2}{2\gamma^2} \Bigg]
\end{align}
  and the remaining term is
\begin{align}
 B_3 = &- \gamma \sin \theta  \Bigg[ \frac{2\eta_\theta}{\gamma^2} \left[ 1 + \alpha \lambda_2^2 \right] \left[ 1 - \frac{1}{\lambda_1^4 \lambda_2^2} \right] - \frac{\mcal{E} \lambda_2^2\eta_\theta }{ 2 \gamma^2} \Bigg] \nonumber \\
 &+ [1 + \gamma \cos \theta] \Bigg[ \frac{2 \eta'}{\gamma^2}  \left[ 2 \alpha \lambda_2 {\lambda_2}_\theta \right] \left[ 1 - \frac{1}{\lambda_1^4 \lambda_2^2} \right] + [1+\alpha\lambda_2^2]\frac{4 \eta_\theta {\lambda_2}_\theta}{\gamma^2 \lambda_1^4 \lambda_2^3}
 - \frac{2\eta_\theta \mcal{E}}{ \gamma^2}\lambda_2 {\lambda_2}_\theta \Bigg] - \frac{P \varrho \varrho_\theta}{\gamma} .
\end{align}

Therefore the above set of coupled ODEs can be rewritten as
\begin{align}
 A_1 \vr_{\theta\theta} + A_2 \eta_{\theta\theta}  + A_3 = 0, \\
 B_1 \vr_{\theta\theta}  + B_2 \eta_{\theta\theta}  + B_3 = 0.
\end{align}
\section{Reformulation of the ODEs arising from the relaxed energy}
\label{appendix:ER_odes}
The governing equations~(31) are now modified as in equations~(50) where the modified energy density function is expressed as:
\begin{align}
\Omega^\ast(\lambda_1,\mcal{E}) = C_1[I^\ast_1 -3]+C_2[I^\ast_2-3] + \beta [ \CGright \disR] \cdot\disR.
\end{align}
As $\Omega^\ast$ is not a function of $\lambda_2$, $\pd{\Omega^\ast}{\vr}$ is vanished. We first use the chain rule to compute the derivatives of $\Omega^\ast$ with respect to $\vr_\theta$ and $\eta_\theta$ as
\begin{align}
&\frac{\partial \Omega^\ast}{\partial \vr_\theta} = \frac{\partial \Omega^\ast}{ \partial\lambda_1}\frac{\partial \lambda_1}{\partial \vr_\theta},  &\frac{\partial \Omega^\ast}{\partial \eta_\theta} = \frac{\partial \Omega^\ast}{ \partial\lambda_1}\frac{\partial \lambda_1}{\partial \eta_\theta} .
\end{align}
The derivatives of $\lambda_1$ with respect to $\vr_\theta$ and $\eta_\theta$ are easy to be computed as
\begin{align}
&\frac{\partial \lambda_1}{\partial \vr_\theta} = \frac{\vr_\theta}{\gamma}[\vr_\theta^2 + \eta_\theta^2]^{-\frac{1}{2}}, & \frac{\partial \lambda_1}{\partial \eta_\theta} = \frac{\eta_\theta}{\gamma}[\vr_\theta^2 + \eta_\theta^2]^{-\frac{1}{2}}.
\end{align}
The derivative of $\Omega^\ast$ with respect to $\lambda_1$ can be decomposed as follows
\begin{align}
 \frac{\partial \Omega^\ast}{ \partial\lambda_1} = C_1 \frac{\partial I^\ast_1}{\partial \lambda_1} + C_2\frac{\partial I^\ast_2}{\partial \lambda_1} + \frac{\partial \left( \beta [ \CGright \disR] \cdot\disR \right)}{\partial \lambda_1} ,
\end{align}
where
\begin{align}
\frac{\partial I^\ast_1}{\partial \lambda_1} = 2\lambda_1 + \frac{\partial ({\lambda^\ast_2}^2)}{\partial \lambda_1} - \frac{2}{\lambda_1^3 {\lambda^\ast_2}^2} - \frac{1}{\lambda_1^2 {\lambda^\ast_2}^4 }\frac{\partial ({\lambda^\ast_2}^2)}{\partial \lambda_1},
\end{align}
\begin{align}
\frac{\partial I^\ast_2}{\partial \lambda_1} = -\frac{2}{\lambda_1^3} - \frac{\frac{\partial ({\lambda^\ast_2}^2)}{\partial \lambda_1}}{{\lambda^\ast_2}^4} +\lambda_1^2 \frac{\partial ({\lambda^\ast_2}^2)}{\partial \lambda_1} +2\lambda_1 {\lambda^\ast_2}^2
\end{align}
 and
{
\begin{align}
\frac{\partial \left( \beta [ \CGright \disR] \cdot\disR \right)}{\partial \lambda_1} =  \frac{C_1\mcal{E}}{4}  \left[ 2\lambda_1 {\lambda^\ast_2}^2 + \lambda_1^2  \frac{\partial ({\lambda^\ast_2}^2)}{\partial \lambda_1}  \right],
\end{align}
}
where
\begin{align}
\frac{\partial ({\lambda^\ast_2}^2)}{\partial \lambda_1} =& -\frac{PH[8\alpha\lambda_1 - 2\mcal{E}\lambda_1]}{R_b[4\lambda_1 + 4\alpha\lambda_1^2 - \mcal{E} \lambda_1^2]^2} \nonumber \\
&+ \frac{\frac{P^2 H^2}{R_b^2}\lambda_1 + 32\alpha^2\lambda_1^3 - 8\mcal{E}\alpha\lambda_1^3 + 32\alpha \lambda_1 - 4 \mcal{E} \lambda_1} { \left[ \sqrt{\frac{P^2 H^2}{R_b^2} {\lambda_1^2}- 4\alpha\mcal{E}\lambda_1^4 + 16\alpha^2\lambda_1^4 + 32\alpha {\lambda_1^2}- 4\mcal{E}{\lambda_1^2}+{16}}\right][4\lambda_1 + 4\alpha\lambda_1^3 - \mcal{E} \lambda_1^3]} \nonumber \\
&- \frac{\left[\frac{P H}{R_b}\lambda_1 + \sqrt{\frac{P^2 H^2}{R_b^2} {\lambda_1^2}- 4\alpha\mcal{E}\lambda_1^4 + 16\alpha^2\lambda_1^4 + 32\alpha {\lambda_1^2}- 4\mcal{E}{\lambda_1^2}+{16}} \right][12\alpha\lambda_1^2 - 3\mcal{E}\lambda_1^2+4]}{[4\lambda_1 + 4\alpha\lambda_1^3 - \mcal{E} \lambda_1^3]^2}.
\end{align}
Then, the total derivatives of the two terms with respect to $\theta$ are calculated as
\begin{align}
\fd{}{\theta} \left( \pd{\Omega^\ast}{\vr_\theta} \right) 
=& \fd{}{\theta}\left(\frac{\vr_\theta}{\gamma}[\vr_\theta^2 + \eta_\theta^2]^{-\frac{1}{2}}\right)  \frac{\partial \Omega^\ast}{ \partial\lambda_1}+ \fd{}{\theta} \left( \frac{\partial \Omega^\ast}{ \partial\lambda_1}\right)\underbrace{\left[\frac{\vr_\theta}{\gamma}[\vr_\theta^2 + \eta_\theta^2]^{-\frac{1}{2}}\right]}_{\mcal{W}_1},\\
\fd{}{\theta} \left( \pd{\Omega^\ast}{\eta_\theta} \right) 
=& \fd{}{\theta}\left(\frac{\eta_\theta}{\gamma}[\vr_\theta^2 + \eta_\theta^2]^{-\frac{1}{2}}\right)  \frac{\partial \Omega^\ast}{ \partial\lambda_1}+ \fd{}{\theta} \left( \frac{\partial \Omega^\ast}{ \partial\lambda_1}\right)\underbrace{\left[\frac{\eta_\theta}{\gamma}[\vr_\theta^2 + \eta_\theta^2]^{-\frac{1}{2}}\right]}_{\mcal{W}_2}.
\end{align}
Upon explicitly computing the first full derivative terms in both equations and separating the coefficients of $\vr_{\theta \theta}$ and $\eta_{\theta \theta}$ yields
\begin{align}
\fd{}{\theta}\left(\frac{\vr_\theta}{\gamma}[\vr_\theta^2 + \eta_\theta^2]^{-\frac{1}{2}}\right) &= \frac{\vr_{\theta\theta}}{\gamma} [\vr_\theta^2 +\eta_\theta^2]^{-\frac{1}{2}} - \frac{\vr_\theta}{\gamma} [\vr_\theta\vr_{\theta\theta} + \eta_\theta\eta_{\theta\theta}] [\vr_\theta^2+\eta_\theta^2]^{-\frac{3}{2}} 
\nonumber \\
& = \underbrace{\left[ \frac{1}{\gamma [\vr_\theta^2 + \eta_\theta^2]^{\frac{1}{2}}} - \frac{\vr_\theta^2}{\gamma [\vr_\theta^2 + \eta_\theta^2]^{\frac{3}{2}}}  \right] }_{\mcal{U}_1}\vr_{\theta \theta} \underbrace{- \frac{\vr_\theta\eta_\theta}{\gamma [\vr_\theta^2 + \eta_\theta^2]^{\frac{3}{2}} }}_{\mcal{U}_2}\eta_{\theta \theta},
\end{align}
and
\begin{align}
\fd{}{\theta}\left(\frac{\eta_\theta}{\gamma}[\vr_\theta^2 + \eta_\theta^2]^{-\frac{1}{2}}\right) &= \frac{\eta_{\theta \theta}}{\gamma} [\vr_\theta^2 +\eta_\theta^2]^{-\frac{1}{2}} - \frac{\eta_\theta}{\gamma} [\vr_\theta\vr_{\theta \theta} + \eta_\theta\eta_{\theta}] [\vr_\theta^2+\eta_\theta^2]^{-\frac{3}{2}}. \nonumber \\
& = \underbrace{\left[ \frac{1}{\gamma [\vr_\theta^2 + \eta_\theta^2]^{\frac{1}{2}}} - \frac{\eta_\theta^2}{\gamma [\vr_\theta^2 + \eta_\theta^2]^{\frac{3}{2}}}  \right]}_{\mcal{V}_2} \eta_{\theta \theta} \underbrace{- \frac{\vr_\theta\eta_\theta}{\gamma [\vr_\theta^2 + \eta_\theta^2]^{\frac{3}{2}} }}_{\mcal{V}_1}\vr_{\theta \theta}.
\end{align}
The remaining full derivative term $\fd{}{\theta} \left( \frac{\partial \Omega^\ast}{ \partial\lambda_1}\right) $ is expressed as
\begin{align}
\fd{}{\theta} \left( \frac{\partial \Omega^\ast}{ \partial\lambda_1}\right) = C_1\fd{}{\theta}\left(\pd{{I_1}^\ast}{\lambda_1}\right) + C_2\fd{}{\theta}\left(\pd{{I_2}^\ast}{\lambda_1}\right) +\fd{}{\theta}\left(\frac{C_1\mcal{E}}{4}  \left[ 2\lambda_1 {\lambda^\ast_2}^2 + \lambda_1^2  \frac{\partial ({\lambda^\ast_2}^2)}{\partial \lambda_1}  \right]\right),
\end{align}
where
\begin{align}
&C_1\fd{}{\theta}\left(\pd{{I_1}^\ast}{\lambda_1}\right) =C_1 \fd{}{\theta} \left(\left[2\lambda_1 + \frac{\partial ({\lambda^\ast_2}^2)}{\partial \lambda_1} - \frac{2}{\lambda_1^3 {\lambda^\ast_2}^2} - \frac{1}{\lambda_1^2 {\lambda^\ast_2}^4 }\frac{\partial ({\lambda^\ast_2}^2)}{\partial \lambda_1}\right]  \right) \nonumber \\
=& 2C_1\lambda_{1\theta} +C_1 \left( \frac{\partial ({\lambda^\ast_2}^2)}{\partial \lambda_1}\right)_\theta +C_1 \left[ \frac{6\lambda_{1\theta}}{\lambda_1^4 {\lambda_2^\ast}^2} + \frac{2({\lambda_2^\ast}^2)_\theta}{\lambda_1^3 {\lambda_2^\ast}^4} \right] \nonumber \\
& - C_1\left[ \frac{1}{\lambda_1^2 {\lambda^\ast_2}^4}\left(\frac{\partial ({\lambda^\ast_2}^2)}{\partial \lambda_1}\right)_\theta - \frac{2 \lambda_{1\theta}}{\lambda_1^3 {\lambda^\ast_2}^4} \frac{\partial ({\lambda^\ast_2}^2)}{\partial \lambda_1} - \frac{2 ( {\lambda^\ast_2}^2)_\theta}{\lambda_1^2 {\lambda^\ast_2}^6} \frac{\partial ({\lambda^\ast_2}^2)}{\partial \lambda_1} \right] \nonumber \\
=&C_1 \Bigg[ \left[   2 + \frac{6}{\lambda_1^4 {\lambda_1^\ast}^2} + \frac{2}{\lambda_1^3 {\lambda_1^\ast}^4} \frac{\partial ({\lambda^\ast_2}^2)}{\partial \lambda_1}  \right] + \left[ \frac{2}{\lambda_1^3 {\lambda_1^\ast}^4} + \frac{2}{\lambda_1^2 {\lambda_1^\ast}^6} \frac{\partial ({\lambda^\ast_2}^2)}{\partial \lambda_1} \right]\frac{\partial ({\lambda^\ast_2}^2)}{\partial \lambda_1} \nonumber \\
 &+ \left[ 1 - \frac{1}{\lambda_1^2 {\lambda^\ast_2}^4} \right] \frac{\partial^2 ({\lambda^\ast_2}^2)}{\partial \lambda_1^2} \Bigg] \lambda_{1\theta},
\end{align}

\begin{align}
&C_2\fd{}{\theta}\left(\pd{{I_2}^\ast}{\lambda_1}\right) =\fd{}{\theta}\left( C_2\left[ -\frac{2}{\lambda_1^3} - \frac{1}{{\lambda^\ast_2}^4} \frac{\partial ({\lambda^\ast_2}^2)}{\partial \lambda_1}+\lambda_1^2 \frac{\partial ({\lambda^\ast_2}^2)}{\partial \lambda_1} +2\lambda_1 {\lambda^\ast_2}^2 \right]\right)\nonumber \\
 =& C_2 \frac{6}{\lambda_1^4} \lambda_{1\theta} - C_2\left[ \frac{1}{{\lambda^\ast_2}^4}\left(\frac{\partial ({\lambda^\ast_2}^2)}{\partial \lambda_1}\right)_\theta - \frac{2({\lambda^\ast_2}^2)_\theta}{{\lambda^\ast_2}^6} \frac{\partial ({\lambda^\ast_2}^2)}{\partial \lambda_1} \right] \nonumber \\
 &+ C_2\left[ \lambda_1^2  \left(\frac{\partial ({\lambda^\ast_2}^2)}{\partial \lambda_1}\right)_\theta + 2\lambda_1\lambda_{1\theta} \frac{\partial ({\lambda^\ast_2}^2)}{\partial \lambda_1}  \right] + C_2\left[
 2 \lambda_1 ({\lambda^\ast_2}^2)_\theta + 2 {\lambda^\ast_2}^2 \lambda_{1\theta}
 \right] \nonumber \\
=& C_2 \Bigg[ \left[ \frac{6}{\lambda_1^4} + 2\lambda_1\frac{\partial ({\lambda^\ast_2}^2)}{\partial \lambda_1}  + 2{\lambda^\ast_2}^2 \right]  +  \left[ \frac{2}{{\lambda^\ast_2}^6}  \frac{\partial ({\lambda^\ast_2}^2)}{\partial \lambda_1} + 2\lambda_1 \right] \frac{\partial ({\lambda^\ast_2}^2)}{\partial \lambda_1}  \nonumber \\
&+ \left[\lambda_1^2 -\frac{1}{{\lambda_2^\ast}^4} \right] \frac{\partial^2 ({\lambda^\ast_2}^2)}{\partial \lambda_1^2} \Bigg] \lambda_{1\theta} ,
\end{align}
 and
\begin{align}
&\fd{}{\theta}\left(\frac{C_1\mcal{E}}{4}  \left[ 2\lambda_1 {\lambda^\ast_2}^2 + \lambda_1^2  \frac{\partial ({\lambda^\ast_2}^2)}{\partial \lambda_1}  \right]\right) \nonumber \\
= & \frac{C_1 \mcal{E}}{4} \left[2{\lambda^\ast_2}^2 \lambda_{1\theta} + 2\lambda_1  { ({\lambda^\ast_2}^2)}_\theta  + 2 \lambda_1  \frac{\partial ({\lambda^\ast_2}^2)}{\partial \lambda_1} \lambda_{1\theta} +  \lambda_1^2 \left(\frac{\partial ({\lambda^\ast_2}^2)}{\partial \lambda_1}\right)_\theta \right] \nonumber \\
=& \frac{C_1 \mcal{E}}{4} \left[2{\lambda^\ast_2}^2  + 4\lambda_1  \frac{ \partial({\lambda^\ast_2}^2)}{\partial \lambda_1}   +  \lambda_1^2 \frac{\partial^2 ({\lambda^\ast_2}^2)}{\partial \lambda_1^2} \right]\lambda_{1\theta}.
\end{align}  
The term 
\begin{align}
\frac{\partial ({\lambda^\ast_2}^2)}{\partial \lambda_1} =& -\frac{PH[8\alpha\lambda_1 - 2\mcal{E}\lambda_1]}{R_b[4\lambda_1 + 4\alpha\lambda_1^3 - \mcal{E} \lambda_1^3]^2} \nonumber \\
&+ \underbrace{ \frac{\frac{P^2 H^2}{R_b^2}\lambda_1 + 32\alpha^2\lambda_1^3 - 8\mcal{E}\alpha\lambda_1^3 + 32\alpha \lambda_1 - 4 \mcal{E} \lambda_1} { \left[ \sqrt{\frac{P^2 H^2}{R_b^2} {\lambda_1^2}- 4\alpha\mcal{E}\lambda_1^4 + 16\alpha^2\lambda_1^4 + 32\alpha {\lambda_1^2}- 4\mcal{E}{\lambda_1^2}+{16}}\right][4\lambda_1 + 4\alpha\lambda_1^3 - \mcal{E} \lambda_1^3]} }_{\mcal{C}} \nonumber \\
&- \underbrace{ \frac{\left[\frac{P H}{R_b}\lambda_1 + \sqrt{\frac{P^2 H^2}{R_b^2} {\lambda_1^2}- 4\alpha\mcal{E}\lambda_1^4 + 16\alpha^2\lambda_1^4 + 32\alpha {\lambda_1^2}- 4\mcal{E}{\lambda_1^2}+{16}} \right][12\alpha\lambda_1^2 - 3\mcal{E}\lambda_1^2+4]}{[4\lambda_1 + 4\alpha\lambda_1^3 - \mcal{E} \lambda_1^3]^2} }_{\mcal{D}},
\end{align} 
and hence
\begin{align}
\frac{ \partial^2({\lambda^\ast_2}^2)}{\partial \lambda_1^2} =  \pd{}{\lambda_1}\left(- \frac{PH[8\alpha\lambda_1 - 2\mcal{E}\lambda_1]}{R_b[4\lambda_1 + 4\alpha\lambda_1^3 - \mcal{E} \lambda_1^3]^2} \right) + \pd{\mcal A}{\lambda_1} + \pd{\mcal B}{\lambda_1},
\end{align}
where
\begin{align}
\pd{ }{\lambda_1}\left( -\frac{PH[8\alpha\lambda_1 - 2\mcal{E}\lambda_1]}{R_b[4\lambda_1 + 4\alpha\lambda_1^2 - \mcal{E} \lambda_1^2]^2} \right)=   -\frac{PH}{R_b}\left[ \frac{8\alpha - 2 \mcal{E}}{[4\lambda_1 + 4 \alpha \lambda_1^2 - \mcal{E}\lambda_1^2]^2} - \frac{2[8\alpha\lambda_1 - 2\mcal{E}\lambda_1]^2}{[4\lambda_1 + 4 \alpha \lambda_1^2 - \mcal{E}\lambda_1^2]^3} \right],
\end{align}
\begin{align}
\pd{\mcal{C}}{\lambda_1} =& \frac{ \frac{P^2 H^2}{R_b^2} + 96\alpha^2\lambda_1^2 - 24\mcal{E}\alpha\lambda_1^2 + 32\alpha  - 4 \mcal{E} }{\left[ \sqrt{\frac{P^2 H^2}{R_b^2} {\lambda_1^2}- 4\alpha\mcal{E}\lambda_1^4 + 16\alpha^2\lambda_1^4 + 32\alpha {\lambda_1^2}- 4\mcal{E}{\lambda_1^2}+{16}}\right] \left[4\lambda_1 + 4\alpha\lambda_1^3 - \mcal{E} \lambda_1^3\right]} \nonumber \\
& - \frac{\left[\frac{P^2 H^2}{R_b^2}\lambda_1 + 32\alpha^2\lambda_1^3 - 8\mcal{E}\alpha\lambda_1^3 + 32\alpha \lambda_1 - 4 \mcal{E} \lambda_1\right]^2}
{\left[ \sqrt{\frac{P^2 H^2}{R_b^2} {\lambda_1^2}- 4\alpha\mcal{E}\lambda_1^4 + 16\alpha^2\lambda_1^4 + 32\alpha {\lambda_1^2}- 4\mcal{E}{\lambda_1^2}+{16}}\right]^3 \left[4\lambda_1 + 4\alpha\lambda_1^3 - \mcal{E} \lambda_1^3\right]} \nonumber \\
&- \frac{\left[\frac{P^2 H^2}{R_b^2}\lambda_1 + 32\alpha^2\lambda_1^3 - 8\mcal{E}\alpha\lambda_1^3 + 32\alpha \lambda_1 - 4 \mcal{E} \lambda_1\right] \left[ 12 \alpha \lambda_1^2 -3\mcal{E}\lambda_1^2 + 4 \right]}{\left[ \sqrt{\frac{P^2 H^2}{R_b^2} {\lambda_1^2}- 4\alpha\mcal{E}\lambda_1^4 + 16\alpha^2\lambda_1^4 + 32\alpha {\lambda_1^2}- 4\mcal{E}{\lambda_1^2}+{16}}\right] \left[4\lambda_1 + 4\alpha\lambda_1^3 - \mcal{E} \lambda_1^3\right]^2},
\end{align}
and
\begin{align}
\pd{\mcal{D}}{\lambda_1}=&\frac{\left[\frac{P H}{R_b}\lambda_1 + \sqrt{\frac{P^2 H^2}{R_b^2} {\lambda_1^2}- 4\alpha\mcal{E}\lambda_1^4 + 16\alpha^2\lambda_1^4 + 32\alpha {\lambda_1^2}- 4\mcal{E}{\lambda_1^2}+{16}} \right]
[(24\alpha -6\mcal{E})\lambda_1]
}{[4\lambda_1 + 4\alpha\lambda_1^3 - \mcal{E} \lambda_1^3]^2} \nonumber \\
&+ \frac{\left[\frac{P H}{R_b} +\frac{\frac{P^2 H^2}{R_b^2}\lambda_1 + 32\alpha^2\lambda_1^3 - 8\mcal{E}\alpha\lambda_1^3 + 32\alpha \lambda_1 - 4 \mcal{E} \lambda_1}{\sqrt{\frac{P^2 H^2}{R_b^2} {\lambda_1^2}- 4\alpha\mcal{E}\lambda_1^4 + 16\alpha^2\lambda_1^4 + 32\alpha {\lambda_1^2}- 4\mcal{E}{\lambda_1^2}+{16}}} \right][12\alpha\lambda_1^2 - 3\mcal{E}\lambda_1^2+4]}{[4\lambda_1 + 4\alpha\lambda_1^3 - \mcal{E} \lambda_1^3]^2} \nonumber \\
&-2 \frac{\left[\frac{P H}{R_b}\lambda_1 + \sqrt{\frac{P^2 H^2}{R_b^2} {\lambda_1^2}- 4\alpha\mcal{E}\lambda_1^4 + 16\alpha^2\lambda_1^4 + 32\alpha {\lambda_1^2}- 4\mcal{E}{\lambda_1^2}+{16}} \right][12\alpha\lambda_1^2 - 3\mcal{E}\lambda_1^2+4]^2}{[4\lambda_1 + 4\alpha\lambda_1^3 - \mcal{E} \lambda_1^3]^3}.
\end{align}
\begin{align}
\fd{}{\theta} \left( \frac{\partial \Omega^\ast}{ \partial\lambda_1}\right) = C_1\mcal{Z}\mcal{W}_1 \vr_{\theta \theta} + C_1\mcal{Z}\mcal{W}_2\eta_{\theta \theta}
\end{align}
where 
\begin{align}
\mcal{Z} =&  \left[ \left[   2 + \frac{6}{\lambda_1^4 {\lambda_1^\ast}^2} + \frac{2}{\lambda_1^3 {\lambda_1^\ast}^4} \frac{\partial ({\lambda^\ast_2}^2)}{\partial \lambda_1}  \right] + \left[ \frac{2}{\lambda_1^3 {\lambda_1^\ast}^4} + \frac{2}{\lambda_1^2 {\lambda_1^\ast}^6} \frac{\partial ({\lambda^\ast_2}^2)}{\partial \lambda_1} \right]\frac{\partial ({\lambda^\ast_2}^2)}{\partial \lambda_1} + \left[ 1 - \frac{1}{\lambda_1^2 {\lambda^\ast_2}^4} \right] \frac{\partial^2 ({\lambda^\ast_2}^2)}{\partial \lambda_1^2} \right] \nonumber \\
&+ \alpha \left[ \left[ \frac{6}{\lambda_1^4} + 2\lambda_1\frac{\partial ({\lambda^\ast_2}^2)}{\partial \lambda_1}  + 2{\lambda^\ast_2}^2 \right]  +  \left[ \frac{2}{{\lambda^\ast_2}^6}  \frac{\partial ({\lambda^\ast_2}^2)}{\partial \lambda_1} + 2\lambda_1 \right] \frac{\partial ({\lambda^\ast_2}^2)}{\partial \lambda_1}  + \left[\lambda_1^2 -\frac{1}{{\lambda_2^\ast}^4} \right] \frac{\partial^2 ({\lambda^\ast_2}^2)}{\partial \lambda_1^2} \right] \nonumber \\
&+\frac{\mcal{E}}{4} \left[2{\lambda^\ast_2}^2  + 4\lambda_1  \frac{ \partial({\lambda^\ast_2}^2)}{\partial \lambda_1} +  \lambda_1^2 \frac{\partial^2 ({\lambda^\ast_2}^2)}{\partial \lambda_1^2} \right] .
\end{align}
Thus
\begin{align}
\fd{}{\theta} \left( \pd{\Omega^\ast}{\vr_\theta} \right) 
=\left[ \mcal{U}_1 \frac{\partial \Omega^\ast}{ \partial\lambda_1}+C_1\mcal{Z} \mcal{W}_1^2\right] \vr_{\theta \theta} + \left[ \mcal{U}_2 \frac{\partial \Omega^\ast}{ \partial\lambda_1}+C_1\mcal{Z}\mcal{W}_2 \mcal{W}_1 \right] \eta_{\theta \theta} \\
\fd{}{\theta} \left( \pd{\Omega^\ast}{\eta_\theta} \right) 
=\left[ \mcal{V}_1 \frac{\partial \Omega^\ast}{ \partial\lambda_1}+C_1\mcal{Z}\mcal{W}_1 \mcal{W}_2\right] \vr_{\theta \theta} + \left[ \mcal{V}_2 \frac{\partial \Omega^\ast}{ \partial\lambda_1}+C_1\mcal{Z}\mcal{W}_2^2 \right] \eta_{\theta \theta}
\end{align}
One introduces a term $\mcal Y$ to cancel out the material property $C_1$:
\begin{align}
\mcal{Y} =  \frac{\partial \Omega^\ast}{ \partial\lambda_1} / C_1.
\end{align}
The governing equations are now written as ODEs:
\begin{align}
[1 + \gamma \cos \theta] \bigg[ \left[ \mcal{U}_1\mcal{Y}+\mcal{Z} \mcal{W}_1^2 \right] \vr_{\theta \theta} + \left[ \mcal{U}_2 \mcal{Y}+\mcal{Z}\mcal{W}_2 \mcal{W}_1 \right] \eta_{\theta \theta} \bigg]  -\sin \theta  {\vr_\theta}[\vr_\theta^2 + \eta_\theta^2]^{-\frac{1}{2}} \mcal{Y}+ \frac{{P}  \vr \eta_\theta}{\gamma} = 0 \\
[1 + \gamma \cos \theta] \bigg[ \left[ \mcal{V}_1\mcal{Y}+\mcal{Z}\mcal{W}_1 \mcal{W}_2\right] \vr_{\theta \theta} + \left[ \mcal{V}_2 \mcal{Y}+\mcal{Z} \mcal{W}_2^2 \right] \eta_{\theta \theta} \bigg] -\sin \theta  {\eta_\theta}[\vr_\theta^2 + \eta_\theta^2]^{-\frac{1}{2}} \mcal{Y} - \frac{{P} \vr\vr_\theta}{ \gamma}  = 0.
\end{align}
The coefficients in ODEs system~(34) are modified as
\begin{align}
A_1^\ast = [1 + \gamma \cos \theta] \left[ \mcal{U}_1\mcal{Y}+\mcal{Z} \mcal{W}_1^2 \right], \nonumber \\
A_2^\ast  = [1 + \gamma \cos \theta] \left[ \mcal{U}_2\mcal{Y}+\mcal{Z} \mcal{W}_1 \mcal{W}_2 \right], \nonumber \\
A_3^\ast  = -\sin \theta  {\vr_\theta}[\vr_\theta^2 + \eta_\theta^2]^{-\frac{1}{2}} \mcal{Y}+ \frac{{P}  \vr \eta_\theta}{\gamma},
\end{align}
and
\begin{align}
B_1^\ast  = [1 + \gamma \cos \theta] \left[ \mcal{V}_1\mcal{Y}+\mcal{Z} \mcal{W}_1\mcal{W}_2 \right], \nonumber \\
B_2^\ast  = [1 + \gamma \cos \theta] \left[ \mcal{V}_2\mcal{Y}+\mcal{Z} \mcal{W}_2^2 \right], \nonumber \\
B_3^\ast  = -\sin \theta  {\eta_\theta}[\vr_\theta^2 + \eta_\theta^2]^{-\frac{1}{2}} \mcal{Y} - \frac{{P} \vr\vr_\theta}{ \gamma}.
\end{align}

\end{document}